\documentclass[aps,prd,10pt,twocolumn,superscriptaddress,floatfix,nofootinbib,amsmath,amssymb,altaffilletter,preprintnumbers]{revtex4-1}
\usepackage[colorlinks=true,pdfstartview=FitV,breaklinks=true]{hyperref}

\usepackage[utf8]{inputenc}
\usepackage{graphicx}
\usepackage{booktabs}
\usepackage{tabularx}
\usepackage{textcomp}
\usepackage{xspace}
\usepackage{slashed}
\usepackage{amsmath}
\usepackage{amsfonts}
\usepackage{amssymb}
\usepackage{upgreek}
\usepackage{comment}
\usepackage[dvipsnames,table]{xcolor}
\usepackage[normalem]{ulem}
\usepackage{csquotes}
\usepackage{bbold}
\hypersetup{urlcolor=BlueViolet,
	    citecolor=Plum,
	    linkcolor=PineGreen}
\usepackage[official]{eurosym}
\definecolor{lightgray}{rgb}{0.9,0.9,0.9}	    
\definecolor{green}{rgb}{0,0.5,0}
\definecolor{red}{rgb}{1,0,0}
\definecolor{blue}{rgb}{0,0,0.5}

\begin{document}

\title{Prospects for relic neutrino detection using nuclear spin experiments}

\author{Yeray Garcia del Castillo}\email{y.garcia\textunderscore del\textunderscore castillo@unsw.edu.au}
\author{Giovanni Pierobon}\email{g.pierobon@unsw.edu.au}
\author{Dipan Sengupta}\email{dipan.sengupta@unsw.edu.au}
\author{Yvonne Y. Y. Wong}\email{yvonne.y.wong@unsw.edu.au}
\affiliation{Sydney Consortium for Particle Physics and Cosmology\\
School of Physics, The University of New South Wales, Sydney NSW 2052,  Australia}

\preprint{CPPC-2025-06}

\date{\today}
\smallskip
\begin{abstract}
Direct detection of the cosmic neutrino background (C$\nu$B) remains one of the most formidable experimental challenges in modern physics. 
In this work, we extend recent studies of C$\nu$B-induced coherent transitions in polarised nuclear spin ensembles. Adopting an open quantum system framework, we model coherent neutrino effects in large spin ensembles using a Lindblad master equation that also incorporates realistic experimental imperfections such as local dephasing and imperfect polarisation.  We solve the Lindblad equation numerically by way of a fast and computationally inexpensive method that can be extended to an arbitrarily large number of spins.  Using our numerical solutions, we forecast the sensitivities of future experiments such as CASPEr to the local C$\nu$B overdensity parameter $\delta_\nu$. Our findings indicate that a CASPEr-like experiment, though primarily aimed at axion dark matter search, could also constrain the C$\nu$B overdensity to $\delta_\nu \sim 10^{13}$ in configurations achievable by currently planned experimental efforts, and down to $\delta_\nu \sim 10^{11}$ in the most optimised scenario. While C$\nu$B detection remains out of reach in the foreseeable future, our results highlight the potential of using quantum sensing to probe fundamental physics.
\end{abstract}

\maketitle

\section{Introduction}


Standard hot big bang cosmology predicts a cosmic neutrino background (C$\nu$B), a relic from the early universe analogous to the cosmic microwave background (CMB), that decoupled from the primordial plasma when it cooled to temperatures of $T\sim 1$~MeV~\cite{ParticleDataGroup:2024cfk}. At decoupling these neutrinos had a nearly thermal spectrum of the relativistic Fermi-Dirac form $n_F = [\exp(p/T_\nu)+1]^{-1}$, which is preserved by cosmological redshifting induced by the spatial expansion of the universe. Together with the minimum neutrino mass values implied by flavour oscillation experiments, we expect the C$\nu$B today 
to be largely non-relativistic, with a temperature of $T_{\nu,0}\simeq 1.7 \times 10^{-4}~{\rm eV} \simeq 1.95$~K, slightly cooler than the $2.73$~K  CMB photons.  Its mean number density is predicted to be $n_\nu = n_{\bar{\nu}} \simeq 56~{\rm cm}^{-3}$  per flavour, but can be enhanced locally, i.e., in the solar neighbourhood, by factors of ${\cal O}(1-10)$, depending on the neutrino mass, through gravitational clustering~\cite{Singh:2002de,Ringwald:2004np,Brandbyge:2010ge, Villaescusa-Navarro:2011loy,LoVerde:2013lta,deSalas:2017wtt,Zhang:2017ljh, Mertsch:2019qjv, Holm:2023rml,Zimmer:2023jbb, Zimmer:2024max, Holm:2024zpr,Worku:2024kwv}. 

Nothwithstanding its abundance, the C$\nu$B  has so far evaded direct detection in the laboratory owing to its extremely low energies and weak interactions.  To date, the only evidence we have of its existence is indirect and based entirely on its gravitational effects in the early universe and subsequent impact on precision cosmological observables.  Specifically, measurements of the primordial light element abundances (e.g.,~\cite{Pitrou:2018cgg}) and the CMB anisotropies (e.g.,~\cite{Planck:2018vyg}) consistently prefer an expansion rate---at minutes and 400,000~years post-big bang, respectively---compatible at the 10\% level with the standard prediction of the C$\nu$B energy density.%
\footnote{In cosmology parlance, the effect of the C$\nu$B energy density on the early universal expansion rate is parameterised by the effective number of neutrinos $N_{\rm eff}$. The inferred value of $N_{\rm eff}$ from primordial light element abundances and the CMB are $N_{\rm eff} = 2.88 \pm 0.27$ (68\% C.I.)~\cite{Pitrou:2018cgg} and $N_{\rm eff} = 2.99 \pm 0.17$~(68\% C.I.)~\cite{Planck:2018vyg}, respectively, consistent with the energy density in three light neutrino flavours.} 
Anisotropic stress constraints from the CMB likewise support the weakly-interacting nature of the C$\nu$B~\cite{Oldengott:2017fhy,Chen:2022idm}.

In the face of ever-improving cosmological/gravity-based constraints, it nonetheless remains of great interest to detect or at least constrain the C$\nu$B in a more direct way, using methods based on particle scattering.  In this connection, a number of possible probes and constraints have been discussed.  Below is a non-exhaustive list.
\begin{itemize}
 \item At leading order in the Fermi constant $G_F$, the so-called Stodolsky effect~\cite{Stodolsky:1974aq} refers to a spin-dependent shift in the energy of a standard-model (SM) fermion sitting in a $CP$ asymmetric bath of neutrinos and antineutrinos.  While cosmological neutrino-antineutrino asymmetries of $\eta_\nu \equiv (n_\nu-n_{\bar{\nu}})/n_\gamma \lesssim {\cal O}(10^{-2})$ are not excluded by observations~(e.g.,~\cite{Oldengott:2017tzj}), it is also not well motivated within the SM, where ${\cal O}(10^{-10})$ particle-antiparticle asymmetries  are generically expected in light of the observed matter-antimatter asymmetry in the universe.  We remark however on a recent claim of a large $\eta_\nu \sim 10^{-2} - 0.25$ being preferred by extragalactic $^4$He observations~\cite{Burns:2022hkq}. Note also discussions of the survival of such asymmetries in the local universe~\cite{Arvanitaki:2022oby,Ruchayskiy:2022eog}.  
 
\item At next-to-leading order ${\cal O}(G_F^2)$, coherent elastic scattering of a C$\nu$B wind with a macroscopic target can produce a mechanical force that causes the target to accelerate~\cite{Shvartsman:1982sn,Langacker:1982ih, Smith:1983jj,Duda:2001hd,Domcke:2017aqj,Shergold:2021evs}. 
This force does not require the presence of a neutrino-antineutrino asymmetry, but works best for Dirac neutrinos.  For Majorana neutrinos, the absence of vector currents causes the interaction cross section to be suppressed by a factor  $v^2 \sim 10^{-6}$, where $v$ is the Earth-C$\nu$B relative speed. Current torsion balance setups cannot measure this acceleration unless the C$\nu$B number density is enhanced by a factor $\delta_\nu \sim 10^{12}$ or more~\cite{Shergold:2021evs}.
Advancement on this front consists in using new techniques, e.g., interferometry~\cite{Domcke:2017aqj}, to enhance the sensitivity to the acceleration.

\item  Another ${\cal O}(G_F^2)$ effect, neutrino capture by $\beta$-decaying nuclei remains unique to C$\nu$B detection~\cite{Weinberg:1962zza,Cocco:2007za,Long:2014zva,Akita:2020jbo} and goes hand-in-hand with direct neutrino mass searches based upon $\beta$-decay end-point spectrum measurements~\cite{KATRIN:2022kkv,KATRIN:2024cdt,Project8:2022wqh}.  The flagship tritium $\beta$-decay neutrino mass experiment KATRIN currently limits the local C$\nu$B overdensity to $\delta_\nu \lesssim 10^{11}$~\cite{KATRIN:2022kkv}.  The proposed PTOLEMY experiment aims to constrain this parameter to $\delta_\nu \sim {\cal O}(1)$~\cite{PTOLEMY:2018jst}.

\item An accelerator-based variant of the above has also been discussed in Ref.~\cite{Bauer:2021uyj}, which proposes to use accelerated beams of ions in ion storage rings to capture C$\nu$B neutrinos at nuclear $\beta$-resonances to enhance the capture cross section.

\item Inelastic ${\cal O}(G_F^2)$ scattering of the C$\nu$B on a target system can also induce transitions between different energy states. These include atomic transitions~\cite{Yoshimura:2014hfa,Huang:2025yqu}, as well as nuclear spin flips~\cite{Bauer:2025hdq,Arvanitaki:2024taq}.  Of the latter, Hydrogen spin flip induced by absorption of a relic neutrino-antineutrino pair may be sensitive to $\delta_\nu \sim 10^{9}$ for neutrino masses $m_\nu \sim 10^{-3}$~eV~\cite{Bauer:2025hdq}.

\item On astrophysical scales, probes of the C$\nu$B range from small-scale (i.e., sub-1~AU) perturbations to planetary orbits (current constraint  $\delta_\nu \lesssim  5.34\times 10^{10}$~\cite{Aliberti:2024udm}) and the cooling of old neutron stars from C$\nu$B capture (potential sensitivity $\delta_\nu \sim 10^9$~\cite{Das:2024thc}), to large-scale  ($\gtrsim {\cal O}(10)$~Mpc) up-scattering of the C$\nu$B by cosmic rays~\cite{Ciscar-Monsalvatje:2024tvm, DeMarchi:2024zer,Herrera:2024upj} (current constraint $\delta_\nu \lesssim 10^{4}$~\cite{Herrera:2024upj}) and features in the energy spectra of ultra-high energy neutrinos due to C$\nu$B scattering (current constraint $\delta_\nu \lesssim 10^{14}$~\cite{Franklin:2024amy}).  Also proposed are resonant absorption dips in the
cosmogenic neutrino energy spectrum when these neutrinos interact with the C$\nu$B on a meson resonance~\cite{Brdar:2022kpu}, a variant of the original $Z$-resonance absorption dips for ultra-high energy neutrinos~\cite{Weiler:1982qy}.

\end{itemize}

We are particularly interested in the proposal of Ref.~\cite{Arvanitaki:2024taq}, which exploits spin ensembles such as in  
nuclear magnetic resonance (NMR) experiments to detect coherent effects induced by the C$\nu$B through inelastic neutrino-spin interactions.  Specifically, Ref.~\cite{Arvanitaki:2024taq} argued that the momentum transfer $q$ in such interactions can be macroscopic in wavelength terms,
causing the collective C$\nu$B-induced spin flip rates across the spin ensemble to scale as $N^2$, where $N$ is the number of spins, if the spin sample's physical size $R$ is smaller than or comparable to~$q^{-1}$  (this situation is to be compared with an $\propto N$ scaling in an incoherent, $qR \gg 1$ setting).  Such macroscopic coherence of the target system can significantly enhance sensitivity to weak interactions and may conceivably yield a constraint on the C$\nu$B overdensity parameter comparable to the current KATRIN bound, $\delta_\nu \lesssim 10^{11}$~\cite{KATRIN:2022kkv}, for specific initial spin configurations~\cite{Arvanitaki:2024taq}.  
Added to its appeal is that planned NMR experiments to search for dark matter axions and axion-like particles, e.g., CASPEr~\cite{Budker:2013hfa, JacksonKimball:2017elr}, are in principle sensitive to such coherent signals. 
Besides opening up a new avenue for ultra-low-threshold searches for fundamental physics, the potential for these experiments to simultaneously search for and/or constrain both the axion and C$\nu$B parameter space is a most tantalising prospect.

In this work, we expand on the study of Ref.~\cite{Arvanitaki:2024taq} as follows: (i) We explore realistic experimental effects such as local dephasing and imperfect initial polarisation of the spin ensemble, both of which can degrade the coherent C$\nu$B signature. (ii) Modelling these effects using an open quantum system framework, we solve the Lindblad master equation numerically, both for the full density matrix of $N$ spins, and using a computationally efficient second-order approximation that tracks only a limited set of observables. Using these solutions, we are able to confirm for large $N$ the validity of the perturbative solutions put forward in Ref.~\cite{Arvanitaki:2024taq}.  (iii) We forecast the sensitivity of realistic future experiments to the C$\nu$B overdensity parameter $\delta_\nu$, especially the reach of the CASPEr experiment~\cite{Budker:2013hfa, JacksonKimball:2017elr,Walter:2025ldb}, by including a more realistic modelling of the instrumental noise, initial polarisation of the spin ensemble, and sample size. 

In addition to the sample size, which restricts the degree of C$\nu$B coherence across the spin ensemble, we find the initial spin polarisation to be a particularly important limiting factor---and also the primary reason why the current generation of NMR axion searches cannot place competitive constraints on $\delta_\nu$.  In the most optimistic case (no instrumental noise, $R \sim 10$~cm, 100\% polarisation), a sensitivity of $\delta_\nu \sim 10^{11}$ may be achievable using $^{129}$Xe nuclei. In more realistic settings (instrumental noise, $R \sim 1$~cm, 25\% polarisation), however, the sensitivity to the C$\nu$B deteriorates to $\delta_\nu \sim 10^{13}$. 

The paper is structured as follows. In Sec.~\ref{sec:nuspin} we introduce the scattering rates of the C$\nu$B with a single two-level system, such as a nuclear spin with $I=1/2$, and with an ensemble of $N$ spins, highlighting a regime where a coherent $\sim N^2$ enhancement can be achieved. In Sec.~\ref{sec:oqs}, we review the Lindblad master equation commonly used to model open quantum systems, and describe the dynamics in the Dicke basis. Sections~\ref{sec:coherentobservables} and \ref{sec:local} discuss, respectively, the emergence of $N^2$ effects in collective observables and the limitations arising from incoherent local effects that eventually degrade the signal.  We show in Sec.~\ref{sec:numerical} how the Lindblad master equation combining coherent and incoherent local effects can be efficiently solved using a numerical approximation. We use these numerical solutions to forecast the sensitivity of currently-planned experiments to the C$\nu$B overdensity parameter $\delta_\nu$, as detailed in Sec.~\ref{sec:futureconstraints}.  Section~\ref{sec:conclusions} contains our conclusions.  The computational details of this work are reported in eight appendices.


\section{Neutrino-spin interaction}\label{sec:nuspin}

Consider a spin-$1/2$ fermion---such as an electron or a nucleus---interacting with neutrinos via the weak interaction. We assume coherence over nuclear distances, allowing us to treat the fermion as a point-like particle and neglect its internal structure. At low energies, this interaction can be described by an effective 4-fermi interaction specified by the Lagrangian 
\begin{equation}
\begin{aligned}
    {\cal L} &= \bar{\Psi}(i\slashed{\partial} - m)\Psi - q\bar{\Psi}\gamma^\mu \Psi A_\mu  + {\cal L}_{\rm int}\, ,\\
    {\cal L}_{\rm int}&=-\sum_f \frac{G_F}{\sqrt{2}} \bar{\Psi} \gamma^\mu (g^f_V - g^f_A \gamma^5) \Psi \, \bar{\nu}_f \gamma_\mu (1 - \gamma^5) \nu_f \,,
    \label{eq:4fermi}
\end{aligned}
\end{equation}
where $\Psi$ denotes the fermion field of mass $m$ and electric charge $q$, $A_\mu = (A_0, \vec{A})$ is  the electromagnetic 4-potential,  $G_F$ the Fermi constant,  $\nu_f$ the neutrino field of flavour $f = e, \mu, \tau$, and $g_V^f$ and $g_A^f$ are the vector and axial-vector couplings of the fermion to $\nu_f$.

The couplings $g_V^f$ and $g_A^f$ generally depend on the neutrino flavour as well as the nature of the fermion.  Here, we focus exclusively on neutrino–nucleus interaction through the neutral-current channel. The corresponding couplings are therefore independent of flavour. The values of $g_V$ and $g_A$ are nucleus-dependent. In what follows, we consider the axial coupling $g_A$, which is only sensitive to the spin of the unpaired nucleon (see, e.g.,~\cite{bohr1998nuclear}). 

It is convenient to express the neutrino fields in the mass basis.
Performing the basis transformation $\nu_f = \sum_i U_{fi} \nu_i$, where $\nu_i$ are the mass eigenstates (for $i = 1,2,3$) and $U_{fi}$ is the Pontecorvo-Maki-Nakagawa-Sakata matrix, the interaction Lagrangian becomes
\begin{equation}
  \mathcal{L}_{\rm int} = -\sum_{ij} \frac{G_F}{\sqrt{2}} \bar{\Psi} \gamma^\mu \left(g^{i,j}_V - g^{i,j}_A \gamma^5\right) \Psi \, \bar{\nu}_i \gamma_\mu (1 - \gamma^5) \nu_j \,,
  \label{eq:lint2}
\end{equation}
where the effective couplings in the mass basis are given by $g^{i,j}_{V(A)} = \sum_f g_{V(A)}^f U_{fi}^* U_{fj}$. In the flavour-independent case under consideration,
the effective couplings are diagonal, i.e., non-zero only for $i = j$, and equal for all mass eigenstates.%
\footnote{If instead of nuclear spins we consider electron spins, then the neutrino-spin interaction can also proceed via the charged-current channel.  In that case, the couplings $g_{V(A)}^{i,j}$ generally contain non-zero off-diagonal terms, allowing the neutrino to change mass eigenstates via the interaction.  This case has been considered in Ref.~\cite{Arvanitaki:2024taq}.}

In a typical NMR setup, a sample of (non-relativistic) nuclear spins is placed in a constant magnetic bias field $\vec{B} = \nabla \times \vec{A}$, which we choose to point in the negative-$z$ direction, i.e., $\vec{B} = - B\hat{z}$. We can therefore take the non-relativistic limit and derive from the relativistic Lagrangian~\eqref{eq:4fermi} the corresponding Hamiltonian for a single spin,
\begin{equation}
\begin{aligned}
    H^{(1/2)}&=H_S^{(1/2)}+H_{\rm int}^{(1/2)} \\
    &=\omega_0J_z+\Sigma_zJ_z+\Sigma_-J_-+\Sigma_+J_+\, .
    \label{h1}
\end{aligned}
\end{equation}
Here, $H_S^{(1/2)} = \omega_0 J_z$ describes the interaction of the spin with the magnetic field, where $\omega_0=\gamma B$ is the energy splitting between adjacent spin states---also called the Larmor frequency, and $\gamma=g\, \mu_N$ is the gyromagnetic ratio specific to the nucleus under consideration, defined via the nuclear magneton $\mu_N=e/2m_p$ and the nucleus-specific $g$-factor. 
The $H_{\rm int}^{(1/2)}$ terms describe interaction of the spin with the C$\nu$B, where the spin raising and lowering operators are defined as $J_\pm \equiv J_x \pm {\rm i} J_y$. In our flavour-independent case the neutrino terms $\Sigma_{\pm,z}$ read 
\begin{equation}
\begin{aligned}
    \Sigma_z&=\sum_i \sqrt{2}G_Fg_A\bar{\nu}_i\gamma^3(1-\gamma^5)\nu_i\, ,\\
    \Sigma_\pm&= \sum_i \frac{G_F}{\sqrt{2}}g_A\bar{\nu}_i(\gamma^1\mp {\rm i}\gamma^2)(1-\gamma^5)\nu_i\, ,
    \label{eq:nuterms}
\end{aligned}
\end{equation}
and they arise exclusively from the axial part of the weak interaction. We refer to the reader to Appendix~\ref{app:ham} for the derivation of the low-energy Hamiltonian, including the flavour-dependent case.

Physically, the term $\Sigma_z J_z$ corresponds to elastic scattering of neutrinos off a spin, which alters the phase evolution of the spin but does not change its energy. In contrast, the $\Sigma_\pm J_\pm$ terms describe inelastic processes, in which the neutrino exchanges energy with the spin, either exciting it~($\Sigma_+$) or de-exciting it~($\Sigma_-$).  The single-spin excitation and de-excitation rates $\gamma_\pm$, defined as the  transition rate of a spin from down to up~($+$) and vice versa~($-$), can be estimated using Fermi's golden rule.  Accounting only for inelastic neutrino-spin scattering, the rates read 
\begin{equation}
\begin{aligned}
    \gamma_{\pm}=&\frac{ 4 G_F^2}{(2 \pi)^3}  g_A^2 \sum_i
    \int_{p^{\rm low}_{i\pm}}^\infty  {\rm d} |\vec{p}|\;|\vec{p}|^2\, p_{i\pm}\, \sqrt{m_i^2+|\vec{p}|^2_{i \pm}}  \\
    &\times\left[(1-N_{i}(p^0_{i\pm})) \, N_{i}(p^0)  +(1-\bar{N}_{i}(p^0_{i\pm}))\, \bar{N}_{i}(p^0)\right]\,,
    \label{eq:gamma+-heuristic}
\end{aligned}
\end{equation}
where $p_{i\pm}=[(\sqrt{|\vec{p}|^2+m_i^2}\mp \omega_0)^2-m_i^2]^{1/2}$, $p^0_{i\pm}=\sqrt{|\vec{p}|^2+m_i^2} \mp \omega_0$, $m_i$ is the mass of the $i$th mass eigenstate, and the lower integration limits are  given by $p^{\rm low}_{i+}=\sqrt{(\omega_0+ m_i)^2-m_i^2}$ and $p^{\rm low}_{i-}=0$.%
\footnote{All neutrino momenta throughout this work are 3-momenta.  We do however use a superscript ``0'', e.g., $p^0 \equiv \sqrt{|\vec{p}|^2+m^2}$ to denote energy.}
 This expression applies to both Dirac and Majorana neutrinos, but with the understanding that the neutrino and antineutrino occupation numbers, $N_i$ and $\bar{N}_i$, in the Dirac case are a factor two smaller than in the Majorana case.%
\footnote{Although at decoupling relativistic Dirac neutrinos (antineutrinos) are exclusively in the left- (right-) helicity state, as they redshift to non-relativistic velocities at late times, the generic expectation is  they should relax into an equal mixture of left- and right-helicity states via gravitational interactions.}
The detailed calculation of $\gamma_{\pm}$ from the low-energy Hamiltonian~\eqref{h1} can  be found in Appendix~\ref{app:rates}. 
Assuming a standard, $CP$-symmetric C$\nu$B with $N_i = \bar{N}_i \sim n_F$, and a spin sample consisting of  $^{129}\mathrm{Xe}$ nuclei with $g_A= 1/2$, the rates~\eqref{eq:gamma+-heuristic} evaluate numerically to
\begin{equation}
 \gamma_-^{\rm SM} \simeq 1.1\times 10^{-48}~{\rm Hz}, \quad \frac{\gamma_+^{\rm SM}}{\gamma_-^{\rm SM}}\simeq 0.994\,, 
 \label{eq:gammanumbers}
\end{equation} 
for  a typical NMR energy splitting of  $\omega_0=1 \times 10^{-8}$~eV and a normally-ordered neutrino mass sum of $\sum m_\nu=0.15$~eV.%
\footnote{We adopt mass values consistent with global neutrino oscillation fits~\cite{Esteban:2020cvm,deSalas:2020pgw,Capozzi:2017ipn}. 
For normal ordering, the masses are fixed by the total mass sum and the measured mass-squared splittings, $\Delta m_{21}^2 \simeq 7.4\times 10^{-5}~\mathrm{eV}^2$ and $\Delta m_{31}^2 \simeq 2.5\times 10^{-3}~\mathrm{eV}^2$, giving $m_1 \simeq 0.0419~\mathrm{eV}$, $m_2 \simeq 0.0428~\mathrm{eV}$, and $m_3 \simeq 0.0652~\mathrm{eV}$ for $\sum m_\nu = 0.15$~eV.}

In the case of $N$ spins, the low-energy Hamiltonian~\eqref{h1} generalises to 
\begin{equation}
H =\sum_\alpha \left( \omega_0 J_z^\alpha +\Sigma_z^\alpha J_z^\alpha +\Sigma_-^\alpha J_-^\alpha+\Sigma_+^\alpha J_+^\alpha \right)\, ,
\label{eq:Nham}
\end{equation}
where $J_{\pm,z}^\alpha$ are understood to operate only on the spin at spatial position $\vec{x}_\alpha$, and the neutrino terms $\Sigma^\alpha_{\pm,z}$ are the same as those given in Eq.~\eqref{eq:nuterms} but now with the neutrino fields specified by $\nu_i(t,\vec{x}_\alpha)$.  The total excitation and de-excitation rates of the ensemble, defined here as the rate at which the total energy of the ensemble increases or decreases by one unit of $\omega_0$, can be written as%
\footnote{Throughout this work, we always denote integrated (ensemble-level) rates by an upper-case $\Gamma$, while single-spin interaction rates---relevant for the Lindblad dynamics discussed in Sec.~\ref{sec:oqs}---are represented by a lower-case $\gamma$.}
\begin{equation}
\begin{aligned}
\Gamma_{\pm}
    =&\;\frac{ 2 G_F^2}{(2 \pi)^3}  g_A^2 \sum_i
    \int_{p^{\rm low}_{i\pm}}^\infty  {\rm d} |\vec{p}|\;|\vec{p}|^2\, p_{i\pm}\, \sqrt{m_i^2+p^2_{i \pm}} \\  
&\times   \left[(1-N_{i}(p^0_{i\pm})) \, N_{i}(p^0)  +(1-\bar{N}_{i}(p^0_{i\pm}))\, \bar{N}_{i}(p^0)\right] \\
& \times 
\int_{-1}^1 {\rm d} \mu \; F_\pm(\vec{p}_{i\pm}-\vec{p})\, ,
\label{eq:biggamma}
    \end{aligned}
\end{equation}
where $\mu \equiv \hat{p}_{i\pm} \cdot \hat{p}$, is the polar angle between the momentum vectors $\vec{p}$ and $\vec{p}\,'$, and 
\begin{equation}
F_\pm(\vec{q}) \equiv \langle {\rm i}_{\rm spin}| \sum_{\alpha,\beta} J_\mp^\beta J_\pm^\alpha e^{-{\rm i} \vec{q}\cdot (\vec{x}_\alpha-\vec{x}_\beta)}|{\rm i}_{\rm spin}\rangle\,
\label{eq:formfactormaintext}
\end{equation} 
are the excitation and de-excitation form factors.

Depending on the initial spin configuration and the momentum transfer $\vec{q} \equiv\vec{p}_{i\pm}-\vec{p}$ from the neutrinos to the ensemble, the form factors
$F_\pm(\vec{q})$ can evaluate to a range of values.  For example, a system prepared in the ground state $\vert G\rangle\equiv\vert\downarrow\rangle^{\otimes N}$ in which all spins align initially with the magnetic field has 
$F^G_+(\vec{q})=N$ and $F^G_-(\vec{q})= 0$, and hence a non-vanishing total excitation rate $\Gamma_+=\gamma_+N$. In contrast, for an initial spin state aligned with the equatorial plane, $\vert P\rangle= \prod_\alpha \frac{1}{\sqrt{2}}(\vert\uparrow\rangle+\vert\downarrow\rangle)$, the form factors evaluate to
\begin{align}
F^P_\pm(\vec{q}) 
&= \frac{N}{2} + \frac{N^2}{4} \frac{9}{q^2 R^2} \left|j_1\left(q R\right) \right|^2,
 \label{eq:formfactorpol}
\end{align}
with $j_1(x)$ the spherical Bessel function of order one.  Clearly, $F^P_\pm(\vec{q})$ can range from ${\cal O}(N)$ in the fully incoherent regime, to ${\cal O}(N^2)$ in the fully coherent regime, discussed in the following subsection.


\begin{figure*}[t]
    \centering
    \includegraphics[width=\textwidth]{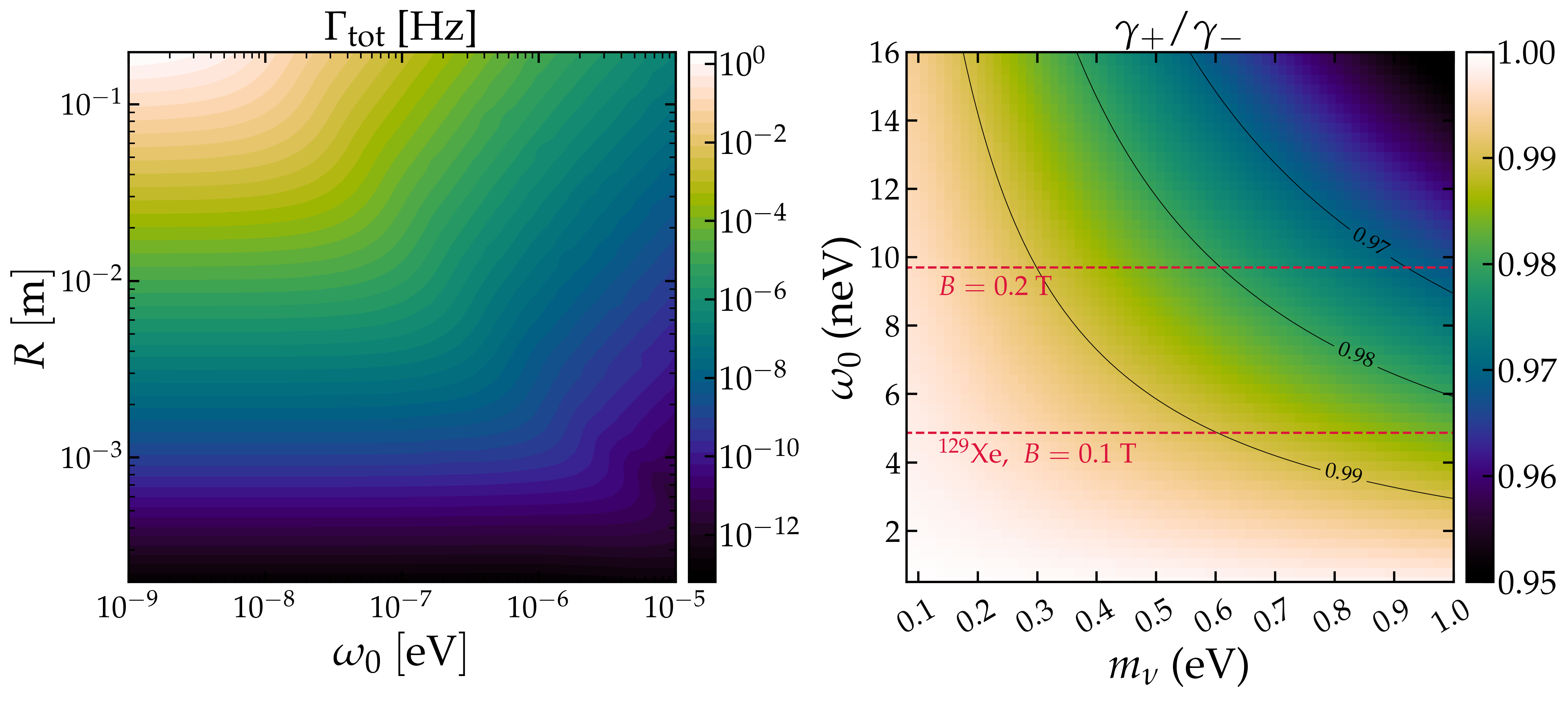}
    \caption{Typical neutrino-spin interaction rates for a single neutrino species at a standard number density $n_\nu \sim 110~{\rm cm}^{-3}$ and with a typical momentum $p_\nu \sim 27~{\rm cm}^{-1}$.
    {\it Left}: The total rate $\Gamma_{\rm tot} \equiv \Gamma_++\Gamma_-$ in the  parameter space of system energy splitting $\omega_0$ and sample size $R$, where we have assumed a spin density of $n_s\sim 3\times 10^{22}$ cm$^{-3}$ and a single neutrino mass of $m_{\nu}=0.1$ eV.
    The regions that exhibit coherent effects are in the top-left corner.  {\it Right}: Dissipation ratio $\gamma_+/\gamma_-$ in the parameter space of neutrino mass $m_\nu$ and system energy splitting $\omega_0$ in units of neV $=10^{-9}$ eV. The ratio quantifies the imbalance between neutrino-induced excitation ($\gamma_+$) and de-excitation ($\gamma_-$) single-spin processes, setting the dynamics of the system in the absence of other interactions. We find $\gamma_+/\gamma_- < 1$ across the entire parameter space shown, indicating a net de-excitation effect from relic neutrinos on the spin ensemble.  The two horizontal red dashed lines denote the splitting energies for an ensemble of $^{129}$Xe spins achieved with two different sub-Tesla magnetic field values.}
    \label{fig:param_space}
\end{figure*}

\subsection{Coherent regime}
An inspection of the form factor~\eqref{eq:formfactorpol} reveals that there are essentially two physical scales that determine the spatial coherence of the neutrino-spin interactions. The first is the characteristic size of the spin ensemble~$R$, which sets the spatial extent over which the spins are distributed. The second is the typical momentum transfer induced by the relic neutrinos $q$. 

Since the incoming neutrinos follow a momentum distribution $N_i(p^0)$, the momentum transfer $\vec{q}=\vec{p}_{i\pm}(\vec{p})-\vec{p}$ is also dictated by the distribution. The coherence properties must therefore be understood in terms of the typical momentum transfer ${}\vec{q}_{\rm typ}$  extracted from this distribution.
When the condition $q_{\rm typ}R\ll 1$ is satisfied, we see in Eq.~\eqref{eq:formfactorpol} that $j_1(q_{\rm typ}R)\approx  q_{\rm typ} R/3$, leading to $F_{\pm}\sim {\cal O}(N^2)$. 
This condition can always be fulfilled if the typical neutrino momentum $p_\nu$  itself is small compared with the inverse system size, i.e., $p_{\nu} R \ll1$. For relic neutrinos with a typical incoming momentum of $p_{\nu} \sim 3.15\, T_{\nu,0} \simeq 5.3 \times 10^{-4}~\mathrm{eV} \sim 27~\mathrm{cm
}^{-1}$, the condition $p_\nu R\ll 1$ therefore requires a very small $R \ll 0.04~\mathrm{cm}$. However, even for the larger-size spin samples to be considered in this work, e.g., $R \sim {\cal O}(1)$~cm, where formally $p_\nu R \gg 1$, it is still possible to fulfil the broader condition $q_{\rm typ} R \ll 1$ and hence maintain coherence across the sample in some limits. Specifically, the typical C$\nu$B kinetic energy is much larger than the energy splittings $\omega_0 \sim 10^{-8}$~eV considered in this work.  In this limit, the typical momentum transfer is $q_{\rm typ}\approx 2p_\nu\sin(\theta/2)$, where $\theta$ is the angle between the incoming and outgoing neutrino.  Thus, we see that coherence can be maintained for those scattering angles satisfying $\theta \ll (p_\nu R)^{-1}$.

In terms of the total excitation and de-excitation rates~\eqref{eq:biggamma}, the $p_\nu R \gg 1$ limit is effectively equivalent to demanding that only  
outgoing momentum  directions falling in the range $\cos ((p_\nu R)^{-1})<\mu<1$ contribute to the $N^2$ enhancement.  Thus, evaluating the $\mu$-integral immediately yields \begin{equation}
\Gamma_\pm \approx \frac{N^2}{16p^2_\nu R^2} \, \gamma_\pm,
\label{eq:partial}
\end{equation}
which we note is suppressed by a factor $\sim (2 p_\nu R)^{-2}$ relative to the na\"{\i}ve coherent rate. The typical momentum transfer in this regime is $q_{\rm typ} \sim \omega_0 m_\nu/p_\nu$.   This will be the case of interest in our study and was first introduced in Ref.~\cite{Arvanitaki:2024taq} as an analogy to the Rayleigh-Gans regime~\cite{1925AnP...381...29G}.  See Appendix~\ref{app:rates} for details.

As a concrete example, consider a spin ensemble composed of liquid $^{129}$Xe  characterised with a spin density of $n_s \sim N/R^3 \sim  10^{22}~\text{cm}^{-3}$. In this case, assuming an energy splitting $\omega_0 = 10^{-8}~\text{eV}$ 
and a standard  C$\nu$B with three normally-ordered masses summing to $\sum m_\nu = 0.15~\text{eV}$,
the total excitation rate of the spin ensemble evaluates to
\begin{equation}
 \Gamma_\pm^{\rm SM} \simeq 1.2 \times 10^{-6}~\text{Hz} \left( \frac{R}{1~\text{cm}} \right)^4,
\end{equation}
where the $R^4$ scaling represents the aforementioned forward limit of the $N^2$ enhancement~\cite{Arvanitaki:2024taq}.
The dependence of the total interaction rate ($\Gamma_{\rm tot}=\Gamma_++\Gamma_-$) on the system parameters, including $\omega_0$ and the sample size $R$, is illustrated in the left panel of Fig.~\ref{fig:param_space}, where regions exhibiting enhanced coherence can be identified. The right panel of the same figure shows the dissipation ratio $\Gamma_+/\Gamma_-\equiv\gamma_+/\gamma_-$ across the relevant parameter space, and highlights a slight dominance of neutrino-induced de-excitation. This dominance arises because de-excitation is always kinematically allowed as it has no energy threshold: any neutrino regardless of its momentum can induce the transition; see Eq.~\eqref{eq:gamma+-heuristic}.  Excitation is on the other hand not always possible.  The fractional difference is approximately $\gamma_+/\gamma_- -1  \sim - 2 m_\nu \omega_0/p_\nu^2$~\cite{Arvanitaki:2024taq} for $\omega_0 \ll p_\nu^2/(2 m_\nu) \ll m_\nu$.  See also Appendix~\ref{app:rates}.


\section{An open quantum system}\label{sec:oqs}

Although our full system comprises a spin ensemble~$S$ and a relic neutrino bath $B$ that interact with each other, in practice we are interested only in the dynamics of the former.  To this end, we adopt the framework of open quantum systems, which assumes unitary evolution of the density matrix of the full system, $\rho = \rho_S \otimes \rho_B$, under a total Hamiltonian
\begin{equation}
    H=H_S+H_B+H_{\rm int}\, ,
\end{equation} 
where $H_S$ and $H_B$ denote, respectively, the free Hamiltonian of the spin ensemble and of the bath, and $H_{\rm int}$ describes their interaction.  Tracing out the bath degrees of freedom, the evolution of the reduced density matrix $\rho_S$ is described by the Lindblad master equation \cite{Lindblad:1975ef,Gorini:1975nb, Manzano:2020yyw}, 
\begin{align}
        \frac{\mathrm{d} \rho_S}{\mathrm{d}t} & =-{\rm i}[H_S+H_{LS}, \rho_S]+\sum_k \gamma_k{\cal D}_{{\cal O}_k}[\rho_S(t)]\, ,
        \label{eq:master1}
\end{align}
where the so-called Lamb-shift Hamiltonian  $H_{LS}$ is a small hermitian correction to $H_S$ arising from $H_{\rm int}$, and
\begin{align}
        {\cal D}_{\cal O}[\rho_S(t)]&={\cal O}\rho_S{\cal O}^{\dagger}-\frac{1}{2}\{{\cal O}^{\dagger}{\cal O}, \rho_S\}\label{eq:calD}
\end{align} 
is a super-operator that acts on the reduced density matrix and describes the \emph{dissipative} effects of the system-bath interaction. Dissipation can occur through multiple channels, each specified by its own operator ${\cal O}_k$ and the associated interaction rate $\gamma_k$. 

The Lindblad equation~\eqref{eq:master1} serves well under the condition of weak coupling of the system to a large bath with no memory and when the Larmor frequency $\omega_0$ is much larger than the interaction rates.  These conditions enter the derivation of Eq.~\eqref{eq:master1} at various stages via the so-called Born, Markovian, and rotating-wave/secular approximation, which are discussed in more detail in Appendix~\ref{app:derlindblad}.   See also, e.g., Ref.~\cite{Manzano:2020yyw} for a general introduction to the Lindblad formalism.


\subsection{The master equation}

In the case of a single spin-$1/2$ interacting with a neutrino bath, the Hamiltonians of the spin system and its interaction with the neutrinos, $H_S^{(1/2)}$ and $H_I^{(1/2)}$, are given in Eq.~\eqref{h1}.  The corresponding Lindblad equation for the spin density matrix in the interaction picture takes the form
\begin{equation}
\begin{aligned}
\frac{{\rm d} \rho_S^{(1/2)}}{{\rm d} t} =& -{\rm i}\Delta[J_z, \rho^{(1/2)}_S]  \\
&~+ \gamma_{-}(J_- \rho^{(1/2)}_S J_+ - \frac{1}{2} \{J_+ J_-,\rho^{(1/2)}_S\} ) \\
&~+\gamma_{+}(J_+ \rho^{(1/2)}_S J_- - \frac{1}{2} \{J_- J_+,\rho^{(1/2)}_S\} ) \\
&~+\gamma_{z}(J_z \rho^{(1/2)}_S J_z - \frac{1}{2} \{J_z^2,\rho^{(1/2)}_S\})\,,
\label{eq:master2}
\end{aligned} 
\end{equation} where we have highlighted $\rho^{(1/2)}_S$ as the reduced density matrix for a single two-level system. We refer the reader to Appendix~\ref{app:derlindblad} for an {\it ab initio} derivation of this equation.

Inspecting Eq.~\eqref{eq:master2}, we see firstly an energy shift due to the $\Delta$, which is a combination of the ${\cal O}(G_F)$, Stodolsky effect that is present only when the neutrino-antineutrino asymmetry is non-zero, and ${\cal O}(G^2_F)$ terms from the Hermitian part of the Lindblad equation.  The $J_{\pm}$ terms and the corresponding coefficients $\gamma_\pm$ 
represent the non-Hermitian interaction of the spin with the neutrinos, causing the spin to either excite or de-excite via inelastic scattering, analogously to absorption and emission  in quantum optics.%
\footnote{While we draw parallels between our system and quantum optics and introduced terminology commonly used in the latter subject, we always use ``excitation'' and ``de-excitation'' to refer to the effects of the C$\nu$B.}
The $J_z$/$\gamma_z$ term is, on the other hand, energy conserving and represents the effect of dephasing. 
The computations of the coefficients $\gamma_{\pm,z}$ can be found in Appendix~\ref{app:gammas}. Their general forms including both neutrino scattering and pair emission/absorption processes can be found in Eqs.~\eqref{exp3}, \eqref{exp4}, and \eqref{eq:gammaz}, and, for $\gamma_\pm$, match the heuristic single-spin excitation and de-excitation rates~\eqref{eq:gamma+-heuristic} under the same settings.

The single-spin equation~\eqref{eq:master2} can be generalised to $N$~spins, where the Hilbert space of the spin ensemble is now $\rho_S=\otimes^N\rho^{(1/2)}$ and Eq.~\eqref{eq:Nham} gives the corresponding Hamiltonian. As we show in Appendix~\ref{app:lindn}, omitting the Hermitian part, the Lindblad equation for the $N$ two-level systems  reads
\begin{equation}
\begin{aligned}
\frac{{\rm d} \rho_S}{{\rm d} t} =& 
\sum_{\alpha \beta} \gamma^{\alpha \beta}_{-}(J^\alpha_- \rho_S J^\beta_+ - \tfrac{1}{2} \{J^\alpha_+ J^\beta_-,\rho_S\})  \\
&+ \sum_{\alpha \beta} \gamma^{\alpha \beta}_{+}(J^\alpha_+ \rho_S J^\beta_- - \tfrac{1}{2} \{J^\alpha_- J^\beta_+,\rho_S\})  \\
&+\sum_{\alpha \beta} \gamma^{\alpha \beta}_{z}(J^\alpha_z \rho_S J^\beta_z - \tfrac{1}{2} \{J^\alpha_z J^\beta_z,\rho_S\})\, ,
\label{gen_lindblad}
\end{aligned} 
\end{equation}
where $\alpha$ and $\beta$ run over all the spins of the system, and the angular operators $J_{\pm,z}^\alpha$ act only on the spin $\alpha$.  The coefficients $\gamma^{\alpha \beta}_{\pm,z}$
again encode information of the spin-bath interaction,  but now correlate the interaction at spatial position $\vec{x}_\alpha$ to that at $\vec{x}_\beta$.  For instance, assuming only inelastic neutrino-spin scattering, the $\gamma_\pm^{\alpha \beta}$ coefficients read
\begin{equation}
\begin{aligned}
\gamma_{\pm}^{\alpha \beta}= & \;2  \, G_F^2 \pi g_A^2\, 
    \int \frac{{\rm d}^3\vec{p}}{(2 \pi)^3 } \int \frac{{\rm d}^3 \vec{p}\,'}{(2\pi)^3} \; 
    e^{-{\rm i} \Delta\vec{x}_{\alpha \beta}\cdot (\vec{p}-\vec{p}\,')} \\
 &\times    \delta^{(1)}(\omega_0 \pm p'^0 \mp p^0)  \\
   & \times \left[(1-N_{k}(p'^0)) \, N_{j}(p^0)  +(1-\bar{N}_{k}(p'^0))\, \bar{N}_{j}(p^0)\right]\,.
    \label{eq:gammaalphabeta}
    \end{aligned}
\end{equation}
Depending on the extent to which correlations can be maintained across spatial distances $\Delta \vec{x}_{\alpha \beta} \equiv \vec{x}_\alpha - \vec{x}_\beta$ of the spin ensemble, qualitatively very different dynamics can arise.  Two contrasting limits are the fully-coherent and the fully-incoherent regimes.


\subsubsection*{Fully-coherent regime}

Full coherence is achieved when $e^{-{\rm i} \Delta \vec{x}_{\alpha \beta} .(\vec{p} - \vec{p}\,')} \approx 1$ across the whole spin ensemble, or, equivalently, when the condition $qR \ll 1$ is satisfied for momentum transfer~$q$ and a sample size~$R$.  In this limit, the interaction coefficients are effectively the same for all spin pairs. The Lindblad equation then reduces to 
\begin{equation}
\begin{aligned}
\frac{{\rm d}\rho_S}{{\rm d} t} = ~&\gamma_{-}({\cal J}_- \rho_S {\cal J}_+ - \tfrac{1}{2} \{{\cal J}_+ {\cal J}_-,\rho_S\}) \\
&+ \gamma_{+}({\cal J}_+ \rho_S {\cal J}_- - \tfrac{1}{2} \{{\cal J}_- {\cal J}_+,\rho_S\})  \\
&+ \gamma_{z}({\cal J}_z \rho_S {\cal J}_z - \tfrac{1}{2} \{{\cal J}_z^2,\rho_S\})\, ,
\label{lin12}
\end{aligned}
\end{equation}
where we have introduced the \emph{collective} spin operators  ${\cal J}_{\pm, z} = \sum_\alpha J_{\pm, z}^{\alpha}$.  This collective Lindblad equation is structurally identical to that for a single spin and the coefficients $\gamma_{\pm,z}$ are the same single-spin interaction rates of Eqs.~\eqref{exp3}, \eqref{exp4}, and \eqref{eq:gammaz}.
In the language of quantum optics, these terms describe, in the order of appearance, collective emission, collective absorption, and collective dephasing (see, e.g., Ref.~\cite{PhysRevA.98.063815}). Since the system Hamiltonian is $H_S = \omega_0 {\cal J}_z$, the collective dephasing term commutes with $H_S$ and has no dynamical effect on the observables of interest; we will therefore neglect it in the following.


\subsubsection*{Incoherent regime}

In the opposite, $qR\gg 1$ limit, the phase differences across the system are large, leading to a complete loss of coherence. This defines the incoherent regime, in which neutrino-spin interactions are fully local. The master equation in this case factorises into $N$ independent Lindblad equations, i.e.,
\begin{equation}
\begin{aligned}
\frac{{\rm d}\rho_S}{{\rm d} t} = &\, 
\sum_\alpha \gamma_{-}(J^\alpha_- \rho_S J^\alpha_+ - \tfrac{1}{2} \{J^\alpha_+ J^\alpha_-,\rho_S\}) \\
&+ \sum_\alpha \gamma_{+}(J^\alpha_+ \rho_S J^\alpha_- - \tfrac{1}{2} \{J^\alpha_- J^\alpha_+,\rho_S\}) \\
&+\sum_\alpha \gamma_{z}(J^\alpha_z \rho_S J^\alpha_z - \tfrac{1}{2} \{(J^\alpha_z)^2,\rho_S\})\, .
\label{eq:locall}
\end{aligned} 
\end{equation}
Here, each spin evolves independently under local interactions with the bath, and the rates $\gamma_{\pm,z}$ are again the single-spin rates. This local Lindblad form is general for any uncorrelated ensemble-bath interaction and applies to a wide class of physical systems for which the Lindblad approximation holds.  In Sec.~\ref{sec:local}, we will use Eq.~\eqref{eq:locall} to model other (non-neutrino) environmental effects that eventually drive the spin ensemble towards thermal equilibrium in a realistic NMR experiment.

\begin{figure*}
    \centering
    \includegraphics[width=\linewidth]{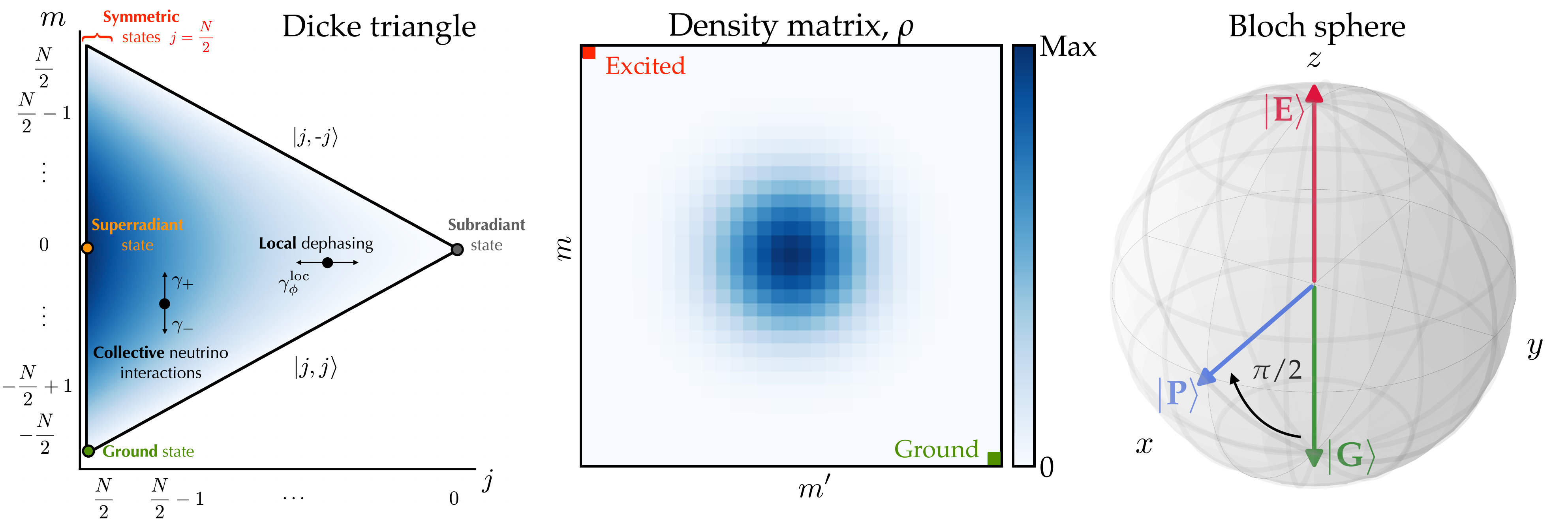}
    \caption{Visual representations of collective spin states in an ensemble of spin-$1/2$ particles. {\it Left}: The Dicke triangle shows the structure of permutational invariant Dicke states $\vert j, m\rangle$, including the ground state $\vert j, -j\rangle$, excited state $\vert j, j\rangle$, and intermediate states such as the superradiant state $\vert j=N/2, 0\rangle$. The vertical coordinate $m$ parameterises the energy of the system, while the horizontal coordinate $j$ represents the cooperativity of the system, i.e., the degree of coherence in the system. Blue shading indicates regions of enhanced collective effects, as determined by the expectation value of the emission rate operator $\langle \Gamma_+ \Gamma_- \rangle$. {\it Centre}: The density matrix for the coherent spins state (or product state) $\vert P\rangle$ in the symmetric subspace basis $\vert N/2, m\rangle\langle N/2, m'\vert$, with $m\in (-N/2, N/2)$ and for $N=30$. We also depict the excited and the ground states, which lie in the opposite extremities of the matrix. {\it Right}: Bloch sphere depiction of collective spin states. The ground and the excited Dicke states point in the negative and positive $z$-directions, respectively.   The product state $\vert P\rangle$ also has a well-defined classical direction: in this figure, it aligns with $+x$ and relates to the ground state by a $\pi/2$ pulse.}
    \label{fig:DickeTriangle}
\end{figure*}


\subsection{Dicke basis and initial states}
\label{sec:dickebasis}

Suppose all $N$ spins are identically prepared and indistinguishable, and experience the same interactions with the environment. It is then convenient to use the framework of Dicke states~\cite{PhysRev.93.99}, which are eigenstates of the collective operators ${\cal J}^2$ and ${\cal J}_z$, and form a basis for the permutationally invariant subspace of the full  Hilbert space.  As permutational invariance is preserved by the dynamics of the scenarios of interest, adoption of the Dicke basis allows us to drastically reduce the computational complexity of the problem, both analytically and numerically: instead of working in the full $2^N$-dimensional Hilbert space, we can restrict the dynamics to a much smaller subspace,
spanned by the Dicke states labelled by the total spin and its projection, $\vert j,m\rangle$, where $j = N/2, N/2 - 1, \ldots, j_{\min}$, with $j_{\min}=0$ if $N$ even or $j_{\min}=1/2$ if $N$ odd, and $\vert m\vert \leq j$. The full density matrix therefore contains 
\begin{equation}
    n_D = \sum_{j_{\rm min}}^{N/2} (2j + 1)^2 = \mathcal{O}(N^3)
\end{equation}
elements. In the presence of purely-coherent interactions, where the dynamics are confined to the fully symmetric subspace (\( j = N/2 \)), this reduces further to $n_D = N + 1$.

The left panel of Fig.~\ref{fig:DickeTriangle} illustrates the intuition provided by the Dicke representation---commonly visualised as the Dicke triangle---which organises the Hilbert space according to total spin $j$, including both the symmetric subspace ($j=N/2$) and the non-symmetric sectors ($j<N/2$) that do not exhibit cooperative or superradiant behaviours.  When the coherence condition is satisfied, the neutrino-spin interaction is represented by collective, ``vertical'' transitions along the Dicke ladder in steps of $\vert N/2, m\rangle\to\vert N/2, m\pm1\rangle$. This behavior contrasts with that of incoherent local processes such as dephasing, which induce only ``horizontal'' motion along the $j$-axis of the Dicke triangle. For more details on the Dicke basis and the effects of local interactions, we refer the reader to Ref.~\cite{PhysRevA.98.063815} and references therein.  See also Ref.~\cite{Gross:1982dkt,Masson:2021eyk} for an exposition on Dicke superradiance.

We are interested in three types of permutation-symmetric initial states, whose density matrices in the symmetric Dicke basis and Bloch sphere visualisations are shown in the centre and right panels of Fig.~\ref{fig:DickeTriangle}.
\begin{itemize}
    \item The ground state $\vert G\rangle\equiv\vert\downarrow\rangle^{\otimes N}=\vert N/2,-N/2\rangle$ represents the case in which all $N$ spins are in the low-energy state of the splitting (pointing down in our notation), such that the polarisation is unity in the Bloch sphere. From an experimental perspective, such a state is called hyperpolarised, as the polarisation far exceeds the thermal equilibrium expectation (see  Sec.~\ref{sec:local}).  Nonetheless, hyperpolarisation can be achieved experimentally for, e.g., $^{129}$Xe-enriched ensembles, via spin exchange optical pumping, as planned in the upcoming stages of the CASPEr experiment~\cite{Walter:2025ldb}. 
    
    \item The fully excited state $\vert E \rangle\equiv\vert\uparrow\rangle^{\otimes N}=\vert N/2,N/2\rangle$ has all $N$ spins in the high-energy state of the splitting (pointing up). Starting from a hyperpolarised state like $\vert G\rangle$, the $\vert E\rangle$ state can be produced by applying a $\pi$ pulse, i.e., applying a weak magnetic field oscillating at the Larmor frequency, until the polarisation is inverted.

    \item The equatorial coherent spin state (CSS), or ``product state'', is defined by $\vert P\rangle\equiv\vert \theta=\pi/2,\varphi=0 \rangle_{\rm CSS}$, where
\begin{align}
\qquad \vert \theta, \varphi \rangle_{\rm CSS}& =\otimes^{N}\left(\cos\frac{\theta}{2}\vert\uparrow\rangle+e^{{\rm i}\varphi}\sin \frac{\theta}{2}\vert\downarrow\rangle\right)\, .
\end{align}
    In the Dicke basis it can be written as $\vert P\rangle=\sum_m c_m\vert j, m\rangle$, with 
$c_m=2^{-j} \binom{2j}{j+m}$.
    This state serves as the standard benchmark in quantum sensing because of its resemblance to the superradiant state $\vert S\rangle\equiv\vert N/2,0\rangle$~\cite{PhysRev.93.99}. It can be prepared by applying a $\pi/2$ pulse to the ground state.
\end{itemize}

The preparation of a hyperpolarised ground state $|G\rangle$ is critical for all three initial configurations.   However, achieving a pure state with near-unity polarisation approximating $|G\rangle$ may not always be possible under realistic experimental settings.  A more faithful description of the spin ensemble's initial configuration should therefore be given in terms of a density matrix, shown here in the Dicke basis in a generic form,%
\footnote{The complete Dicke basis should in principle include a degeneracy label for each $|j,m\rangle$ state, i.e., $|j,m,\alpha\rangle$, with $1\leq \alpha \leq d_j^N$, where $d_j^N$ can be found in, e.g., Eq.~(5) of Ref.~\cite{Shammah_2017}, 
in order to recover the full Hilbert space.  For simplicity, however, we have omitted this label, since it does not affect our arguments.}
\begin{equation}
\rho=\sum_{j,m} \sum_{j',m'} f(j,j',m,m') ~\vert j, m\rangle\langle j', m'\vert \, .
\label{eq:dickerho}
\end{equation} 
The initial polarisation of the spin ensemble {\it prior to the application of any pulse} can then be trivially specified via the expectation value of the collective operator ${\cal J}_z$, i.e., 
\begin{equation}
p \equiv - \frac{2}{N}\langle {\cal J}_z\rangle =-\frac{2}{N}\sum_{j,m} m  f(j,m)\, ,
\label{eq:pol}
\end{equation}
which always evaluates to $|p| < 1$ if the configuration is not purely the ground ($m=-N/2$) or purely the excited ($m=N/2$) state.

We caution however that the definition~\eqref{eq:pol} of $p$ alone does not completely specify the effect of partial polarisation on the spin ensemble. To see this, consider also the expectation values
\begin{equation}
\begin{aligned}
\langle {\cal J}_z^2 \rangle & = \sum_{j,m} m^2  f(j,m) \leq \frac{N^2}{4}\, , \\
\langle {\cal J}^2 \rangle &= \sum_{j,m} j(j+1)  f(j,m) \leq \frac{N}{2} + \frac{N^2}{4}\, ,
\end{aligned}
\end{equation} 
where the upper bounds correspond to the expectations for the ground  ($j=N/2,m=-N/2$)  and the excited  ($j=N/2,m=N/2$) states.   Clearly, for $\langle {\cal J}_z^2 \rangle$ to be suppressed relative to the optimal value we must have admixtures of $|m| < N/2$ states (i.e., states along the Dicke ladder in Fig.~\ref{fig:DickeTriangle}), while for  $\langle {\cal J}^2 \rangle$ to become suboptimal the ensemble must contain $j<N/2$ states (i.e., states along the $j$-axis in the Dicke triangle).
Crucially, neither $\langle {\cal J}_z^2 \rangle$ nor $\langle {\cal J}^2 \rangle$ can be expressed in general in terms of $p$ as defined in Eq.~\eqref{eq:pol}.

We shall return to the issue of polarisation in Sec.~\ref{sec:2ndorder}, where we numerically evaluate in the Lindblad equation and connect our solutions with the hyperpolarisation protocols and level of polarisation achievable in current and future experimental setups.


\section{Coherent observables}
\label{sec:coherentobservables}

Given an initial state, the Lindblad equation~\eqref{gen_lindblad} can be solved to find the evolution of the spin ensemble.  Several observables are of interest.  The most common first-order observable is the collective magnetisation, whose expectation value can be constructed from $\langle {\cal J}_z\rangle={\rm tr}(\rho_S {\cal J}_z)$, and is related to the ensemble's energy.  At second order, relevant observables include $\langle {\cal J}^2_z\rangle$ and $\langle {\cal J}^2_x\rangle$, where the latter represents the squared transverse spin component. 

Exact analytical solutions for these observables generally do not exist.  Numerical solutions, on the other hand, can become prohibitive for large values of $N$.  We therefore consider first approximate analytical solutions in different limits and under different settings in order to map out the limiting behaviours of the system, before attempting a numerical treatment.  We discuss first in this section approximate analytical solutions in the presence of coherent neutrino effects only.  Section~\ref{sec:local} considers the effects of non-neutrino ``noise'' from local interactions.   
We present numerical solutions including both neutrino-induced coherent and local noise effects in Sec.~\ref{sec:numerical}.

Before we delve into the solutions, we first define two timescales that characterise the evolution of the spin ensemble in the presence of coherent neutrino interactions.
\begin{itemize}
\item The superradiant timescale,
\begin{equation}
t_{\rm SR}\equiv \frac{1}{N^2\gamma_{\pm}}\sim 8.7\times 10^{-3}~{\rm s}~\left(\frac{10^{25}}{N}\right)^2, 
\end{equation} 
characterises the timescale for a single collective spin-flip, i.e., the transition $\vert j, m\rangle\to \vert j, m\pm1\rangle$. The numerical estimate follows from setting $\gamma_\pm=\gamma_\pm^{\rm SM}$, where $\gamma_\pm^{\rm SM}$ is given in Eq.~\eqref{eq:gammanumbers}. 

\item The dissipative timescale, 
\begin{equation}
t_{\rm diss}\equiv \frac{1}{N\gamma_{\pm}}\sim 8.7\times 10^{22}~{\rm s}~\left(\frac{10^{25}}{N}\right), 
\label{eq:tdiss}
\end{equation} 
is the dynamical timescale of the system, defined as the duration over which an ${\cal O}(1)$ change in a collective observable such as $\langle {\cal J}_z\rangle$ becomes apparent. This timescale also marks the end of the coherent superradiant regime, and limits the validity of the perturbative solution discussed below.  The numerical estimate assumes again $\gamma_\pm=\gamma_\pm^{\rm SM}$ from Eq.~\eqref{eq:gammanumbers}.

\end{itemize}


\subsection{Perturbative solution}

Considering the weakness of the neutrino-spin interactions, Ref.~\cite{Arvanitaki:2024taq} proposed a perturbative/iterative solution to Eq.~\eqref{lin12} around the initial state.  Assuming only coherent neutrino interactions are present and to linear order in $t$, the solution reads 
\begin{equation} 
\rho_S(t) \approx 
\rho_S(0)  +t\gamma_{-}{\cal D}_{{\cal J}_-}[\rho_S(0)] + t\gamma_{+}{\cal D}_{{\cal J}_+}[\rho_S(0)]\, ,
\label{eq:perturb} 
\end{equation} 
recalling the notation of ${\cal D}_{\cal O}[\rho]$ from Eq.~\eqref{eq:calD}, and we have again omitted the collective dephasing term.

In this perturbative limit, the evolution of $\langle {\cal J}_z\rangle$ for an ensemble initially prepared in the $\vert P\rangle$ state is given by
\begin{equation}
\begin{aligned}
    \langle {\cal J}_z\rangle_{\vert P\rangle}& \approx -\left(\frac{N}{2}+1\right)\left(\frac{N}{2}+\frac{1}{2}\right)\gamma_{\rm net}t \\
    & \approx -\frac{N^2}{4}\gamma_{\rm net}t \, , 
    \label{eq:pert1}
    \end{aligned}
\end{equation} 
with $\gamma_{\rm net} \equiv \gamma_--\gamma_+$. As already noted in Ref.~\cite{Arvanitaki:2024taq}, $\langle {\cal J}_z\rangle$ benefits from an $N^2$ enhancement when the ensemble is initialised in the $|P \rangle$ state. However, this enhancement is partially offset by a suppression arising from the near cancellation between the $\gamma_+$ and $\gamma_-$ coefficients---typically 
$\gamma_{\rm net} \sim 2 (m_\nu \omega_0/p_\nu^2) \gamma_\pm \sim {\cal O}(10^{-2}-10^{-3})\gamma_\pm$, and can be even more strongly suppressed for smaller values of $\omega_0$ and $m_{\nu}$, as illustrated in Fig.~\ref{fig:param_space}. 

In a similar vein, the second-order observables evolve in the perturbative limit according to
\begin{align}
   \langle {\cal J}^2_z\rangle_{\vert P\rangle}&\approx \frac{N}{4}+\gamma_{\rm tot}N(N+1)~t\, ,\label{eq:pert2} \\
   \langle {\cal J}^2_x\rangle_{\vert G\rangle}&\approx \frac{N}{4}+2\gamma_{+}N(N+1)~t\,,
   \label{eq:pert3} 
\end{align} 
where $\gamma_{\rm tot}=\gamma_-+\gamma_+$.  Note that in the case of $\langle {\cal J}^2_x\rangle_{|G\rangle}$, we have assumed the ensemble to be prepared in the ground state: this observable can be advantageous, as it captures the same neutrino-induced signal as $\langle {\cal J}_z^2 \rangle_{\vert P\rangle}$ 
(for the $m_\nu$ and $\omega_0$ values considered here, $2\gamma_+\simeq \gamma_{\rm tot}$) but 
without the need to rotate the spins to the equatorial plane.%
\footnote{As pointed out in Ref.~\cite{Boyers:2025qgc}, pulse imperfections can lead to errors in the manipulation of the collective spin.}
It is also a quantity directly accessed by experiments like CASPEr that measure the transverse magnetisation. 
The perturbative solutions~\eqref{eq:pert1}-\eqref{eq:pert3} should remain valid as long as the state of the system does not deviate too much from the initial state. For the states considered in this work, we therefore expect the perturbative solutions to hold up to $t \lesssim t_{\rm diss}$, even in the case of the superradiant state.%
\footnote{In the case of the superradiant state, although perturbation theory formally breaks down at times $t\gtrsim t_{\rm SR}$~\cite{Arvanitaki:2024taq}, the coherent effect persists up to  the dissipation timescale $t_{\rm diss}$, i.e., until the system has evolved significantly away from the initial state $\vert N/2, 0\rangle$. This suggests that, even though the perturbative solution no longer accurately describes the full density matrix, it can still capture the expectation values of the relevant observables within $t_{\rm diss}$.}

Inspecting the solutions~\eqref{eq:pert1}-\eqref{eq:pert3}, we note first of all that the well-known momentum-transfer condition $qR \lesssim 1$ is not, by itself, sufficient to guarantee a coherent effect. One must also select an observable suited to the given initial state. For a first-order observable, measuring $\langle{\cal J}_{z}\rangle$ is advantageous when starting from the $\vert P\rangle$ state, as the quantum states before and after scattering are not fully orthogonal. This requirement was already emphasised in Ref.~\cite{Arvanitaki:2024taq}. Nevertheless, coherence can also build up dynamically through the action of collective neutrino-spin interactions, even if the initial state has no transverse coherence at $t=0$, such as in the ground state $\vert G\rangle$. This is evident from Eqs.~\eqref{eq:pert2} and~\eqref{eq:pert3}, where we have shown that measuring $\langle{\cal J}^2_x\rangle_{\vert G\rangle}$ produces the same coherent effect as measuring $\langle{\cal J}^2_z\rangle_{\vert P\rangle}$.

\begin{figure}
    \centering
    \includegraphics[width=\linewidth]{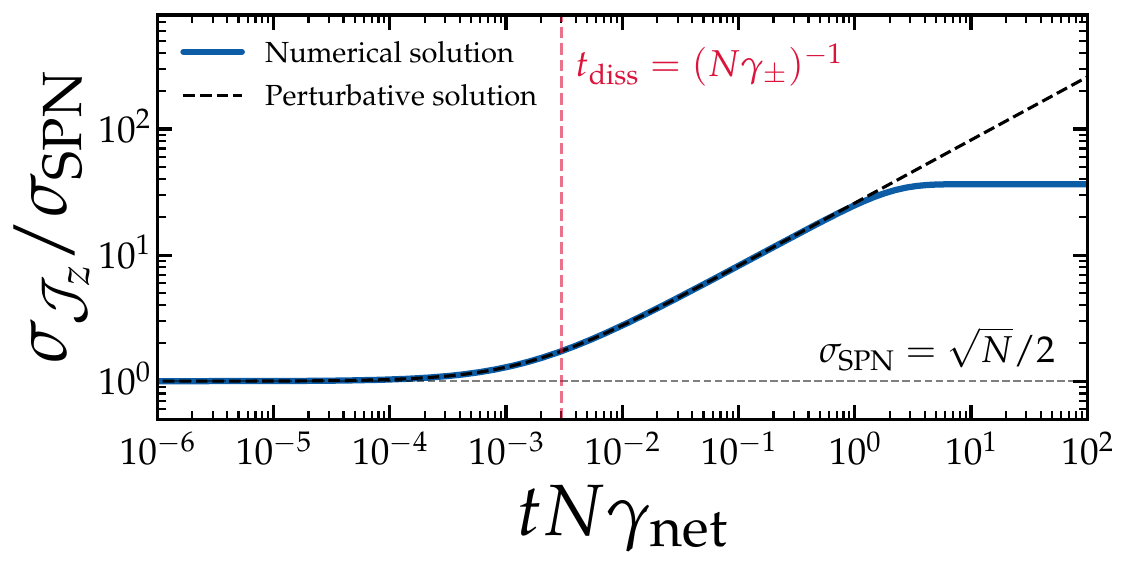}
    \caption{Time evolution of the standard deviation of the first order observable ${\cal J}_z$, as defined in Eq.~\eqref{eq:variance_z}, for $N=10^8$ and $\gamma_+/\gamma_-=0.997$. The perturbative solution is taken from the approximation in Eq.~\eqref{eq:variance_z}, $\sigma_{{\cal J}_z}=\sqrt{N/4+\gamma_{\rm tot}N^2t}$, whereas the numerical solution is obtained with the second-order approximation method, introduced in Sec.~\ref{sec:2ndorder}. We highlight the departure from the spin projection noise at $t\gtrsim t_{\rm diss}$ and the breakdown of the perturbative solution at $t\gtrsim (N\gamma_{\rm net})^{-1}$.}
    \label{fig:sigma_z}
\end{figure}

Secondly, while Eqs.~\eqref{eq:pert1}-\eqref{eq:pert3} clearly demonstrate that both the first- and second-order observables benefit from an $N^2$ enhancement, to assess the degree of observability we need also to consider the intrinsic quantum noise of the states as quantified by the variance of the observable.   For the first-order observable $\langle {\cal J}_z\rangle_{\vert P\rangle}$, the variance is
\begin{equation}
\begin{aligned}
\sigma_{{\cal J}_z}^2 & \equiv  \langle {\cal J}^2_z\rangle_{\vert P\rangle} - \langle {\cal J}_z\rangle_{\vert P\rangle}^2\\
& \approx  \frac{N}{4} +\gamma_{\rm tot} N(N+1) t + {\cal O}(t^2)\,,
\label{eq:variance_z}
\end{aligned}
\end{equation}
where we have supplied at the second equality the perturbative expression. The first, $N/4$ term in the second line is commonly known as the spin projection noise (SPN), and respresents  (uncorrelated) noise in the initial state.
Figure~\ref{fig:sigma_z} displays the perturbative solution for $\sigma_{{\cal J}_z}$ alongside its numerical solution (see Sec.~\ref{sec:2ndorder}) and illustrates that deviations from the SPN value of $N/4$ emerge around the dissipative timescale $t_{\rm diss}$. Therefore, at $t \lesssim t_{\rm diss}$, the signal and noise scale as $\langle {\cal J}_z\rangle \sim \gamma_{\rm net} N^2$ and $\sigma_{\mathcal{J}_z} \sim \sqrt{N}$, respectively, leading to a highly favourable signal-to-noise ratio scaling of  $\sim \gamma_{\rm net}N^{3/2}$.

On the other hand, at $t\lesssim t_{\rm diss}$, the second-order observables $\langle {\cal J}_z^2 \rangle_{\vert P\rangle}$ and $\langle {\cal J}_x^2 \rangle_{\vert G\rangle}$  have an intrinsic variance of $\sigma^2_{{\cal J}^2}\sim N^2$, such that the signal-to-noise ratio scales as $\sim \gamma_{\rm tot}N$. Thus,  while higher-order observables may appear at first glance to contain more information, they do not necessarily imply an improvement in sensitivity to the C$\nu$B signal unless the intrinsic noise of the observable can be reduced, for instance, through squeezing~\cite{Arvanitaki:2024taq}. Thus, we expect the $\langle J_z\rangle_{\vert P\rangle}$ observable to yield the best constraint on the C$\nu$B overdensity because of its favourable scaling with the number of spins.


\subsection{Steady-state solution}

At $t \gg t_{\rm diss}$, we expect the ensemble to reach a steady state characterised by saturation of the observable.  This steady state is not captured by the perturbative solution~\eqref{eq:pert1}--\eqref{eq:pert3}, but can be found by solving the Lindblad equation~\eqref{lin12} in the limit $\dot{\rho}_S=0$. Writing the density matrix in the Dicke basis as $\rho_S = \sum_{mm'}\rho^{mm'}_S\vert j,m\rangle\langle j,m'\vert$ and assuming the stationary solution to be unique (i.e., $\rho^{mm'}_S=0$ for $m\neq m'$), a recursive relation for the diagonal elements $\rho^{m,m}_{S}\equiv\rho^{m}_{S}$ can be obtained,
\begin{equation}
\begin{aligned}
&\quad~\gamma_-\left(\frac{N}{2}+m+1\right)\left(\frac{N}{2}-m \right)\rho_{S}^{m+1}\\
&+\gamma_+\left(\frac{N}{2}-m+1\right)\left(\frac{N}{2}+m\right)\rho_S^{m-1}  \\
&-\gamma_-\left(\frac{N}{2}+m\right)\left(\frac{N}{2}-m+1\right)\rho_S^m \\
&-\gamma_+\left(\frac{N}{2}-m\right)\left(\frac{N}{2}+m+1\right)\rho_S^m=0\,,
\label{eq:tridiag}
\end{aligned}
\end{equation}
for $m\in [-N/2-1, N/2+1]$, and $\rho_S^{-N/2-1}$=$\rho_S^{N/2+1}=0$. Equation~\eqref{eq:tridiag} is effectively a linear equation for the entries of the diagonal with a tridiagonal matrix, which can be solved iteratively for $\rho^{m}_{t\to\infty}$, from which we can compute the steady-state values of the observables via 
\begin{equation}
\begin{aligned}
    \langle {\cal J}_z\rangle_{\vert P\rangle}^{t\to\infty}&=\sum_m m~\rho^m_{t\to\infty}\,,\\
    \langle {\cal J}^2_z\rangle_{\vert P\rangle}^{t\to\infty}&=\sum_m m^2~\rho^m_{t\to\infty}\,,\\
    \langle {\cal J}^2_x\rangle_{\vert G\rangle}^{t\to\infty}&=\sum_m \langle j, m \vert {\cal J}^2_x\vert j ,m\rangle\rho^m_{t\to\infty}\,.
    \label{eq:steadyJ}
\end{aligned}
\end{equation}
Note that, since we consider only coherent neutrino interactions, the final state of the spin ensemble  implied by Eqs.~\eqref{eq:tridiag} and~\eqref{eq:steadyJ} is not a thermal equilibrium state one would generically expect the system to tend to: for this to happen local interactions are required, to be discussed in Sec.~\ref{sec:local}.

\begin{figure}
    \centering
    \includegraphics[width=\linewidth]{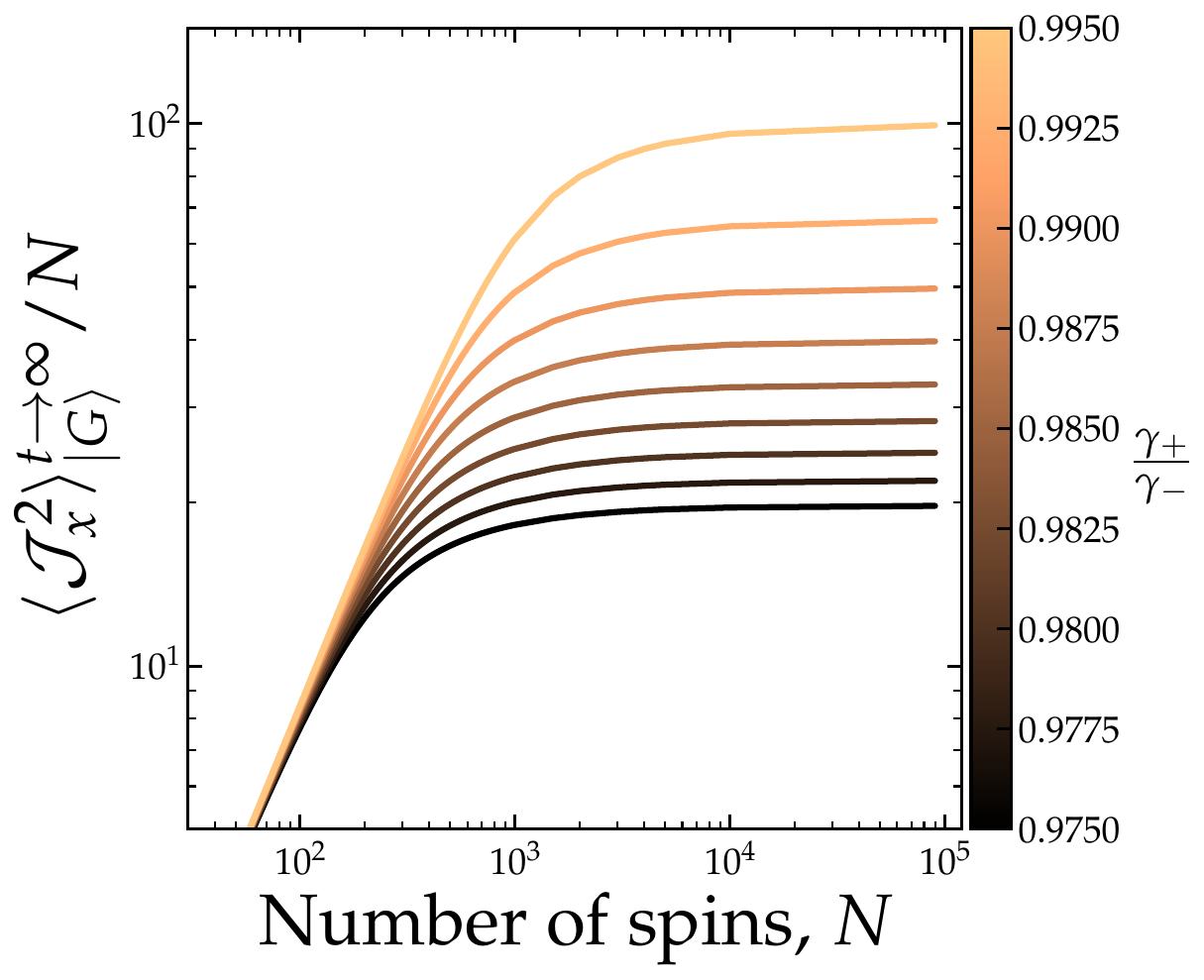}
    \caption{Steady-state scaling of the collective expectation value $\langle  {\cal J}^2_x\rangle_{|G\rangle}$ as a function of the number of spins $N$, in the presence of neutrino collective interactions only. The nine curves shown here correspond to different values of the dissipation ratio $\gamma_+ / \gamma_-$ as indicated by the colour scale. For a given $\gamma_+ / \gamma_-$, the observable exhibits convergence at $N \gtrsim \mathcal{O}(10^3)$, with the convergence point shifting to larger values of $N$ as $\gamma_+/\gamma_-$ moves closer to unity.
    }
    \label{fig:steadystate}
\end{figure}

Figure \ref{fig:steadystate} shows the late-time values of the normalised collective observable $\sigma_x \equiv (\langle {\cal J}^2_x\rangle_{\vert G\rangle})^{1/2}$, as a function of the number of spins $N$. Although not shown here, we also find that the steady-state value of $\langle {\cal J}_z\rangle_{|P\rangle}$ approaches the ground-state expectation $-N/2$ around $t\sim (N\gamma_{\rm net})^{-1}$ in the large $N$ limit, following a similar scaling behaviour in $N$ as $\sigma_x$. For all observables convergence is reached for $N \gtrsim \mathcal{O}(10^3)$, but at larger values of $N$ in those cases where a near cancellation between excitation and de-excitation, i.e., $\gamma_+\approx\gamma_-$, exists.

\section{Local noise effects}\label{sec:local}

Our analysis so far has focused exclusively on the coherent interaction between the spin ensemble and the C$\nu$B.  Realistic experimental setups, however, are inevitably subject to other, non-neutrino environmental interactions that relax the system, i.e., drive it to a state of thermal equilibrium,  by locally exciting or de-exciting the spins, as well as locally dephasing the spin precessions. 
Amongst these, local dephasing plays a particularly important role. 

This form of noise originates from random, uncorrelated fluctuations acting independently on each spin, due to, e.g., spin-spin interactions and magnetic field fluctuations, and, phenomenologically, can be  described by the operator $\sqrt{\gamma_{\phi}^{\rm loc}}J_z^{\alpha}$, where $\alpha$ labels the individual spin. As illustrated in Fig.~\ref{fig:DickeTriangle}, the key feature of local dephasing is that it acts incoherently and forces the system to move into lower-$j$ sectors where collective effects such as superradiance are suppressed or entirely lost. The relevant timescale is $t_{\phi}\sim 1/\gamma_{\phi}^{\rm loc}$, which is easily much shorter than the dissipative timescale $t_{\rm diss}$ from Eq.~\eqref{eq:tdiss}.  Consequently, local dephasing can wash out coherent signatures induced by the neutrino-spin interaction, effectively setting a limit on the observable impact of the C$\nu$B.

In a purely phenomenological approach, let us assume that relaxation is dominated by local (single-particle) effects, i.e., we do not model spin-spin interactions. This simplification allows us to ``repurpose'' the Lindblad equation~\eqref{eq:locall} in the incoherent regime to model the relaxation characteristics of a realistic experimental sample.  Identifying $\gamma_z$ in Eq.~\eqref{eq:locall} with the local dephasing rate $\gamma^{\rm loc}_{\phi}$ and the rates $\gamma_\pm$ with the local  emission and absortion rates $\gamma^{\rm loc}_{\pm}$, we can tune  $\gamma_{\pm,\phi}^{\rm loc}$ to match the experimental longitudinal and transverse relaxation times, $T_1$ and $T_2$, and the final equilibrium state.

Explicitly, starting from Eq.~\eqref{eq:locall}, we take the trace ${\rm tr}[\rho_S {\cal J}_i]$, with $i=x,y,z$, on both sides of the equation, in order to obtain an evolution equation for the collective magnetisation $\langle {\cal J}_i\rangle$ in the $i$th direction,
\begin{equation}
\begin{aligned}
\frac{\mathrm{d} }{\mathrm{d}t}\langle {\cal J}_i \rangle & =  
N \gamma^{\rm loc}_- ~\text{tr}[J_iJ_- \rho_S^{(1/2)} J_+ - \frac{1}{2} J_i\{J_+ J_-,\rho_S^{(1/2)}\} ] \\
& + N \gamma^{\rm loc}_+~\text{tr}[J_iJ_+ \rho_S^{(1/2)} J_- - \frac{1}{2} J_i\{J_- J_+,\rho_S^{(1/2)}\} ] \\
& + N\gamma_{\phi}^{\rm loc}~\text{tr}[J_iJ_z \rho_S^{(1/2)} J_z - \frac{1}{2} J_i\{J_z J_z,\rho_S^{(1/2)}\} ]\, ,
\end{aligned}
\end{equation}
where $J_{\pm, z,i}$ are single-spin operators, and $\rho_S^{(1/2)}$  is a single-spin density matrix.
Solving this equation analytically, we find for the longitudinal magnetisation 
\begin{equation}
    \langle {\cal J}_z (t)\rangle  =\langle {\cal J}_z \rangle^{\rm eq}-[\langle {\cal J}_z \rangle^{\rm eq}-\langle {\cal J}_z \rangle (0)]~e^{-t(\gamma^{\rm loc}_++\gamma^{\rm loc}_-)}\, ,
    \label{eq:fidmz}
\end{equation}    
with $\langle {\cal J}_z \rangle(0)=0,-N/2$ for the $\vert P\rangle$ and the $\vert G\rangle$ state, respectively.  The longitudinal magnetisation tends to an equilibrium value of
\begin{equation}
    \langle {\cal J}_z \rangle^{\rm eq} =\frac{N}{2}\frac{\gamma^{\rm loc}_{+}-\gamma^{\rm loc}_{-}}{\gamma^{\rm loc}_{+}+\gamma^{\rm loc}_{-}}\,,
    \label{eq:mzeq}
\end{equation}
at a rate that can be matched to the longitudinal relaxation time $T_1$ via
 \begin{equation}
    T_1^{-1} = \gamma^{\rm loc}_++\gamma^{\rm loc}_-\,.
    \label{eq:t1a}
\end{equation}
Similarly, starting from the $\vert P\rangle$ state, we find the magnetisation in the transverse direction  to be
\begin{equation}
    \langle {\cal J}_x (t) \rangle=\frac{N}{2}~e^{-t(\gamma^{\rm loc}_++\gamma^{\rm loc}_-+\gamma^{\rm loc}_{\phi})/2}\, ,
\end{equation} 
yielding a match 
 \begin{equation}
    T_2^{-1} = \frac{1}{2}(\gamma^{\rm loc}_++\gamma^{\rm loc}_-+\gamma^{\rm loc}_{\phi})\, 
    \label{eq:t2a}
\end{equation}
to the transverse relaxation time $T_2$.

Observe in Eq.~\eqref{eq:mzeq} that the equilibrium magnetisation $\langle {\cal J}_z \rangle^{\text{eq}}$ depends only on the ratio $\gamma^{\text{loc}}_+ / \gamma^{\text{loc}}_-$.  This is a consequence of the detailed balance condition $\gamma^{\text{loc}}_+/\gamma^{\text{loc}}_- = e^{-\beta \omega}$, which ensures that the ensemble approaches the well-known thermal steady state
\begin{equation}
\langle {\cal J}_z \rangle^{\rm eq} = \frac{N}{2} \tanh\left(\frac{\beta \omega_0}{2}\right)\, ,
\label{eq:mzeq2}
\end{equation}
as long as all interaction channels couple the system to a thermal reservoir at the same temperature $T= 1/\beta$. 
Equation~\eqref{eq:mzeq2} also highlights the inherent difficulty in achieving a strong initial polarisation: even under cryogenic temperatures and extremely large magnetic fields, the polarisation remains small. Reinstating all physical constants, the thermal polarisation takes the form 
\begin{equation}
\begin{aligned}
    p&=\tanh\left(\frac{\hbar\gamma B}{2k_BT}\right)\approx \frac{\hbar\gamma B}{2k_BT}\\
    &\simeq 10^{-3}\left(\frac{B}{10~ {\rm T}}\right)\left(\frac{4\, {\rm K}}{T}\right)\, ,
    \label{eq:thermalpol}
\end{aligned}
\end{equation}
where the numerical estimate assumes the gyromagnetic ratio $\gamma$ of $^{129}$Xe nuclei. At ${\cal O}(10^{-3})$, this estimate demonstrates that thermal polarisation alone does not suffice to reach the strongly polarised regime, necessitating the use of advanced techniques to hyperpolarise the spin ensemble.\looseness=-1

As established in Eqs.~\eqref{eq:t1a} and~\eqref{eq:t2a}, the local relaxation times, $T_1$ and $T_2$, depend on the absolute values of the local rates $\gamma^{\text{loc}}_{\pm,\phi}$. In typical solid-state NMR systems, the longitudinal relaxation timescale is much larger than the transverse relaxation timescale, i.e., $T_1\gg T_2$. This implies $\gamma^{\rm loc}_{\pm} \ll \gamma^{\rm loc}_{\phi}$, and the transverse relaxation time $T_2$ is well-approximated by the effective dephasing time set by $1/\gamma^{\rm loc}_{\phi}$.%
\footnote{In standard NMR terminology, the experimentally-observed dephasing time is denoted $T_2^*$ and includes contributions from magnetic field inhomogeneities and chemical shifts. In contrast, $T_2$ refers to the intrinsic dephasing time due to dipolar spin-spin interactions, and typically satisfies $T_2 \gg T_2^*$. For our purposes, we do not distinguish between $T_2$ and $T_2^*$, but adopt the convention $T_2 = T_2^* = \gamma_{\phi}^{-1}$, which we always use to refer to the actual (measured) dephasing time.}
Note however that in a generic (i.e., not specific to solid state) NMR setup, $T_1$ can be as small as $T_2$. In what follows, we model the $\gamma^{\rm loc}_{\pm}$ and $\gamma_\phi^{\rm loc}$ rates according to the ratio $T_1/T_2$, with the additional stipulation that the ratio $\gamma^{\rm loc}_+/\gamma^{\rm loc}_-$ should reproduce the equilibrium magnetisation achievable at cryogenic temperatures.


\section{Numerical solutions}\label{sec:numerical}

Having discussed both coherent neutrino and incoherent local effects, we are now in a position to explore a more complete, numerical solution of a master equation that accounts for both effects.
The strong separation of timescales established above, $T_2 \ll T_1$, implies that coherent effects such as those induced by neutrino-spin interactions are primarily limited by local dephasing.  However, because our observable of interest, $\langle {\cal J}_z \rangle_{|P\rangle}$, experiences longitudinal relaxation, we must also account for the effects of $T_1$, particularly on the correlations of the signal between different times. Therefore, a minimum master equation that should capture the essential behaviours of the system would read
\begin{equation}
\begin{aligned}
    \frac{\mathrm{d} \rho_S}{\mathrm{d}t} =\; &  \gamma_-{\cal D}_{{\cal  J}_-}[\rho_S]+\gamma_+{\cal D}_{{\cal J}_+}[\rho_S] \\
    &  +\gamma_{-}^{\rm loc} \sum_{\alpha}{\cal D}_{J^\alpha_-}[\rho_S] +\gamma_{+}^{\rm loc} \sum_{\alpha}{\cal D}_{J^\alpha_+}[\rho_S] \\
    &+\gamma_{\phi}^{\rm loc} \sum_{\alpha}{\cal D}_{J^\alpha_z}[\rho_S]\\
 \equiv\; & {\cal L} \{ \rho_S(t)\}\, ,
    \label{eq:lindnum}
    \end{aligned}
\end{equation} 
where ${\cal L}$ is the Lindblad super-operator of this system, and we remind the reader that $\gamma_\pm$ are the C$\nu$B-induced excitation and de-excitation rates, while $\gamma_\pm^{\rm loc}$ and $\gamma_{\phi}^{\rm loc}$ are the local absorption/emission and dephasing rates.

\begin{figure}
    \centering
    \includegraphics[width=\linewidth]{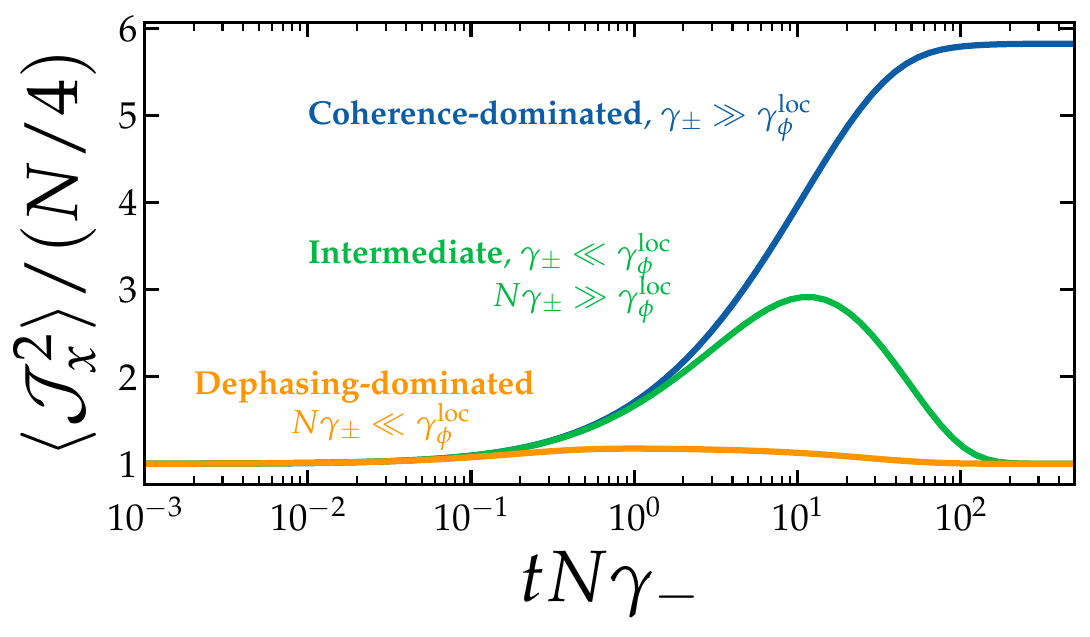}
    \caption{
Time evolution of collective spin variance $\langle {\cal J}_x^2 \rangle$. We highlight three distinct regimes: (i) coherence-dominated ($\gamma_\pm \gg \gamma_\phi^{\mathrm{loc}}$), where collective dynamics are preserved; (ii)~intermediate ($\gamma_\pm \ll \gamma_\phi^{\mathrm{loc}}$, but $N\gamma_\pm \gg \gamma_\phi^{\mathrm{loc}}$), where partial coherence remains; and (iii) dephasing-dominated ($N\gamma_\pm \ll \gamma_\phi^{\mathrm{loc}}$), where local noise suppresses collectivity. Time is scaled by $N\gamma_-$, and the spin variance is shown in units of $N/4$ for $N=100$.}
    \label{fig:regimes}
\end{figure}

Given Eq.~\eqref{eq:lindnum} and assuming the C$\nu$B-induced excitation and de-excitation rates are tunable by an overdensity factor $\delta_\nu$, i.e.,%
\footnote{The overdensity parameter $\delta_\nu$ is strictly speaking not an independent parameter, in the sense that other characteristics of the C$\nu$B must also change when $\delta_\nu$ is varied.  For example, enhancement of $\delta_\nu$ due to local gravitational clustering generally results in a larger typical momentum~\cite{Ringwald:2004np}. For excessively large values of $\delta_\nu$ (e.g., $\sim 10^{11}$, such as the current KATRIN $\delta_\nu$ limit~\cite{KATRIN:2022kkv}), physically consistent modelling can in fact lead to a signal that is completely different from the $\delta_\nu \sim {\cal O}(1)$ case~\cite{Bondarenko:2023ukx}.  We nonetheless adopt the definition~\eqref{eq:deltanudef}, firstly in order to facilitate comparison with constraints and forecasted sensitivities in the literature that have been derived under similar assumptions.  Secondly, for extremely large values of $\delta_\nu$, there is also no model-independent way to consistently relate $\delta_\nu$ to other C$\nu$B properties.  Consequently, the interpretation of $\delta_\nu$ must be limited to a mere characterisation of how away an experimental technique is from achieving detection, rather than a necessarily physically meaningful measure of the local C$\nu$B overdensity.\label{footnote:deltanu}}
\begin{equation}
\gamma_\pm \approx \delta_\nu \gamma_\pm^{\rm SM}\,,
\label{eq:deltanudef}
\end{equation} we can now identify three distinct dynamical regimes defined by the competition between collective neutrino-induced interactions and local dephasing,  in the limit where $T_1\gg T_2$. 

\begin{itemize}
    \item \emph{Coherence-dominated regime:} When $\gamma_{\pm} \gg \gamma^{\rm loc}_{\phi} \approx T_2^{-1}$, coherent effects dominate even at the single-spin level, keeping the dynamics well confined to the symmetric subspace. The system exhibits collective behaviours with enhancement scaling as $\sim N^2$, providing optimal conditions for signal-to-noise amplification. This regime is depicted in blue in Fig.~\ref{fig:regimes} and is unattainable in realistic experiments under realistic dephasing, unless $\delta_\nu$ is very large.

    \item \emph{Intermediate regime:} In the case $\gamma_{\pm} \ll \gamma^{\rm loc}_{\phi}$, but $N\gamma_{\pm} \gg \gamma^{\rm loc}_{\phi}$ is still maintained, individual spins are dominated by local dephasing. Yet collective effects are stronger in the global dynamics. Thus, collective effects remain observable, even while symmetry is partially broken by local noise.
    This regime is depicted in Fig.~\ref{fig:regimes} in green, highlighting a complete suppression of the coherent signal only at $t\sim\gamma^{-1}_{\pm}$.   The intermediate regime is also the most relevant experimentally for detecting weak collective signatures under realistic dephasing, although an enhanced $\delta_{\nu}\gg {\cal O}(1)$ will still be necessary. 
    
    \item \emph{Dephasing-dominated regime:} When $N\gamma_{\pm} \ll \gamma^{\rm loc}_{\phi}$, incoherent processes dominate over the collective interaction. In this regime, the system rapidly departs from the symmetric subspace, the signal-to-noise ratio saturates well below unity, and sensitivity to the neutrino–spin interaction is effectively lost. This region, shown in orange in Fig.~\ref{fig:regimes}, corresponds to the case of a standard C$\nu$B, where $\delta_\nu \sim {\cal O}(1)$ is the generic expectation. 
   
\end{itemize}

In the following, we first solve Eq.~\eqref{eq:lindnum} in Sec.~\ref{sec:2ndorder}
numerically by way of an efficient method that tracks explicitly the observables of interest: we call it the ``second-order approximation'' because of a truncation that needs to be applied at second order.  This is to be compared to a complete numerical solution for the full density matrix $\rho_S$, whose details can be found in Appendix~\ref{sec:fullsol}.  In Sec.~\ref{sec:twotimemain} we apply the same method to investigate the evolution of the two-time correlation functions under Eq.~\eqref{eq:lindnum}, which will be used later in Sec.~\ref{sec:futureconstraints} to form the covariance matrix that quantifies the theoretical uncertainties in the observables.


\subsection{Second-order approximation}\label{sec:2ndorder}

While it is possible to numerically solve the Lindblad master equation~\eqref{eq:lindnum} at the level of the density matrix for a system of $N$ spins, it can quickly become impractical or even intractable for values of $N$ larger than about 100 (see Appendix~\ref{sec:fullsol}).
Rather than dealing with the full density matrix, then, we shift the focus to tracking just a handful  physically meaningful quantities.  

Specifically, as was done in, e.g., Refs.~\cite{Shammah_2017,carmichael1}, for the case of Dicke superradiance and also touched upon in Sec.~\ref{sec:local}, we construct  differential equations from the master equation~\eqref{eq:lindnum} directly for the expectation values of the relevant observables, namely, the collective magnetisation $\langle {\cal J}_z\rangle$, $\langle {\cal J}_z^2 \rangle$, and  $\langle {\cal J}_x^2 \rangle$.  This is akin to tracking the moments of a continuous system, which is widely used in many different fields.
However, differential equations thus constructed generally do not form a closed system, in the sense that the equation of motion for, e.g., $\langle {\cal J}_z \rangle$ depends on a higher moment $\langle {\cal J}_z^2 \rangle$, whose own equation of motion depends in turn on  $\langle {\cal J}_z^3 \rangle$, and so on, giving rise an infinite hierarchy of coupled equations.
To solve the system, therefore, one needs to truncate the hierarchy at a chosen order~\cite{Shammah_2017}.

Here, we opt to truncate the hierarchy at second-order.  For the observables $\langle {\cal J}_z \rangle$ and $\langle {\cal J}_z^2 \rangle$, this means we adopt the closure conditions 
 $\langle {\cal J}_z^3 \rangle \approx \langle {\cal J}_z \rangle \langle {\cal J}_z^2 \rangle$ and $\langle {\cal J}_z {\cal J}^2\rangle\approx\langle {\cal J}_z \rangle\langle {\cal J}^2\rangle$ to arrive at~\cite{Shammah_2017}
 \begin{equation}
\begin{aligned}
    \frac{\rm d}{{\rm d}t} \langle {\cal J}_z \rangle = \; & \sum_\pm \pm \gamma_\pm  \Big[ \langle {\cal J}^2 \rangle - \langle {\cal J}_z^2 \rangle \mp \langle {\cal J}_z \rangle \Big] \, \\
    & + \sum_{\pm}\pm\gamma_{\pm}^{\rm loc} \Bigg[\mp\langle {\cal J}_z \rangle + \frac{N}{2} \Bigg] \, ,\\
    \frac{\rm d}{{\rm d}t} \langle {\cal J}^2 \rangle = \; &
    -\sum_{\pm} \gamma_\pm^{\rm loc} \Big[\langle {\cal J}^2 \rangle +\langle {\cal J}_z^2 \rangle \mp (N-1) \langle {\cal J}_z \rangle -N \Big] \\
\; &- \gamma_{\phi}^{\rm loc} \Bigg[ \langle {\cal J}^2 \rangle - \langle {\cal J}_z^2 \rangle - \frac{N}{2} \Bigg]\, ,\\ 
    \frac{\rm d}{{\rm d}t} \langle {\cal J}_z^2 \rangle = \; & 
\sum_{\pm}\gamma_{\pm} \Big[ \langle {\cal J}^2 \rangle - 3 \langle {\cal J}_z^2 \rangle\mp \langle {\cal J}_z \rangle 
     \\
    & \qquad \quad \mp 2 \langle {\cal J}_z \rangle \langle {\cal J}_z^2 \rangle 
    \pm 2 \langle {\cal J}_z \rangle \langle {\cal J}^2 \rangle \Big] \\
& - \sum_{\pm}\gamma_\pm^{\rm loc} \Bigg[ 2 \langle {\cal J}_z^2 \rangle\mp(N-1) \langle {\cal J}_z \rangle  -\frac{N}{2}\Bigg]\, .
    \label{eq:secondorderapprox}
\end{aligned}
\end{equation}
Note that our closure conditions are by no means unique and other choices exist. For example, closure can be achieved alternatively by demanding that all cumulants (i.e., connected pieces of the moments) vanish at third order and above, e.g., $\langle {\cal J}_z^3 \rangle = 3 \langle {\cal J}_z^2\rangle \langle {\cal J}_z\rangle + 2 \langle {\cal J}_z\rangle^3 + \langle {\cal J}_z^3 \rangle_c$, where $\langle {\cal J}_z^3 \rangle_c$ is the connected piece to be set to zero.
We observe however that all closure-dependent terms in equation~\eqref{eq:secondorderapprox} are proportional to $\gamma_{\rm net}$ and vanish in the limit $\gamma_{\rm net} \to 0$.  Indeed, given the small $\gamma_{\rm net} \ll \gamma_\pm$ in our scenario, we have verified that adopting a different closure condition  (e.g., the cumulant truncation scheme mentioned above) has no discernible impact on our numerical solutions within in the time range of interest (i.e., $t \lesssim T_2$) even for extremely large $N$ values.

\subsubsection*{Initial conditions including partial polarisation}

As discussed in Sec.~\ref{sec:dickebasis}, in experimental settings spin ensembles are prepared in a hyperpolarised configuration that approximates the ground state and, if need be, rotated to the desired orientation by applying an appropriate pulse.  Common hyperpolarisation protocols such as spin-exchange optical pumping achieve close-to-unit polarisation by way of spin excitation and de-excitation processes that can be modelled as local interactions. See Appendix~\ref{app:hyperpolarised} for details.  Thus, in order to establish the initial conditions of a partially-polarised spin ensemble prepared in the aforementioned manner, 
we can simply take Eq.~\eqref{eq:secondorderapprox} assuming only the local rates $\gamma_\pm^{\rm loc}$ are non-zero, re-interpret these rates as the excitation and de-excitation rates $\gamma_\pm^{\rm hyper}$ of the hyperpolarisation technique in use, and compute the corresponding steady-state solutions for $\langle {\cal J}_z\rangle$ and $\langle {\cal J}_z^2 \rangle$.

These solutions then serve as initial conditions for the partially-polarised ``ground state'', which read
\begin{equation}
\begin{aligned}
    \langle {\cal J}_z(0) \rangle_{\vert G\rangle} & =  \frac{Np}{2}\, , \quad
    \langle {\cal J}_z^2(0) \rangle_{\vert G\rangle}  =  \frac{N}{4} +\frac{N(N-1)}{4}p^2 \, ,\\
    \langle {\cal J}^2(0) \rangle & = \frac{3N}{4} + \frac{N(N-1)}{4} p^2 \, ,
    \label{eq:pinit2} 
    \end{aligned}
\end{equation}
where we have used $\langle {\cal J}^2 \rangle=\langle {\cal J}_x^2 \rangle+\langle {\cal J}_y^2 \rangle+\langle {\cal J}_z^2 \rangle=N/2+\langle {\cal J}_z^2 \rangle$, and  
\begin{equation}
p=\frac{\gamma^{\rm hyper}_- -\gamma^{\rm hyper}_+ }{\gamma^{\rm hyper}_- + \gamma^{\rm hyper}_+}\, 
\label{eq:polhyper}
\end{equation}
is the initial polarisation of the ensemble expressed in terms of the rates of the hyperpolarisation method.
See also Eq.~\eqref{eq:mzeq}.  If on the other hand we wish to initialise in a product state-like configuration, a $\pi/2$ rotation of the coordinates  should be applied to  Eq.~\eqref{eq:pinit2} to get
\begin{equation}
\begin{aligned}
\langle {\cal J}_z(0) \rangle_{\vert P\rangle} &=  0\, , \qquad  \langle {\cal J}_z^2(0) \rangle_{\vert P\rangle}   =  \frac{N}{4}\, ,
\label{eq:pinit}
\end{aligned}
\end{equation}
while $\langle {\cal J}^2 \rangle$ remains the same under rotation.
Then, together, Eqs.~\eqref{eq:secondorderapprox} and~\eqref{eq:pinit2} or \eqref{eq:pinit} form a closed set of equations that completely determine $\langle {\cal J}_z\rangle, \langle {\cal J}^2_z\rangle$, and~$\langle {\cal J}^2\rangle$.

To obtain approximate numerical solutions for the transverse observable $\langle {\cal J}^2_x\rangle$, we supplement the differential equations~\eqref{eq:secondorderapprox} with an additional equation of motion for  the non-hermitian variables $\langle {\cal J}_\pm \rangle \equiv \langle {\cal J}_x \rangle  \pm {\rm i} \, \langle {\cal J}_y \rangle$, i.e.,
\begin{equation}
\begin{aligned}
\frac{\rm d}{{\rm d}t} \langle {\cal J}_{\pm} \rangle = \; & \gamma_-  \langle {\cal J}_z \rangle \langle {\cal J}_{\pm} \rangle
- \gamma_+ \langle {\cal J}_z \rangle \langle {\cal J}_{\mp}\rangle\\
\; & - \frac{1}{2} \left(\gamma_-^{\rm loc}  +\gamma_+^{\rm loc} + \gamma_{\phi}^{\rm loc} \right) \langle {\cal J}_{\pm} \rangle \, ,
\label{eq:2nd_5}
\end{aligned}
\end{equation} 
where we have applied the closure conditions $\langle {\cal J}_+{\cal J}_z\rangle\approx\langle {\cal J}_+ \rangle\langle {\cal J}_z \rangle$ and $\langle {\cal J}_z{\cal J}_-\rangle\approx\langle {\cal J}_z \rangle\langle {\cal J}_- \rangle$. Here, we are interested in a spin ensemble initially prepared in the ground state $\vert G\rangle$, such that the initial conditions are $\langle {\cal J}_\pm(0)\rangle_{\vert G\rangle}=0$. Then, a solution for $\langle {\cal J}^2_x\rangle$ can be constructed from
\begin{equation}
 \langle {\cal J}^2_x\rangle=  2\langle {\cal J}^2\rangle-2\langle {\cal J}_z^2\rangle+\langle {\cal J}_+\rangle\langle {\cal J}_-\rangle+\langle {\cal J}_-\rangle\langle {\cal J}_+\rangle\, ,
\end{equation} 
where $\langle {\cal J}_z^2\rangle$, $\langle {\cal J}^2\rangle$, and $\langle {\cal J}_{\pm} \rangle$ are solutions of Eqs.~\eqref{eq:secondorderapprox} and~\eqref{eq:2nd_5}.

Figures~\ref{fig:jz_t}-\ref{fig:sigmaz_p} show the numerical solutions obtained from the above equations of motion under different settings, which we discuss in more detail below.


\subsubsection*{The ideal case: no local noise and perfect polarisation}

\begin{figure}
    \centering
    \includegraphics[width=\linewidth]{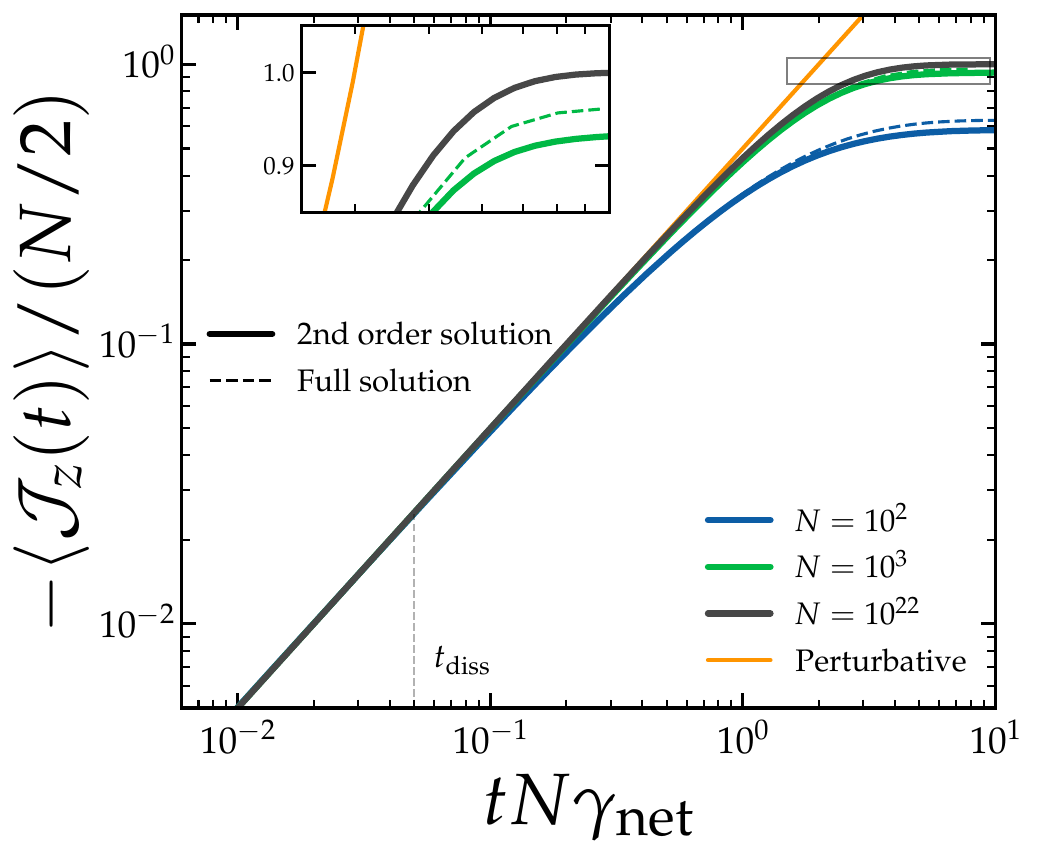}
    \caption{Time evolution of $\langle {\cal J}_z\rangle$ from the initial state $\vert P\rangle$, in the presence of neutrino collective interaction only, with $\gamma_{+}/\gamma_-=0.95$.  The solid blue, green, and gray lines denote our numerical solutions of the Lindblad equation~\eqref{eq:lindnum} using the second-order approximation~\eqref{eq:secondorderapprox} for $N=10^2, 10^3$, and $10^{22}$. In the case of $N=10^2, 10^3$ (blue and green), the second-order solutions are compared with the full density matrix solutions  displayed in long-dashed lines of the same colours, and show excellent agreement. The perturbative solution~\eqref{eq:pert1}, represented here by the red short-dashed line, reproduces the large $N$ numerical solution well at $t\lesssim(N\gamma_{\rm net})^{-1}$.}
    \label{fig:jz_t}
\end{figure}

We consider first the case in which there is no local noise and the polarisation of the initial state is perfect, i.e., $p=1$.
Figure~\ref{fig:jz_t} shows the solutions for the collective observable $\langle {\cal J}_z\rangle$ under these settings for several choices of~$N$.  These are contrasted with numerical solutions of the full density matrix for those choices of $N$ values where a full numerical solution is tractable.  Details of the full density  matrix solution can be found in Appendix~\ref{sec:fullsol}.

Remarkably, where the full density matrix solution can be found, the second-order solution is in excellent agreement with it at all times. This agreement follows from the small $\gamma_{\rm net} \ll \gamma_{\pm}$ of our system, where, as discussed above, the hierarchy of equations~\eqref{eq:secondorderapprox} becomes exact and independent of the truncation scheme in the limit $\gamma_{\rm net} \to 0$,
and indicates that the second-order method is indeed a reliable way to obtain predictions for the collective observables. 
Importantly, the method remains valid for arbitrarily large $N$ values---$N=10^{22}$ in this figure---at no additional cost in terms of computational runtime or memory relative to the small $N$ case. In contrast, the computational cost of numerically solving the full density matrix scales as $\propto N^3$, and becomes prohibitively expensive even on modern computers when $N \gtrsim 10^4$. For completeness, we also show in the same figure the perturbative solution~\eqref{eq:pert1}.  Evidently, our numerical solutions are well approximated by the perturbative solution at $t\lesssim (N\gamma_{\rm net})^{-1}$, especially for large $N$ values.


\subsubsection*{With local noise and partial polarisation}

Having established that the second-order method is reliable, we now use it to investigate more realistic cases in which local noise is present and the polarisation of the initial state may only be partial.

\begin{figure}
    \centering
    \includegraphics[width=\linewidth]{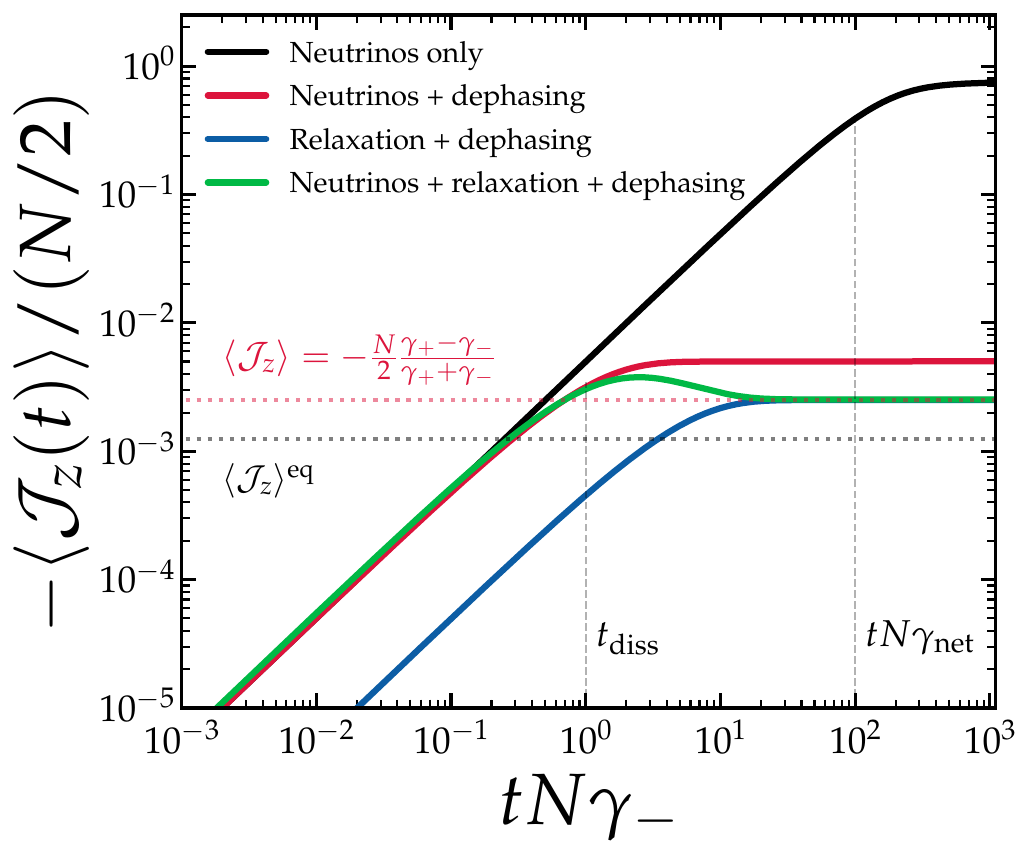}
    \caption{Time evolution of $\langle {\cal J}_z \rangle$ from the initial state $\vert P \rangle$ in the presence of neutrinos and local interactions. 
    The black curve shows the evolution with neutrinos only, as in Fig.~\ref{fig:jz_t}. The red curve includes local dephasing effects with $\gamma_{\phi}^{\rm loc} = N\gamma_-$, such that $t_{\rm diss}$ coincides with $T_2$. 
    In this case, dephasing suppresses the superradiant neutrino interaction, resulting in a neutrino-induced relaxation characterised by $\gamma_+ / \gamma_- = 0.99$. 
    The green curve includes both dephasing and relaxation effects specified by  $\gamma^{\rm loc}_+ / \gamma^{\rm loc}_- = 0.995$ and $T_1 = 10~T_2$ (i.e., $\gamma_+^{\rm loc}+\gamma_-^{\rm loc}\simeq 0.05~\gamma_\phi^{\rm loc}$).  The latter effect leads to the equilibrium magnetisation $\langle {\cal J}_{z}\rangle^{\rm eq}$ defined in Eq.~\eqref{eq:mzeq}.
 Finally, the blue curve represents the evolution of $\langle {\cal J}_z\rangle$ in the absence of neutrinos. 
    All solutions have been computed for $N = 10^3$ using the second-order approximation.}
    \label{fig:jz_t_2}
\end{figure}

Figure~\ref{fig:jz_t_2} illustrates the impact of non-zero dephasing on the observable $\langle {\cal J}_z \rangle$ for a large system size $N$ (red curve). As shown, dephasing suppresses the growth of $\langle {\cal J}_z \rangle$ at late times. In this example, the suppression begins around the dissipative timescale $t_{\rm diss}$ due to our specific choice of local dephasing rate, $\gamma_\phi^{\rm loc} = N\gamma_-$. Different choices for this rate would lead to alternative behaviours, as previously demonstrated in Fig.~\ref{fig:regimes}. While dephasing does not directly alter the dynamics of $\langle {\cal J}_z \rangle$, it inhibits the coherent dissipation mediated by neutrino interactions. This dissipation then drives the system toward its own equilibrium value, given by $\langle {\cal J}_z \rangle = -(N/2) \gamma_{\rm net}/\gamma_{\rm tot}$ in the limit $T_1 \to \infty$.

Finite $T_1$ effects are introduced via the local emission and absorption rates $\gamma_\pm^{\rm loc}$ in the master equation. We select a ratio $\gamma_+^{\rm loc} / \gamma_-^{\rm loc} = 0.995$, corresponding to an equilibrium polarisation of approximately $p_{\rm eq} \sim 10^{-3}$—a value achievable at cryogenic temperatures. The strengths of the local relaxation processes are chosen such that $T_1$ is ten times longer than $T_2$, and we allow this ratio to vary in the forecasting analysis in Sec.~\ref{sec:futureconstraints}. The blue and green curves  in Fig.~\ref{fig:jz_t_2} show the effects of local relaxation and dephasing, with and without superradiant neutrino interactions, respectively.

\begin{figure*}
    \centering
    \includegraphics[width=0.49\linewidth]{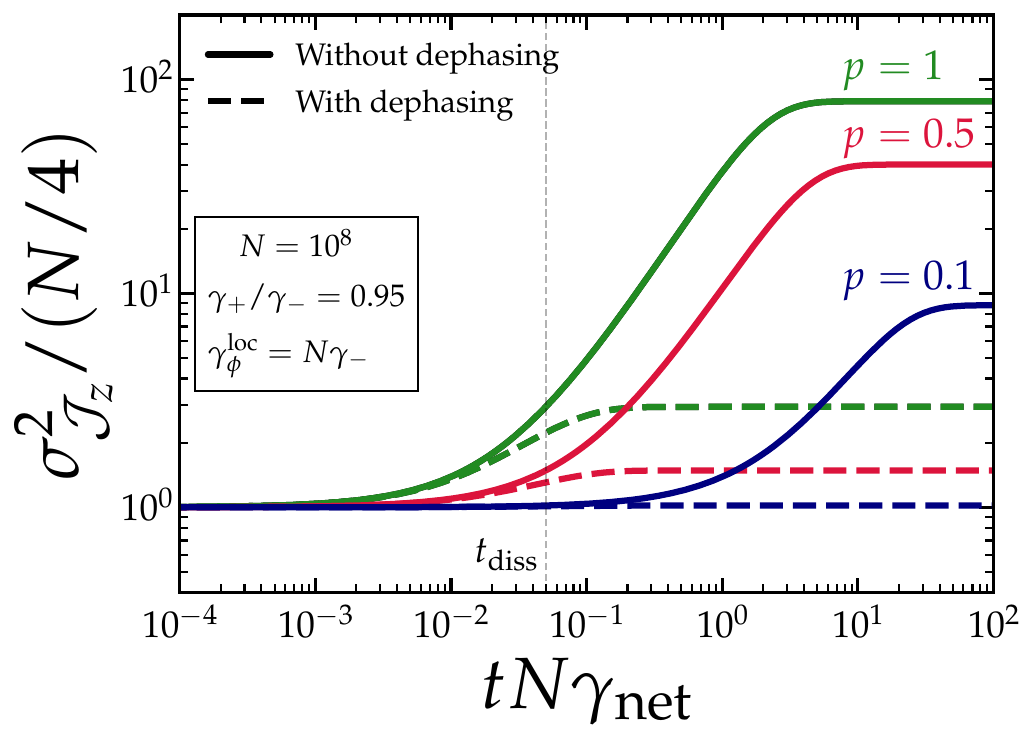}
    \includegraphics[width=0.49\linewidth]{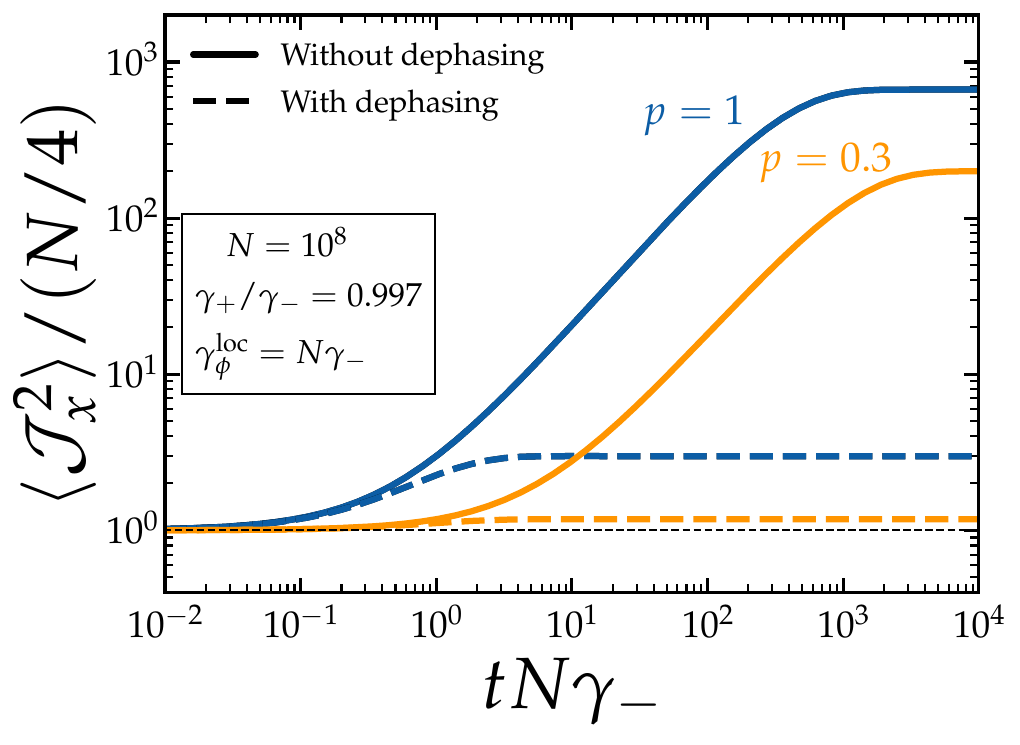}
    \caption{\emph{Left}: Time evolution of $\sigma^2_{{\cal J}_z}$ starting from the product state $\vert P\rangle$ (green lines), and from a state with an initial 10/50\% polarisation in the $z$-axis (blue/red lines), with $\gamma_{+}/\gamma_-=0.95$. In all cases, we illustrate in dashed lines the impact of local dephasing at a rate $\gamma_{\phi}^{\rm loc}=N\gamma_-$. \emph{Right}: Similar to $\sigma^2_{{\cal J}_z}$, but for $\langle J_x^2\rangle$ starting from the fully polarised ground state $\vert G\rangle$ (blue lines), and from a state with 30\% polarisation in the $z$-axis (orange lines).  We have set $\gamma_{+}/\gamma_-=0.997$, and assumed local dephasing at a rate $\gamma_{\phi}^{\rm loc}=N\gamma_-$ for the dashed lines.}
    \label{fig:sigmaz_p}
\end{figure*}

Figure~\ref{fig:sigmaz_p} shows the variance $\sigma_{{\cal J}_z}^2$ of ${\cal J}_z$ and the transverse observable $\langle J_x^2\rangle_{\vert G\rangle}$. Here, in addition to demonstrating the effects of local dephasing, we also show how partial polarisation affects the time evolution of the observables. Clearly, in the presence of dephasing, the variance $\sigma_{{\cal J}_z}^2$ shows minimal growth unless $p > 0.1$.  In the case of $\langle J_x^2\rangle_{\vert G\rangle}$, the polarisation requirement is even stronger---$p>0.3$---although, as discussed in Sec.~\ref{sec:coherentobservables}, we do not expect this observable to be competitive against $\langle J_z \rangle_{|P\rangle}$.
This result therefore highlights the importance of polarisation for the purpose of C$\nu$B detection. For instance, a liquid methanol-based setup~\cite{Walter:2025ldb}, with a polarisation of order $p \sim 10^{-5}$, is currently insufficient for NMR experiments to target C$\nu$B effects. More generally, thermal polarisation levels achievable at typical experimental conditions---${\cal O}(10^{-3})$ for $^{129}$Xe from Eq.~\eqref{eq:thermalpol}---remain too low to yield detectable signals.  Advanced polarisation techniques are thus essential to this venture.

\subsection{Two-time correlation functions}
\label{sec:twotimemain}

We envisage a future experiment that measures an observable, e.g., $\langle{\cal J}_z \rangle_{|P\rangle}$, at a series of time points.  To estimate the sensitivity to the C$\nu$B overdensity parameter of such an experiment, besides the observable's variance $\sigma^2(t)$ at each time $t$, we will also need its covariance $\sigma^2(t,t+\tau)$ at two different times.  This covariance can be constructed from the two-time correlation function, generally written as
 $\langle {\cal O}_1(t) {\cal O}_2(t+\tau)\rangle$, where ${\cal O}_1$ and ${\cal O}_2$ are two operators that act only on the spins.  
 
 Following the quantum regression theorem~\cite{QRT}, the two-time correlation function can be computed as
\begin{equation}
\langle {\cal O}_1(t) {\cal O}_2(t+\tau) \rangle \equiv {\rm tr}[\rho_{S;{\cal O}_1}(t+\tau) {\cal O}_2(0)]\, ,
\end{equation}
where the modified density matrix $\rho_{S;{\cal O}_1}$ is defined at $t$ via $\rho_{S;{\cal O}_1}(t) = \rho_S(t) {\cal O}_1(0)$.  The evolution of $\rho_{S;{\cal O}_1}(t+\tau)$ from $t$ to $t+\tau$  for $\tau \geq 0$ is governed by  the same Lindblad master equation~\eqref{eq:lindnum} used to compute our observables but for a change of the time variable $t \to \tau$, i.e.,
\begin{equation}
 \frac{{\rm d}}{{\rm d} \tau} \rho_{S:{\cal O}_1} (t+\tau)  = {\cal L}\{\rho_{S;{\cal O}_1}(t+\tau) \}\, ,  
 \label{eq:lindcorr}
\end{equation}
where ${\cal L}$ is the Lindblad super-operator of Eq.~\eqref{eq:lindnum}.
A more detailed discussion about this equation of motion and the construction of two-time correlation functions can be found in Appendix~\ref{sec:twotime}.

Since our main observable is $\langle {\cal J}_z \rangle_{|P\rangle}$, we are particularly interested in the two-time correlation function $\langle {\cal J}_z(t) {\cal J}_z(t+\tau) \rangle$.  Following the second-order method of Sec.~\ref{sec:2ndorder}---or, equivalently, using the correspondence between Eqs.~\eqref{eq:corr1} and \eqref{eq:corr2}) in Appendix~\ref{sec:twotime}---a hierarchy of differential equations can be constructed from Eq.~\eqref{eq:lindcorr} to describe its time evolution, namely,
\begin{widetext}
\begin{equation}
\begin{aligned}
\frac{{\rm d}}{{\rm d} \tau} \langle {\cal J}_z (t) {\cal J}_z(t+\tau) \rangle
 = \; & \sum_\pm \pm \gamma_\pm \Big[\langle {\cal J}_z(t) {\cal J}^2(t+\tau)  \rangle - \langle  {\cal J}_z(t){\cal J}_z^2(t+\tau) \rangle  \mp \langle {\cal J}_z(t) {\cal J}_z(t+\tau) \rangle
\Big] \\
\; &  + \sum_\pm \pm \gamma_\pm^{\rm loc} \Bigg[\mp\langle {\cal J}_z (t) {\cal J}_z(t+\tau) \rangle + \frac{N}{2}\langle {\cal J}_z(t)\rangle \Bigg] \, , \\
\frac{\rm d}{{\rm d}\tau} \langle {\cal J}_z(t)   {\cal J}^2(t+\tau) \rangle =
\; &- \sum_{\pm}\gamma_\pm^{\rm loc} \Big[\langle  {\cal J}_z(t) {\cal J}^2(t+\tau)\rangle +\langle {\cal J}_z(t) {\cal J}_z^2 (t+\tau)\rangle \mp (N-1) \langle {\cal J}_z (t) {\cal J}_z(t+\tau) \rangle  - N \langle {\cal J}_z(t) \rangle\Big]  \\
\; & - \gamma_{\phi}^{\rm loc} \Bigg[ \langle {\cal J}_z(t)  {\cal J}^2(t+\tau)  \rangle - \langle {\cal J}_z(t) {\cal J}_z^2(t+\tau)   \rangle  - \frac{N}{2} \langle{\cal J}_z(t)\rangle \Bigg]\, ,\\
\frac{\rm d}{{\rm d}\tau} \langle  {\cal J}_z(t)  {\cal J}_z^2(t+\tau) \rangle  
=\; & \sum_{\pm}\gamma_{\pm} \Big[ \langle  {\cal J}_z(t)  {\cal J}^2(t+\tau)  \rangle  - 3 \langle  {\cal J}_z(t) {\cal J}_z^2(t+\tau) \rangle \mp \langle {\cal J}_z(t) {\cal J}_z(t+\tau) \rangle     \\
\; & \qquad \qquad \mp  2 \langle  {\cal J}_z(t)  {\cal J}_z^2(t+\tau) \rangle \langle {\cal J}_z(t+\tau)  \rangle \pm 2 \langle {\cal J}_z(t)  {\cal J}^2(t+\tau)   \rangle \langle  {\cal J}_z(t+\tau) \rangle  \Big]\\
\; &  - \sum_\pm\gamma_\pm^{\rm loc} \Bigg[2 \langle  {\cal J}_z(t) {\cal J}_z^2(t+\tau) \rangle \mp(N-1) \langle {\cal J}_z(t) {\cal J}_z(t+\tau) \rangle -\frac{N}{2} \langle  {\cal J}_z(t)\rangle\Bigg] \, ,
    \label{eq:secondorderapproxCorr}
\end{aligned}
\end{equation}
\end{widetext}
where we have truncated the hierarchy by approximating $\langle {\cal J}_z(t)  {\cal J}_z^3(t+\tau)  \rangle \approx \langle {\cal J}_z(t)  {\cal J}_z^2(t+\tau)  \rangle \langle {\cal J}_z (t+\tau)\rangle$  and $\langle {\cal J}_z(t) {\cal J}_z(t+\tau) {\cal J}^2(t+\tau)   \rangle \approx \langle {\cal J}_z(t)  {\cal J}^2(t+\tau)   \rangle \langle {\cal J}_z(t+\tau)\rangle$,
with  $\langle {\cal J}_z(t+\tau)\rangle \equiv {\rm tr}[\rho_S(t+\tau){\cal J}_z]$.  We also remind the reader that the time variable here is $\tau$, and the function $\langle {\cal J}(t) \rangle$ which appears in all three equations above does not evolve with $\tau$.

Equation~\eqref{eq:secondorderapproxCorr} can be solved from any time point $t$ in the duration of an experiment, given the ``initial conditions'' at $\tau=0$, 
$\langle {\cal J}_z(t) \rangle$, $\langle {\cal J}_z^2(t) \rangle$, $\langle {\cal J}^2(t) {\cal J}_z(t)  \rangle$, and 
$\langle {\cal J}_z^3(t) \rangle$.  For a spin ensemble originally prepared in the $|P\rangle$ state, these initial conditions amount at $t=0$ to
\begin{equation}
\begin{aligned}
& \langle {\cal J}_z^2(0) \rangle_{|P\rangle} = \frac{N}{4}\,, \\
& \langle {\cal J}_z(0) \rangle_{|P\rangle} = \langle {\cal J}^2(0) {\cal J}_z(0)  \rangle_{|P\rangle} = \langle {\cal J}_z^3(0) \rangle_{|P\rangle}=0\, .
\end{aligned}
\end{equation}
For $t >0$, on the other hand, $\langle {\cal J}_z(t)\rangle$ and $\langle {\cal J}_z^2(t) \rangle$ can be obtained directly from the solutions to Eq.~\eqref{eq:secondorderapprox}, while we use the same solutions to approximate $\langle {\cal J}^2(t) {\cal J}_z(t)  \rangle \approx \langle {\cal J}^2(t)\rangle \langle {\cal J}_z(t)  \rangle $ and 
$\langle {\cal J}_z^3(t) \rangle\approx \langle {\cal J}_z(t) \rangle \langle {\cal J}_z^2(t) \rangle$.

We are particularly interested in the covariance of the observable $\langle {\cal J}_z \rangle_{|P\rangle}$ at different times, i.e, the connected piece of the two-time correlation function, defined as
\begin{equation}
\begin{aligned}
\sigma^2_{{\cal J}_z}(t,t+\tau) \equiv\; & \langle {\cal J}_z(t) {\cal J}_z(t+\tau) \rangle - \langle {\cal J}_z(t) \rangle \langle {\cal J}_z(t+\tau) \rangle \\
= \; & \sigma^2_{{\cal J}_z}(t+\tau,t) \, ,
\label{eq:connected_piece}
\end{aligned}
\end{equation}
which we will use in Sec.~\ref{sec:futureconstraints} to characterise the theoretical uncertainties in the observable.  This connected piece can be obtained from the solutions of Eqs.~\eqref{eq:secondorderapprox} and~\eqref{eq:secondorderapproxCorr} for every pair of $(t, t+\tau)$.  It is however also possible---and instructive---to construct an equation of motion directly for $\sigma^2_{{\cal J}_z}(t,t+\tau)$ from Eqs.~\eqref{eq:secondorderapprox} and~\eqref{eq:secondorderapproxCorr}, which reads
\begin{equation}
\begin{aligned}
\frac{{\rm d}}{{\rm d} \tau} \sigma^2_{{\cal J}_z} = & - \gamma_{\rm net} \Big[\langle {\cal J}_z(t) {\cal J}^2(t+\tau)\rangle - \langle {\cal J}_z(t) \rangle \langle {\cal J}^2(t+\tau) \rangle \\
& \qquad -\langle {\cal J}_z(t) {\cal J}_z^2(t+\tau)\rangle + \langle {\cal J}_z(t) \rangle \langle {\cal J}_z^2(t+\tau) \rangle \Big]\\
& -(\gamma_{\rm tot} + \gamma_+^{\rm loc}+\gamma_-^{\rm loc} ) \, \sigma^2_{{\cal J}_z} (t,t+\tau)\, .\label{eq:mom_sigma22}
\end{aligned}
\end{equation}
Where the $\propto \gamma_{\rm net}$ terms are suppressed relative to the $\propto (\gamma_{\rm tot}+\gamma_-^{\rm loc}  +\gamma_+^{\rm loc})$ term, this equation of motion is solved approximately by
\begin{equation}
\begin{aligned}
\sigma^2_{{\cal J}_z} (t,t+\tau) = \sigma^2_{{\cal J}_z} (t+\tau,t) \approx \; & \sigma^2_{{\cal J}_z}(t)\, e^{-(\gamma_{\rm tot} + \gamma_-^{\rm loc} +\gamma_+^{\rm loc})\, \tau}   \\
\approx \; & \sigma^2_{{\cal J}_z}(t)\, e^{-(\gamma_-^{\rm loc}+\gamma_+^{\rm loc})\, \tau}\, ,
\label{eq:sigma2off}
\end{aligned}
\end{equation}
where we have used $\gamma_\pm \ll \gamma_\pm^{\rm loc}$ at the second equality, and 
$\sigma^2_{{\cal J}_z}(t)\equiv \langle {\cal J}_z^2 (t) \rangle -\langle {\cal J}_z(t) \rangle^2$ is the usual variance at time $t$ also defined earlier in Eq.~\eqref{eq:variance_z}.

\begin{figure}
    \centering
    \includegraphics[width=\linewidth]{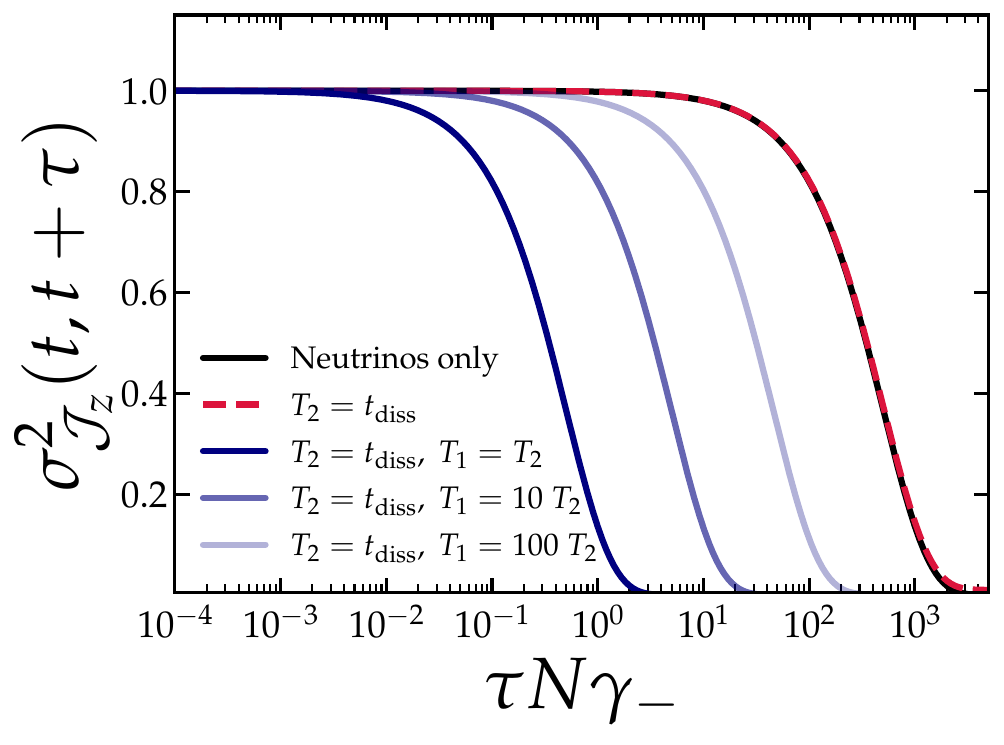}
    \caption{Time evolution of the covariance of ${\cal J}_z$, i.e., the connected piece of the two-time correlation function, $\sigma^2_{{\cal J}_z}(t, t+\tau)$, defined in Eq.~\eqref{eq:connected_piece}, for $t=0$. The correlation decays away on a timescale specified by $T_1 = 1/(\gamma_-^{\rm loc}+\gamma_+^{\rm loc})$ as predicted in Eq.~\eqref{eq:sigma2off},
     while the dephasing timescale $T_2$ does not play a role.  All solutions have been obtained from solving Eqs.~\eqref{eq:secondorderapproxCorr} and~\eqref{eq:mom_sigma22} for $N=10^3$.  We observe however that they agree with the full density matrix solution at the $0.1\%$ level.}
    \label{fig:sigma_corr}
\end{figure}

Figure~\ref{fig:sigma_corr} shows the time evolution of the two-time correlation function $\sigma^2_{{\cal J}_z}(t, t+\tau)$, obtained from solving Eq.~\eqref{eq:secondorderapproxCorr}, for $N=10^3$ and $t=0$.
In the case $\gamma_-^{\rm loc}=N\gamma_-$ (dark blue solid line), we see an exponential decay at a characteristic time $\tau\sim 1/(2 N\gamma_-)\approx 1/(\gamma_-^{\rm loc}+\gamma_+^{\rm loc})$, while in the absence of local emission/absorption the correlation decays at $\tau\sim 1/(2\gamma_-)\approx 1/\gamma_{\rm tot}$ due to neutrino dissipation alone, confirming the analytical estimate~\eqref{eq:sigma2off}.
Observe also that  local dephasing effects do not influence the decay in the longitudinal correlation.
This is primarily a consequence of the small $\gamma_{\rm net}$ in our scenario, which inhibits the effects of $\gamma_\phi^{\rm loc}$ on the evolution of $\sigma_{{\cal J}_z}^2$.
Lastly, while Fig.~\ref{fig:sigma_corr} presents the numerical solutions for $t=0$, the two-time correlation for other choices of $t$ exhibits the same decay behaviour, as suggested by Eq.~\eqref{eq:sigma2off}.

\section{Future constraints on the C\texorpdfstring{$\nu$}{nu}B}
\label{sec:futureconstraints}

To assess the sensitivity of future experiments to collective effects of the C$\nu$B, we consider an idealised NMR experiment using polarised $^{129}$Xe nuclei, as envisioned in future stages of the CASPEr experiment~\cite{Walter:2025ldb}. 
$^{129}$Xe is a noble gas isotope with a nuclear spin of $I = 1/2$, making it naturally suited to modelling as a spin-$1/2$ system. In liquid form Xenon has a density of $\rho = 2.94$ g/cm$^3$, which gives 
\begin{equation}
    n_{\rm Xe}\simeq 1.35 \times 10^{22}~{\rm cm}^{-3}\, 
\end{equation} 
as the number density of Xenon nuclei given an atomic weight of $A_r({\rm Xe})=131.29$~\cite{Prohaska}. 
Assuming 100\% isotopic enrichment of $^{129}$Xe can be achieved, we use the same number to approximate  the $^{129}$Xe spin density~$n_s$. 

The gyromagnetic ratio of $^{129}$Xe is approximately $\gamma_{\text{Xe}} \simeq 11.8~\text{MHz/T}$, which gives a maximum energy splitting of $\omega_0 \simeq 5.8 \times 10^{-7}~\text{eV}$  for a maximum magnetic field strength of $B_{\rm max} = 12$~T. 
For a fixed neutrino mass, the C$\nu$B-induced net interaction rate $\gamma_{\rm net} \propto 1-\gamma_+/\gamma_- \sim 2 m_\nu \omega_0/p_\nu^2$
 generally increases with $\omega_0$, as also shown in Fig.~\ref{fig:param_space}.
  However, $\omega_0$ cannot be made arbitrarily large: for the  typical momentum transfer $q_{\rm typ} \sim \omega_0 m_{\nu}/p_\nu$, the coherence condition must eventually break down  when $q_{\rm typ}$ reaches $\sim R^{-1}$.  Then, setting the coherence limit at $q_{\rm typ}R < 1$ immediately yields an upper bound on the energy splitting $\omega_0 < p_{\nu}/m_{\nu} R$. Thus, in our modelling, the optimal energy splitting is determined by the smaller of the maximum value achievable in the laboratory and the coherence limit, namely, 
\begin{equation}
\omega_0^{\star} = \min\left( \gamma_{\text{Xe}} B_{\text{max}}, \frac{p_{\nu}}{m_1 R} \right),    \label{eq:wo_opt}
\end{equation}
where we have set $m_{\nu}=m_1$.  Assuming the normal hierarchy, this choice of $m_\nu$ gives the largest possible value of $\omega^\star_0$, leading to the best sensitivity to $\delta\nu$.

As already mentioned in Sec.~\ref{sec:dickebasis},
the preparation of the ground state, with all spins aligned along the magnetic field direction (i.e., the $-z$-direction in this work), can be achieved via spin-exchange optical pumping.  This technique  is commonly used to hyperpolarise noble gases like Xenon~\cite{PhysRevLett.87.067601,RevModPhys.69.629} and can reach polarisations up to 90\%~\cite{doi:10.1021/jp501493k}. Notably, hyperpolarised $^{129}$Xe exhibits exceptionally long transverse relaxation times, with values estimated to be up to $T_2\sim 1000$~s~\cite{PhysRevLett.87.067601, Walter:2025ldb}.  
After the nuclear spin ensemble has attained the desired polarisation,
a coherent spin (product) state can be prepared by applying a resonant $\pi/2$ radio-frequency  pulse at the Larmor frequency. This effectively rotates the collective spin vector into the equatorial plane (see Fig.~\ref{fig:DickeTriangle}), thereby allowing us to use the $\langle {\cal J}_z \rangle_{|P \rangle}$ observable to constrain collective neutrino effects. Under certain conditions, the longitudinal relaxation time  $T_1$ can also be long, e.g., $T_1\gtrsim $ hours for solid-state $^{129}$Xe~\cite{PhysRevB.94.094309}. In the following we take $T_1/T_2=10$, which corresponds to a value $T_1\simeq 2.8$ hours for $T_2=1000$ s, modelled according to Eqs.~\eqref{eq:t1a} and~\eqref{eq:t2a}.


\subsection{\texorpdfstring{$\chi^2$}{chi} analysis}

To estimate the sensitivity to the C$\nu$B overdensity parameter $\delta_\nu$, we 
 simulate a mock experimental data set ${\cal \hat{J}}_z(t_i)$ of a fiducial model that does {\it not} contain the coherent neutrino signal,%
\footnote{A precessing transverse magnetisation can in principle induce a current in the detection coil and generate a feedback magnetic field that accelerates the decay of the magnetisation.  This effect is commonly known as radiation damping and may exhibit a collective behaviour scaling.  In this work, we adopt a fiducial model that does not include such collective background effects, which can be justified under certain conditions~\cite{Galanis:2025amc}. A full modeling of the experimental environment lies beyond the scope of this study.}
comprising $N_{\text{shots}}$ measurements at $n_t$ discrete time points sampled from $t_1 = 1/f_s$ to $t_{n_t} = T_2$. The sampling frequency is fixed to $f_s = 14.3$~kHz, in line with the most recent CASPEr setup~\cite{Walter:2025ldb}.  
Then, assuming the likelihood function (i.e., probability of data given a theory) to be gaussian, we can define a $\chi^2$-statistics via
\begin{equation}
\chi^2  (\delta_\nu)= N_{\rm shots} \sum_{ij}^{n_t} d_i \left[C^{-1} \right]_{ij} d_j\, ,
\label{eq:chi2}
\end{equation}
where $C$ is the covariance matrix to be defined below, and 
\begin{equation}
d_i \equiv     \langle {\cal J}_z(t_i; \delta_{\nu})\rangle_{|P\rangle} -\langle {\cal \hat{J}}_z(t_i)\rangle_{|P \rangle}
\end{equation}
is theoretical prediction minus the mock data at time~$t_i$.  

The theoretical prediction  $\langle {\cal J}_z(t_i; \delta_{\nu})\rangle_{|P\rangle}$ {\it includes} neutrino effects tuned by the overdensity parameter~$\delta_\nu$. To ensure consistency, we take $\langle {\cal J}_z(t_i; \delta_{\nu}) \rangle_{|P\rangle}$ from the numerical solutions of Sec.~\ref{sec:numerical}, with the time variable rescaled by a factor $(2 p_\nu R)^{2}$  to account for partial coherence in the  $2 p_\nu R > 1$ regime (note that the local rates $\gamma_{\pm,\phi}^{\rm loc}$ must also be simultaneously rescaled by the same factor in order to preserve their effects). 
For the mock data ${\cal \hat{J}}_z(t_i)$, on the other hand, we assume for simplicity that they coincide with the expectation value $\langle {\cal \hat{J}}_z(t_i) \rangle_{|P \rangle}$ 
under the fiducial model (i.e., {\it without} neutrino effects), 
as given by Eq.~\eqref{eq:fidmz} with $\gamma_-^{\rm loc} \simeq 0.026 \gamma_\phi^{\rm loc}$
and $\gamma_+^{\rm loc}/\gamma_-^{\rm loc}=0.995$ (i.e., blue line in Fig.~\ref{fig:jz_t_2}).

The covariance matrix $C_{ij}=C_{ji}$ is given by
\begin{equation}
C_{ij} = C_{ij}^{\rm th} + \sigma^2_{\rm SQUID} \delta_{ij}\, ,   
\end{equation}
and includes contributions from the total instrumental noise~$\sigma_{\rm SQUID}^2$, as well as 
the theoretical covariance of the model including neutrinos, 
\begin{equation}
\begin{aligned}
C_{ij}^{\rm th} =\; & \begin{cases}
                    \sigma^2_{{\cal J}_z} (t_i;\delta_\nu)\, , & i=j\, , \\
                    \sigma^2_{{\cal J}_z}(t_i; \delta_\nu)\,  e^{-(\gamma_- ^{\rm loc}+\gamma_+^{\rm loc})|(t_j-t_i)}\, , & i < j\, ,\\
                     \sigma^2_{{\cal J}_z}(t_j; \delta_\nu) \, e^{-(\gamma_- ^{\rm loc}+\gamma_+^{\rm loc})|(t_i-t_j)}\, , & i > j\, ,\\
                    \end{cases}   
\end{aligned}
\end{equation}
where $\sigma^2_{{\cal J}_z}(t_i;\delta_\nu)$ is the theoretical variance of ${\cal J}_z$, and we have used the approximation~\eqref{eq:sigma2off} for the two-time correlation functions $\sigma^2_{{\cal J}_z}(t_i, t_j; \delta_\nu)$ in the off-diagonal ($i\neq j$) entries.  Again, we take  $\sigma^2_{{\cal J}_z}(t_i;\delta_\nu)$ from the numerical solutions of Sec.~\ref{sec:numerical}, with a rescaling factor $(2 p_\nu R)^{2}$ applied to the time scale and the local rates $\gamma_{\pm,\phi}^{\rm loc}$.

The total instrumental noise, which we refer to as ``SQUID noise'', includes intrinsic SQUID contributions, thermal (Johnson–Nyquist) noise from the input circuit, as well as other noise sources originating from, e.g., the readout strategy. These noises are assumed to be random, uncorrelated between independent measurements, and uncorrelated with the signal. We model their total variance as  
\begin{equation}
 \sigma_{\rm SQUID}^2 = \frac{N}{4}\, \mathcal{F}_{\rm SQUID} \,,
\end{equation} so that the ratio \begin{equation}
\mathcal{F}_{\rm SQUID} \approx \frac{\mathcal{P}_{\rm SQUID}}{\mathcal{P}_{\rm SPN}}
\label{eq:fsquid}
\end{equation} quantifies the ratio of instrumental noise relative to the spin projection noise. The SQUID and SPN power spectral densities, $\mathcal{P}_{\rm SQUID}(f)$ and $\mathcal{P}_{\rm SPIN}(f)$, are specific to the experimental setup.  For the CASPEr experiment~\cite{Walter:2025ldb}, we estimate that  $\mathcal{F}_{\rm SQUID} \sim 85$.%
\footnote{Using the specifications given in Ref.~\cite{Walter:2025ldb}, we estimate that at $f = 1.2$~MHz (corresponding to $B = 0.1$~T for $^{129}$Xe), the SQUID spectral density is approximately independent of frequency and takes on the value of $\mathcal{P}_{\rm SQUID} \sim 10^{-12}~\Phi_0^2/\mathrm{Hz}$. The SPN power is on the other hand $\mathcal{P}_{\rm SPN} \sim 1.2 \times 10^{-14}~\Phi_0^2/\mathrm{Hz}$.  Combining these numbers per Eq.~\eqref{eq:fsquid} yields $\mathcal{F}_{\rm SQUID} \sim 85$.}
In the following analysis, however, we shall retain $\mathcal{F}_{\rm SQUID}$ as a free parameter.


\subsection{Forecasted constraints}

By construction, the $\chi^2$ of Eq.~\eqref{eq:chi2} is automatically zero when $\delta_\nu=0$.  Then, to determine an upper limit on $\delta_\nu$ we simply need to find the smallest value of $\delta_\nu$ that satisfies
\begin{equation}
    \chi^2(\delta_\nu) > \Delta\chi^2\, ,
\end{equation} 
where $\Delta\chi^2 = 2.71~(3.84)$ corresponds to a 90\% (95\%) confidence level for one degree of freedom. 

\begin{figure}
    \centering
    \includegraphics[width=\linewidth]{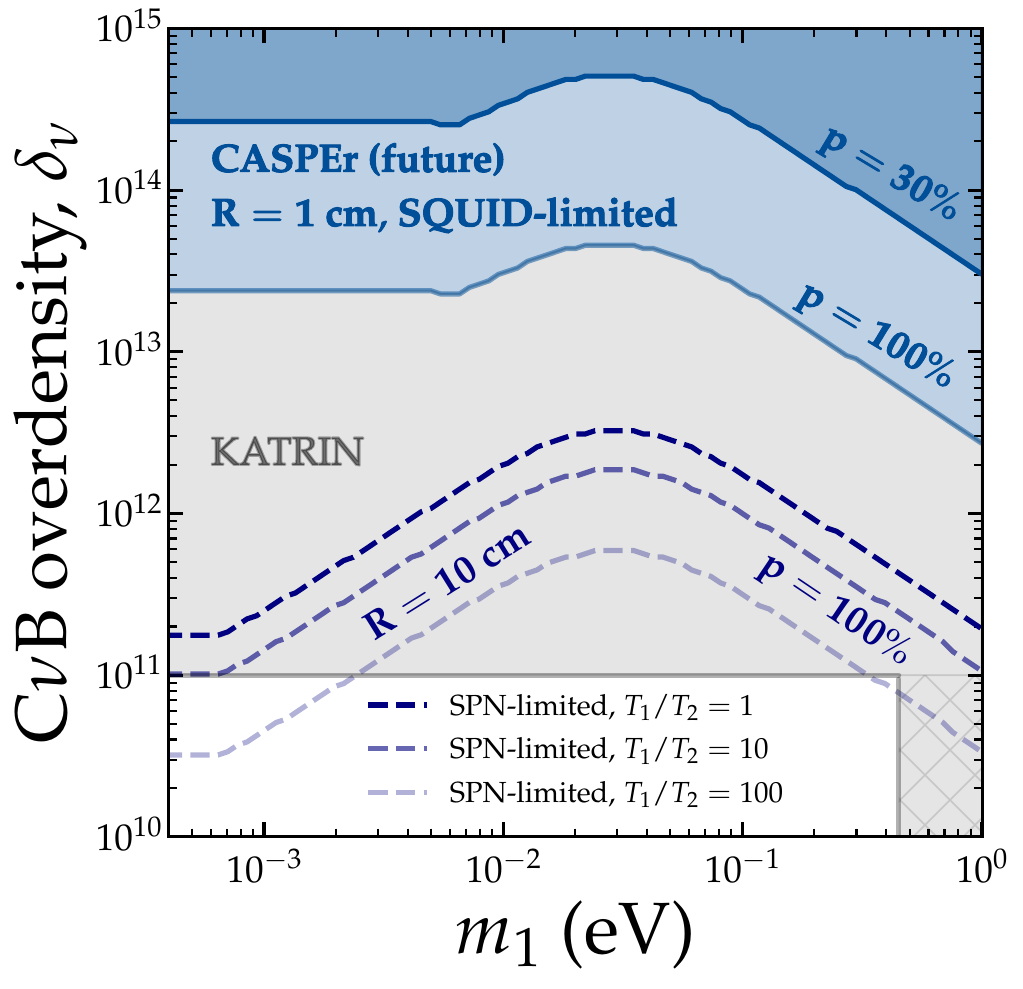}
    \caption{Projected sensitivity to the C$\nu$B overdensity parameter $\delta_\nu$ as a function of the lightest neutrino mass $m_1$, assuming normal ordering. The two blue solid lines correspond to realistic experimental setups planned by the CASPEr experiment~\cite{Walter:2025ldb}, with partial polarisations $p = 0.3$ and $p=1$, a sample radius $R= 1~\mathrm{cm}$, a SQUID noise-to-SPN ratio $\mathcal{F}_{\rm SQUID} = 50$, and the fiducial value $T_1/T_2=10$. 
    The blue dashed lines represent the sensitivity to $\delta_\nu$ of an idealised spin-projection-noise-limited experiment, working at the standard quantum limit, with perfect initial polarisation ($p = 1$) using a 10 cm-radius spherical sample of $^{129}$Xe, for several choices  of the ratio $T_1/T_2$. In all cases we assume an effective coherence time of $T_2 = 10^3$~s, $N_{\rm shots}=10^3$ repeated measurements, and an optimal energy splitting $\omega^\star_0$ according to Eq.~\eqref{eq:wo_opt} for all values of $m_1$. All exclusion projections have been derived at 90\% confidence level. We additionally show the regions excluded by the 90\% C.L.\ KATRIN bounds on the C$\nu$B overdensity~\cite{KATRIN:2022kkv} (horizontal grey region) and the neutrino mass~\cite{KATRIN:2024cdt} (vertical grey region).}
    \label{fig:delta_nu}
\end{figure}

We consider first the best possible constraint on $\delta_\nu$.  This can be achieved by assuming perfect polarisation $p=1$ and an optimal splitting energy $\omega^\star_0$ as defined in Eq.~\eqref{eq:wo_opt}.  In this idealised scenario, we find that a 10 cm-radius spherical sample of $^{129}$Xe nuclei would be able to constrain the C$\nu$B overdensity at 
\begin{equation}
\delta^{\star}_{\nu}\lesssim 1.7 \times 10^{12}\,, \quad 90\%~{\rm C.L.}\, , 
\label{eq:delta_best}
\end{equation}
for the normally-ordered $\sum m_\nu=0.15$ eV benchmark scenario. The limit assumes a coherence time of $T_2=10^3$~s, $T_1 = 10~T_2$, and $N_{\rm shots}=10^3$ along the lines of Ref.~\cite{Arvanitaki:2024taq}, and is comparable to the estimated sensitivity quoted therein (which assumes hydrogen spins) within an order of magnitude. At face value, such an optimistic sensitivity appears uncompetitive against the current KATRIN limit on the neutrino overdensity of $\delta_\nu \lesssim 10^{11}$ for the benchmark scenario (but may  be marginally better for a small range of neutrino masses, depending on the ratio $T_1/T_2$; see Fig.~\ref{fig:delta_nu}). Note however that this conclusion may  change if the system achieves at least an $\mathcal{O}(10)$ value in squeezing~\cite{Galanis:2025amc}.

Realistic NMR experiments face even more limitations: imperfect initial polarisation, smaller sample volumes, and instrumental noise. 
Assuming an experimental setup likely to be realised by CASPEr---instrumental-noise-limited and using a smaller sample volume of $V\sim R^3 \sim {\rm cm}^{3}$---the sensitivity to $\delta_\nu$ reduces significantly to 
\begin{equation}
    \delta_{\nu}^{\rm CASPEr}\lesssim 5.3\times 10^{13}~\left(\frac{0.5}{p}\right)^2\, ,
    \label{eq:realisticlimit}
\end{equation}
at 90\% confidence level, again for the normally-ordered $\sum m_\nu=0.15$ eV benchmark.

Figure~\ref{fig:delta_nu} shows the dependence of the projected $\delta_\nu$ sensitivities on the lightest neutrino mass $m_1$, assuming normal ordering. From $m_1 \simeq 0.05$~eV up to the KATRIN neutrino mass limit $<0.45$~eV~\cite{KATRIN:2024cdt}, future setups following the specifications of CASPEr will be able to set limits of $\delta_\nu \lesssim 10^{12}-10^{14}$,  depending on the SQUID noise floor and provided that the initial sample polarisation reaches ${\cal O}(1)$ values. Because the sensitivity to $\delta_\nu$ increases with neutrino mass for $m_1 \gtrsim 0.05$~eV, our forecasted CASPEr constraints are generally better than the benchmark value~\eqref{eq:realisticlimit} for larger neutrino masses.

For $m_1\lesssim 0.05$~eV, the mass dependence in the rates is lost, and the sensitivities to $\delta_\nu$ asymptote to $\delta_\nu \sim 10^{13}$. On the other hand, while the $m_1$ dependence also drops out in the optimistic ($R=10$ cm) constraint at very small masses,  we still observe a somewhat favourable scaling in the range $m_1\in (2\times 10^{-3}, 3\times 10^{-2})$ eV. This arises because, for a larger sample size $R = 10$~cm, the optimal energy splitting can be tuned to higher values and only reaches the ceiling imposed by the maximum magnetic field $B_{\rm max}$ at around $m_1\sim 10^{-3}$~eV. The effect of this ceiling can be clearly discerned in Fig.~\ref{fig:delta_nu}, where the optimistic sensitivity plateaus at a value of $\delta_\nu\sim 5\times 10^{10}$ for $m_1\lesssim 10^{-3}$~eV.

\begin{figure*}
    \centering
    \includegraphics[width=0.49\linewidth]{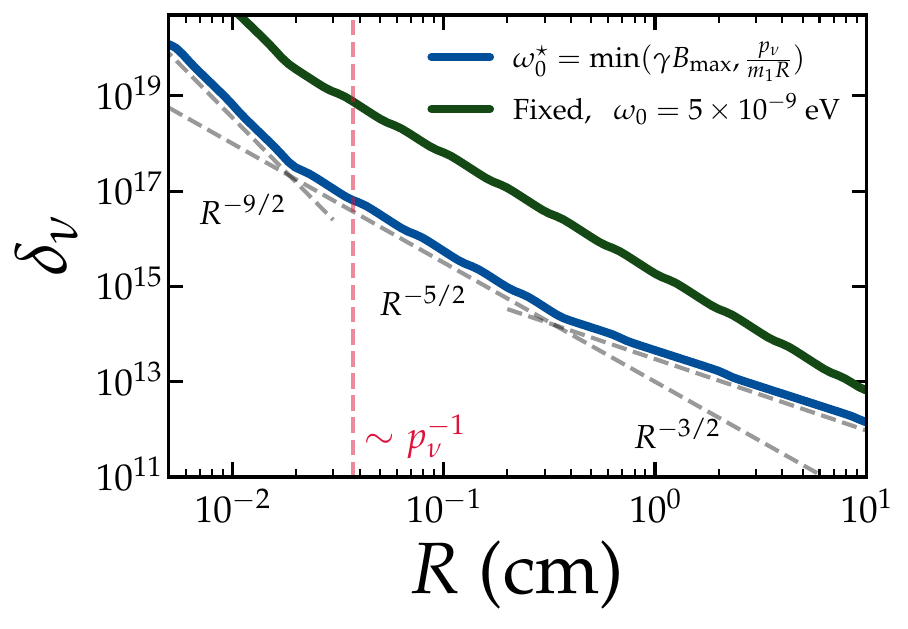}
    \includegraphics[width=0.48\linewidth]{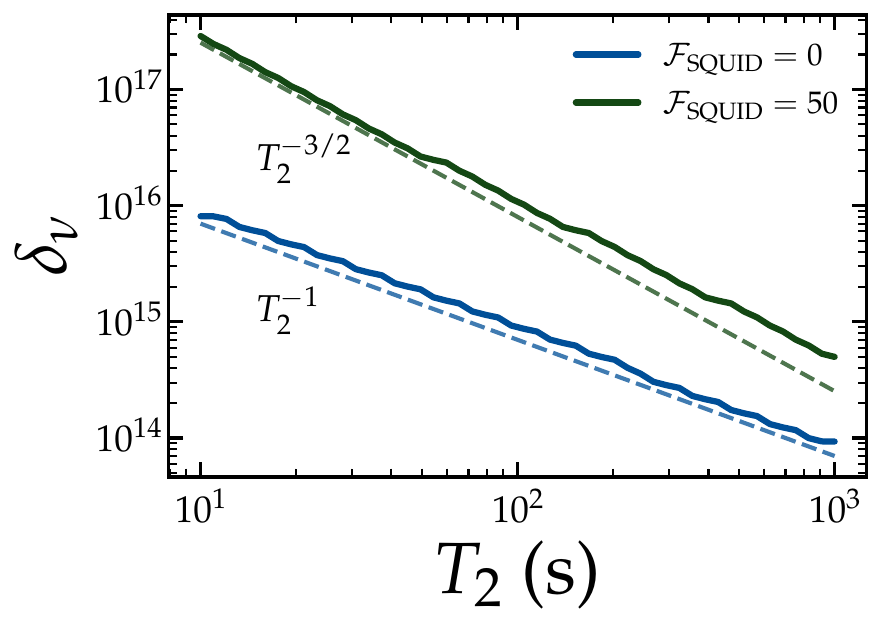}
    \caption{Parametric dependence of the experimental sensitivity to the C$\nu$B overdensity parameter $\delta_\nu$.  {\it Left}: Dependence on the spin ensemble radius $R$.  Here, we illustrate that, despite a weaker scaling at large $R$, choosing the optimal energy splitting $\omega^\star_0$, as given by Eq.~\eqref{eq:wo_opt} for a fixed radius $R$, yields a better sensitivity to $\delta_\nu$ than choosing a fixed splitting with $B<1$~T. See the main text for a detailed explanation of the scaling behaviours with $R$.  {\it Right}: Dependence on the transverse coherence time $T_2$, for both a quantum-limited experiment and a SQUID-noise-dominated experiment with ${\cal F}_{\rm SQUID}=50$.}
    \label{fig:deltaR_scaling}
\end{figure*}

We display in Fig.~\ref{fig:deltaR_scaling} the parametric dependence of the experimental sensitivity to $\delta_\nu$ on the system size $R$ and the transverse coherence time $T_2$ in the SQUID-noise-dominate regime (i.e., ${\cal F}_{\rm SQUID} \gg 1$).  As can be seen, when the condition $p_{\nu} R < 1$ is satisfied, the system is in the fully coherent regime, where the interaction rates  $\Gamma_\pm$ scale as $\propto R^6$. Taking into account the scaling of the total uncertainty, $\sigma \propto N^{1/2} \propto R^{3/2}$, this leads to a scaling of $R^{-9/2}$ for the $\delta_\nu$ sensitivity.  This behaviour applies to both a fixed energy splitting and the optimal case, since the optimal splitting here is effectively constant and determined by the maximum magnetic field achievable in the experiment. As the radius increases, coherence is eventually lost around $R \sim p_\nu^{-1}$ except in the forward direction, resulting in a reduced interaction rate that scales as $\Gamma_\pm \propto R^4$ and hence a $\delta_\nu$ constraint scaling of $R^{-5/2}$. This scaling remains the same even at $p_\nu R \gg 1$ in the case of a fixed energy splitting.

In the case of optimal splitting, however, a change in the scaling behaviour occurs around $R \sim 0.5$~cm for $^{129}$Xe spins. This transition arises because, at these radii, values of the magnetic field required to tune  $\gamma B_{\rm max}$ to the optimal value $p_\nu/(m_1 R)$ are achievable experimentally.
As such, the energy splitting $\omega_0$ becomes dependent on $R$, which leads the net rate $\gamma_{\rm net} \approx  (2 m_1 \omega_0/p_\nu^2) \gamma_\pm$ itself to directly scale with $R^{-1}$ at the optimal $\omega_0$. Combined with the loss of coherence at $p_\nu R \gg 1$, the collective net interaction rate here scales as $\propto R^3$.
Consequently, the sensitivity to $\delta_\nu$ scales as $R^{-3/2}$ in this regime.
Note that, had we used a different target material with a different gyromagnetic ratio, the transition in the scaling behaviour would occur at a different value of $R$. 

In the right panel of Fig.~\ref{fig:deltaR_scaling}, we present the $T_2$ scalings. In the standard quantum-limited scenario, where SQUID noise can be suppressed, the sensitivity to $\delta_\nu$ scales as~$T_2^{-1}$. For a fixed total integration time $t$, this is equivalent to the standard sensitivity scaling of $\sqrt{t T_2}$. In contrast, when uncorrelated SQUID noise dominates the uncertainties, we observe a more favourable scaling of $\delta_\nu \propto T_2^{-3/2}$.  In the case of $T_1$, we find a scaling of $\delta_\nu \propto (T_1/T_2)^{-1/2}$.
Scaling with other parameters can be straightforwardly deduced from Eq.~\eqref{eq:chi2}.  Combining all information, we find that the sensitivity to $\delta_\nu$ in the SQUID-noise-dominated regime scales as 
\begin{equation}
\delta_\nu \propto \, n_s^{-3/2} \, T_2^{-3/2}  \, N^{-1/2}_{\rm shots}\, p^{-2} \,  {\cal F}_{\rm SQUID}^{1/2} ~r, \,
\end{equation} 
where $r$ denotes the $R$ dependence from Fig.~\ref{fig:deltaR_scaling}, which can be parameterised as 
\begin{equation}
    r = \begin{cases}
    R^{-9/2}, & R\lesssim p^{-1}_\nu\, ,\\
    R^{-5/2}, & R\gtrsim p^{-1}_\nu\, , ~~~{\rm if}~\omega^\star_0=\gamma_{\rm Xe}B_{\rm max}\, , \\
    R^{-3/2}, & R\gtrsim p^{-1}_\nu\, , ~~~{\rm if}~\omega^\star_0=p_{\nu}/(m_{\nu} R)\, .
\end{cases}
\end{equation} 
This shows, as expected, that the sensitivity to $\delta_\nu$ improves for larger sample sizes and higher spin densities, as well as longer coherence and measurement times. The sensitivity is, however, expected to degrade quadratically with imperfect polarisation and is further limited by instrumental noise. 

Finally, it is worth mentioning that one could also constrain $\delta_{\nu}$ using the second-order observable $\langle {\cal J}_x^2 \rangle_{\vert G\rangle}$. While this observable has the advantage of probing a C$\nu$B-induced signal without requiring spin rotation into the equatorial plane, its signal-to-noise ratio scales as $\sim \gamma_{\rm tot} N$, as pointed out in Sec.~\ref{sec:coherentobservables}. As a result, it provides a weaker sensitivity to $\delta_\nu$ compared to $\langle {\cal J}_z \rangle_{|P\rangle}$ for spin ensembles with $N \gtrsim (\gamma_{\rm tot}/\gamma_{\rm net})^2 \simeq 10^6$. Indeed, even in the optimisitc $R=10$ cm scenario, the sensitivity of $\langle {\cal J}_x^2 \rangle_{\vert G\rangle}$ to the CMB overdensity is only  $\delta_\nu\gtrsim 10^{20}$, about ten orders of magnitude weaker than what can be achieved with $\langle {\cal J}_z \rangle_{\vert P\rangle}$.


\section{Conclusions}
\label{sec:conclusions}

We have investigated in this work the dynamics of large spin ensembles interacting with the cosmic neutrino background, with a particular focus on identifying the conditions under which coherent collective effects---scaling as $N^2$, where $N$ is the number of spins---can amplify the signal from an otherwise extremely weak neutrino-spin coupling. Such collective neutrino-spin interactions are expected to contribute to dissipating  a spin ensemble and can potentially be exploited as a competitive way to set limits on the local C$\nu$B overdensity, as first put forward in Ref.~\cite{Arvanitaki:2024taq}.

After briefly reviewing the interaction rates of the C$\nu$B with a system of $N$ two-level systems (e.g., spin-$1/2$ particles) and the Lindblad master equation in the open quantum system framework, 
we reproduced the perturbative solutions presented in Ref.~\cite{Arvanitaki:2024taq}, valid when the spin ensemble has not deviated too much from its initial state.  We then extend the analysis by considering steady-state analytical solutions as well as numerical simulations of the Lindblad master equation 
for different initial spin configurations.  Notably, our numerical simulations account for realistic noise sources that can degrade the collective C$\nu$B signal, such as local dephasing as well as imperfect initial spin polarisation. We make use of the Dicke basis---illustrated in Fig.~\ref{fig:DickeTriangle}---to facilitate this analysis, which provides an intuitive representation of how collective and local interactions influence the spin ensemble's evolution under the Lindblad master equation.

We have furthermore presented a fast and computationally inexpensive method to derive approximate numerical solution from the Lindblad equation. Specifically, the method solves directly for the observables and correlation functions of interest while capturing both coherent neutrino effects and local interactions, and 
is able to reproduce the full density matrix solution to ${\cal O}(0.01)\%$ accuracy. This fast method enables us to use statistical inference methods to forecast the sensitivity of future NMR experiments to the C$\nu$B overdensity, while incorporating the complete quantum evolution of the spin ensemble.

Considering the projected experimental specifications of the forthcoming CASPEr experiment~\cite{Budker:2013hfa,JacksonKimball:2017elr,Walter:2025ldb} and nuclear-spin-based experiments more generally, we compute future experimental sensitivities to the C$\nu$B overdensity parameter $\delta_\nu$ using a $\chi^2$ analysis.
We find that achievable near-future setups along the lines of CASPEr could constrain the C$\nu$B overdensity parameter at the level of $\delta_\nu \sim 10^{13}$ for neutrino masses currently within the KATRIN neutrino mass bound~\cite{KATRIN:2024cdt}, with larger masses yielding a more favourable sensitivity.
Notably, in addition to the scaling with the sample volume and the obvious dependence on instrumental noise, we find that the initial polarisation $p$ of the spin ensemble plays a prominent role in our projected sensitivities: scaling as $\delta_\nu \propto p^{-2}$, a competitive constraint requires at least $p \sim 0.25$.  This requirement also means that thermal polarisation levels achievable at typical experimental conditions---$p \sim 10^{-3}$ for $^{129}$Xe at $T\sim 4$~K and $B\sim 10$~T---are not suited to C$\nu$B detection. Advanced polarisition techniques are needed to reach the desired~$p\gtrsim 0.25$.

In optimistic experimental scenarios---with a sample size up to $R = 10$~cm, and 100\% polarisation---the constraint may improve to $\delta_\nu \sim 10^{11}$ for a small range of neutrino mass values. This number is largely consistent with the estimate of Ref.~\cite{Arvanitaki:2024taq},  and also at face value comparable to the current KATRIN limit on the C$\nu$B overdensity~\cite{KATRIN:2022kkv}.

Our results indicate that C$\nu$B direct detection is likely beyond the reach of current NMR technology, even in the most optimistic scenario. Indeed, even meeting the ``realistic'' projected $\delta_\nu$ constraints represent significant experimental challenges.  We emphasise however much effort is currently being devoted to overcoming these challenges---in the case of CASPEr, in the context of dark matter axion searches. Should the desired experimental configurations materialise, we have shown that, by means of a pulsed-NMR measurement such as performed recently in~\cite{Walter:2025ldb},
the potential exists for such axion searches to simultaneously set potentially competitive constraints on the C$\nu$B for free. Constraining the C$\nu$B with nuclear spin experiments therefore remains an attractive possibility.


\acknowledgments

We thank Caterina Braggio, Giovanni Carugno, Jan Hamann, Alex Sushkov, and Oleg Sushkov for enlightening discussions, and Marios Galanis for insightful comments on the manuscript. YG, GP, and Y$^3$W are supported by the Australian Research Council's Discovery Project (DP240103130) scheme. 

\section*{Data Availability Statement}

The data that support the findings of this article are openly available at \url{https://github.com/gpierobon/OpenNu}.
\bibliography{oqs.bib}

\providecommand{\href}[2]{#2}\begingroup\raggedright\begin{thebibliography}{10}

\bibitem{ParticleDataGroup:2024cfk}
{\scshape Particle Data Group} Collaboration, S.~Navas et~al., \emph{{Review of particle physics}}, \href{https://doi.org/10.1103/PhysRevD.110.030001}{\emph{Phys. Rev. D} {\bfseries 110} (2024) 030001}.

\bibitem{Singh:2002de}
S.~Singh and C.-P. Ma, \emph{{Neutrino clustering in cold dark matter halos : Implications for ultrahigh-energy cosmic rays}}, \href{https://doi.org/10.1103/PhysRevD.67.023506}{\emph{Phys. Rev. D} {\bfseries 67} (2003) 023506} [\href{https://arxiv.org/abs/astro-ph/0208419}{{\ttfamily astro-ph/0208419}}].

\bibitem{Ringwald:2004np}
A.~Ringwald and Y.~Y.~Y. Wong, \emph{{Gravitational clustering of relic neutrinos and implications for their detection}}, \href{https://doi.org/10.1088/1475-7516/2004/12/005}{\emph{JCAP} {\bfseries 12} (2004) 005} [\href{https://arxiv.org/abs/hep-ph/0408241}{{\ttfamily hep-ph/0408241}}].

\bibitem{Brandbyge:2010ge}
J.~Brandbyge, S.~Hannestad, T.~Haugb{\o}lle and Y.~Y.~Y. Wong, \emph{{Neutrinos in Non-linear Structure Formation - The Effect on Halo Properties}}, \href{https://doi.org/10.1088/1475-7516/2010/09/014}{\emph{JCAP} {\bfseries 09} (2010) 014} [\href{https://arxiv.org/abs/1004.4105}{{\ttfamily 1004.4105}}].

\bibitem{Villaescusa-Navarro:2011loy}
F.~Villaescusa-Navarro, J.~Miralda-Escud{\'e}, C.~Pe{\~n}a-Garay and V.~Quilis, \emph{{Neutrino Halos in Clusters of Galaxies and their Weak Lensing Signature}}, \href{https://doi.org/10.1088/1475-7516/2011/06/027}{\emph{JCAP} {\bfseries 06} (2011) 027} [\href{https://arxiv.org/abs/1104.4770}{{\ttfamily 1104.4770}}].

\bibitem{LoVerde:2013lta}
M.~LoVerde and M.~Zaldarriaga, \emph{{Neutrino clustering around spherical dark matter halos}}, \href{https://doi.org/10.1103/PhysRevD.89.063502}{\emph{Phys. Rev. D} {\bfseries 89} (2014) 063502} [\href{https://arxiv.org/abs/1310.6459}{{\ttfamily 1310.6459}}].

\bibitem{deSalas:2017wtt}
P.~F. de~Salas, S.~Gariazzo, J.~Lesgourgues and S.~Pastor, \emph{{Calculation of the local density of relic neutrinos}}, \href{https://doi.org/10.1088/1475-7516/2017/09/034}{\emph{JCAP} {\bfseries 09} (2017) 034} [\href{https://arxiv.org/abs/1706.09850}{{\ttfamily 1706.09850}}].

\bibitem{Zhang:2017ljh}
J.~Zhang and X.~Zhang, \emph{{Gravitational clustering of cosmic relic neutrinos in the Milky Way}}, \href{https://doi.org/10.1038/s41467-018-04264-y}{\emph{Nature Commun.} {\bfseries 9} (2018) 1833} [\href{https://arxiv.org/abs/1712.01153}{{\ttfamily 1712.01153}}].

\bibitem{Mertsch:2019qjv}
P.~Mertsch, G.~Parimbelli, P.~F. de~Salas, S.~Gariazzo, J.~Lesgourgues and S.~Pastor, \emph{{Neutrino clustering in the Milky Way and beyond}}, \href{https://doi.org/10.1088/1475-7516/2020/01/015}{\emph{JCAP} {\bfseries 01} (2020) 015} [\href{https://arxiv.org/abs/1910.13388}{{\ttfamily 1910.13388}}].

\bibitem{Holm:2023rml}
E.~B. Holm, I.~M. Oldengott and S.~Zentarra, \emph{{Local clustering of relic neutrinos with kinetic field theory}}, \href{https://doi.org/10.1016/j.physletb.2023.138073}{\emph{Phys. Lett. B} {\bfseries 844} (2023) 138073} [\href{https://arxiv.org/abs/2305.13379}{{\ttfamily 2305.13379}}].

\bibitem{Zimmer:2023jbb}
F.~Zimmer, C.~A. Correa and S.~Ando, \emph{{Influence of local structure on relic neutrino abundances and anisotropies}}, \href{https://doi.org/10.1088/1475-7516/2023/11/038}{\emph{JCAP} {\bfseries 11} (2023) 038} [\href{https://arxiv.org/abs/2306.16444}{{\ttfamily 2306.16444}}].

\bibitem{Zimmer:2024max}
F.~Zimmer, G.~Franco~Abell{\'a}n and S.~Ando, \emph{{Effects of primordial fluctuations on relic neutrino simulations}}, \href{https://doi.org/10.1088/1475-7516/2024/10/098}{\emph{JCAP} {\bfseries 10} (2024) 098} [\href{https://arxiv.org/abs/2407.14582}{{\ttfamily 2407.14582}}].

\bibitem{Holm:2024zpr}
E.~B. Holm, S.~Zentarra and I.~M. Oldengott, \emph{{Local clustering of relic neutrinos: comparison of kinetic field theory and the Vlasov equation}}, \href{https://doi.org/10.1088/1475-7516/2024/07/050}{\emph{JCAP} {\bfseries 07} (2024) 050} [\href{https://arxiv.org/abs/2404.11295}{{\ttfamily 2404.11295}}].

\bibitem{Worku:2024kwv}
K.~Worku, N.~Sabti and M.~Kamionkowski, \emph{{Rapid methods for modeling overdensities of massive neutrinos and other noncold relics}}, \href{https://doi.org/10.1103/PhysRevD.112.023538}{\emph{Phys. Rev. D} {\bfseries 112} (2025) 023538} [\href{https://arxiv.org/abs/2410.08267}{{\ttfamily 2410.08267}}].

\bibitem{Pitrou:2018cgg}
C.~Pitrou, A.~Coc, J.-P. Uzan and E.~Vangioni, \emph{{Precision big bang nucleosynthesis with improved Helium-4 predictions}}, \href{https://doi.org/10.1016/j.physrep.2018.04.005}{\emph{Phys. Rept.} {\bfseries 754} (2018) 1} [\href{https://arxiv.org/abs/1801.08023}{{\ttfamily 1801.08023}}].

\bibitem{Planck:2018vyg}
{\scshape Planck} Collaboration, N.~Aghanim et~al., \emph{{Planck 2018 results. VI. Cosmological parameters}}, \href{https://doi.org/10.1051/0004-6361/201833910}{\emph{Astron. Astrophys.} {\bfseries 641} (2020) A6} [\href{https://arxiv.org/abs/1807.06209}{{\ttfamily 1807.06209}}]. [Erratum: Astron.Astrophys. 652, C4 (2021)].

\bibitem{Oldengott:2017fhy}
I.~M. Oldengott, T.~Tram, C.~Rampf and Y.~Y.~Y. Wong, \emph{{Interacting neutrinos in cosmology: exact description and constraints}}, \href{https://doi.org/10.1088/1475-7516/2017/11/027}{\emph{JCAP} {\bfseries 11} (2017) 027} [\href{https://arxiv.org/abs/1706.02123}{{\ttfamily 1706.02123}}].

\bibitem{Chen:2022idm}
J.~Z. Chen, I.~M. Oldengott, G.~Pierobon and Y.~Y.~Y. Wong, \emph{{Weaker yet again: mass spectrum-consistent cosmological constraints on the neutrino lifetime}}, \href{https://doi.org/10.1140/epjc/s10052-022-10518-3}{\emph{Eur. Phys. J. C} {\bfseries 82} (2022) 640} [\href{https://arxiv.org/abs/2203.09075}{{\ttfamily 2203.09075}}].

\bibitem{Stodolsky:1974aq}
L.~Stodolsky, \emph{{Speculations on Detection of the Neutrino Sea}}, \href{https://doi.org/10.1103/PhysRevLett.34.110}{\emph{Phys. Rev. Lett.} {\bfseries 34} (1975) 110}. [Erratum: Phys.Rev.Lett. 34, 508 (1975)].

\bibitem{Oldengott:2017tzj}
I.~M. Oldengott and D.~J. Schwarz, \emph{{Improved constraints on lepton asymmetry from the cosmic microwave background}}, \href{https://doi.org/10.1209/0295-5075/119/29001}{\emph{EPL} {\bfseries 119} (2017) 29001} [\href{https://arxiv.org/abs/1706.01705}{{\ttfamily 1706.01705}}].

\bibitem{Burns:2022hkq}
A.-K. Burns, T.~M.~P. Tait and M.~Valli, \emph{{Indications for a Nonzero Lepton Asymmetry from Extremely Metal-Poor Galaxies}}, \href{https://doi.org/10.1103/PhysRevLett.130.131001}{\emph{Phys. Rev. Lett.} {\bfseries 130} (2023) 131001} [\href{https://arxiv.org/abs/2206.00693}{{\ttfamily 2206.00693}}].

\bibitem{Arvanitaki:2022oby}
A.~Arvanitaki and S.~Dimopoulos, \emph{{Cosmic neutrino background on the surface of the Earth}}, \href{https://doi.org/10.1103/PhysRevD.108.043517}{\emph{Phys. Rev. D} {\bfseries 108} (2023) 043517} [\href{https://arxiv.org/abs/2212.00036}{{\ttfamily 2212.00036}}].

\bibitem{Ruchayskiy:2022eog}
O.~Ruchayskiy, V.~Syvolap and R.~Wursch, \emph{{Lepton number survival in the cosmic neutrino background}}, \href{https://doi.org/10.1103/PhysRevD.108.123503}{\emph{Phys. Rev. D} {\bfseries 108} (2023) 123503} [\href{https://arxiv.org/abs/2212.01038}{{\ttfamily 2212.01038}}].

\bibitem{Shvartsman:1982sn}
B.~F. Shvartsman, V.~B. Braginsky, S.~S. Gershtein, Y.~B. Zeldovich and M.~Y. Khlopov, \emph{{Possibility of detecting relict massive neutrinos}}, {\emph{JETP Lett.} {\bfseries 36} (1982) 277}.

\bibitem{Langacker:1982ih}
P.~Langacker, J.~P. Leveille and J.~Sheiman, \emph{{On the Detection of Cosmological Neutrinos by Coherent Scattering}}, \href{https://doi.org/10.1103/PhysRevD.27.1228}{\emph{Phys. Rev. D} {\bfseries 27} (1983) 1228}.

\bibitem{Smith:1983jj}
P.~F. Smith and J.~D. Lewin, \emph{{Coherent interactino of galactic neutronos with material targets}}, \href{https://doi.org/10.1016/0370-2693(83)90873-0}{\emph{Phys. Lett. B} {\bfseries 127} (1983) 185}.

\bibitem{Duda:2001hd}
G.~Duda, G.~Gelmini and S.~Nussinov, \emph{{Expected signals in relic neutrino detectors}}, \href{https://doi.org/10.1103/PhysRevD.64.122001}{\emph{Phys. Rev. D} {\bfseries 64} (2001) 122001} [\href{https://arxiv.org/abs/hep-ph/0107027}{{\ttfamily hep-ph/0107027}}].

\bibitem{Domcke:2017aqj}
V.~Domcke and M.~Spinrath, \emph{{Detection prospects for the Cosmic Neutrino Background using laser interferometers}}, \href{https://doi.org/10.1088/1475-7516/2017/06/055}{\emph{JCAP} {\bfseries 06} (2017) 055} [\href{https://arxiv.org/abs/1703.08629}{{\ttfamily 1703.08629}}].

\bibitem{Shergold:2021evs}
J.~D. Shergold, \emph{{Updated detection prospects for relic neutrinos using coherent scattering}}, \href{https://doi.org/10.1088/1475-7516/2021/11/052}{\emph{JCAP} {\bfseries 11} (2021) 052} [\href{https://arxiv.org/abs/2109.07482}{{\ttfamily 2109.07482}}].

\bibitem{Weinberg:1962zza}
S.~Weinberg, \emph{{Universal Neutrino Degeneracy}}, \href{https://doi.org/10.1103/PhysRev.128.1457}{\emph{Phys. Rev.} {\bfseries 128} (1962) 1457}.

\bibitem{Cocco:2007za}
A.~G. Cocco, G.~Mangano and M.~Messina, \emph{{Probing low energy neutrino backgrounds with neutrino capture on beta decaying nuclei}}, \href{https://doi.org/10.1088/1475-7516/2007/06/015}{\emph{JCAP} {\bfseries 06} (2007) 015} [\href{https://arxiv.org/abs/hep-ph/0703075}{{\ttfamily hep-ph/0703075}}].

\bibitem{Long:2014zva}
A.~J. Long, C.~Lunardini and E.~Sabancilar, \emph{{Detecting non-relativistic cosmic neutrinos by capture on tritium: phenomenology and physics potential}}, \href{https://doi.org/10.1088/1475-7516/2014/08/038}{\emph{JCAP} {\bfseries 08} (2014) 038} [\href{https://arxiv.org/abs/1405.7654}{{\ttfamily 1405.7654}}].

\bibitem{Akita:2020jbo}
K.~Akita, S.~Hurwitz and M.~Yamaguchi, \emph{{Precise Capture Rates of Cosmic Neutrinos and Their Implications on Cosmology}}, \href{https://doi.org/10.1140/epjc/s10052-021-09133-5}{\emph{Eur. Phys. J. C} {\bfseries 81} (2021) 344} [\href{https://arxiv.org/abs/2010.04454}{{\ttfamily 2010.04454}}].

\bibitem{KATRIN:2022kkv}
{\scshape KATRIN} Collaboration, M.~Aker et~al., \emph{{New Constraint on the Local Relic Neutrino Background Overdensity with the First KATRIN Data Runs}}, \href{https://doi.org/10.1103/PhysRevLett.129.011806}{\emph{Phys. Rev. Lett.} {\bfseries 129} (2022) 011806} [\href{https://arxiv.org/abs/2202.04587}{{\ttfamily 2202.04587}}].

\bibitem{KATRIN:2024cdt}
{\scshape KATRIN} Collaboration, M.~Aker et~al., \emph{{Direct neutrino-mass measurement based on 259 days of KATRIN data}}, \href{https://doi.org/10.1126/science.adq9592}{\emph{Science} {\bfseries 388} (2025) adq9592} [\href{https://arxiv.org/abs/2406.13516}{{\ttfamily 2406.13516}}].

\bibitem{Project8:2022wqh}
{\scshape Project 8} Collaboration, A.~A. Esfahani et~al., \emph{{The Project 8 Neutrino Mass Experiment}},  in \emph{{Snowmass 2021}}, 3, 2022, \href{https://arxiv.org/abs/2203.07349}{{\ttfamily 2203.07349}}.

\bibitem{PTOLEMY:2018jst}
{\scshape PTOLEMY} Collaboration, E.~Baracchini et~al., \emph{{PTOLEMY: A Proposal for Thermal Relic Detection of Massive Neutrinos and Directional Detection of MeV Dark Matter}},  \href{https://arxiv.org/abs/1808.01892}{{\ttfamily 1808.01892}}.

\bibitem{Bauer:2021uyj}
M.~Bauer and J.~D. Shergold, \emph{{Relic neutrinos at accelerator experiments}}, \href{https://doi.org/10.1103/PhysRevD.104.083039}{\emph{Phys. Rev. D} {\bfseries 104} (2021) 083039} [\href{https://arxiv.org/abs/2104.12784}{{\ttfamily 2104.12784}}].

\bibitem{Yoshimura:2014hfa}
M.~Yoshimura, N.~Sasao and M.~Tanaka, \emph{{Experimental method of detecting relic neutrino by atomic de-excitation}}, \href{https://doi.org/10.1103/PhysRevD.91.063516}{\emph{Phys. Rev. D} {\bfseries 91} (2015) 063516} [\href{https://arxiv.org/abs/1409.3648}{{\ttfamily 1409.3648}}].

\bibitem{Huang:2025yqu}
G.-y. Huang and S.~Zhou, \emph{{Probing Cosmic Neutrino Background through Parametric Fluorescence}},  \href{https://arxiv.org/abs/2507.10868}{{\ttfamily 2507.10868}}.

\bibitem{Bauer:2025hdq}
M.~Bauer, J.~Perez-Soler and J.~D. Shergold, \emph{{Dark matter pair absorption}},  \href{https://arxiv.org/abs/2507.14287}{{\ttfamily 2507.14287}}.

\bibitem{Arvanitaki:2024taq}
A.~Arvanitaki, S.~Dimopoulos and M.~Galanis, \emph{{Superradiant interactions of the cosmic neutrino background, axions, dark matter, and reactor neutrinos}}, \href{https://doi.org/10.1103/PhysRevD.111.055015}{\emph{Phys. Rev. D} {\bfseries 111} (2025) 055015} [\href{https://arxiv.org/abs/2408.04021}{{\ttfamily 2408.04021}}].

\bibitem{Aliberti:2024udm}
S.~R. Aliberti, G.~Lambiase and T.~K. Poddar, \emph{{Limits on dark matter, ultralight scalars, and cosmic neutrinos with gyroscope spin and precision clocks}}, \href{https://doi.org/10.1088/1475-7516/2025/03/049}{\emph{JCAP} {\bfseries 03} (2025) 049} [\href{https://arxiv.org/abs/2412.09575}{{\ttfamily 2412.09575}}].

\bibitem{Das:2024thc}
S.~Das, P.~S.~B. Dev, T.~Okawa and A.~Soni, \emph{{Old neutron stars as a new probe of relic neutrinos and sterile neutrino dark matter}}, \href{https://doi.org/10.1103/PhysRevD.111.055035}{\emph{Phys. Rev. D} {\bfseries 111} (2025) 055035} [\href{https://arxiv.org/abs/2408.01484}{{\ttfamily 2408.01484}}].

\bibitem{Ciscar-Monsalvatje:2024tvm}
M.~C{\'\i}scar-Monsalvatje, G.~Herrera and I.~M. Shoemaker, \emph{{Upper limits on the cosmic neutrino background from cosmic rays}}, \href{https://doi.org/10.1103/PhysRevD.110.063036}{\emph{Phys. Rev. D} {\bfseries 110} (2024) 063036} [\href{https://arxiv.org/abs/2402.00985}{{\ttfamily 2402.00985}}].

\bibitem{DeMarchi:2024zer}
A.~G. De~Marchi, A.~Granelli, J.~Nava and F.~Sala, \emph{{Relic neutrino background from cosmic-ray reservoirs}}, \href{https://doi.org/10.1103/PhysRevD.111.023023}{\emph{Phys. Rev. D} {\bfseries 111} (2025) 023023} [\href{https://arxiv.org/abs/2405.04568}{{\ttfamily 2405.04568}}].

\bibitem{Herrera:2024upj}
G.~Herrera, S.~Horiuchi and X.~Qi, \emph{{Diffuse boosted cosmic neutrino background}}, \href{https://doi.org/10.1103/PhysRevD.111.063016}{\emph{Phys. Rev. D} {\bfseries 111} (2025) 063016} [\href{https://arxiv.org/abs/2405.14946}{{\ttfamily 2405.14946}}].

\bibitem{Franklin:2024amy}
J.~Franklin, I.~Martinez-Soler, Y.~F. Perez-Gonzalez and J.~Turner, \emph{{Probing the cosmic neutrino background and new physics with TeV-scale astrophysical neutrinos}}, \href{https://doi.org/10.1016/j.physletb.2025.139615}{\emph{Phys. Lett. B} {\bfseries 867} (2025) 139615} [\href{https://arxiv.org/abs/2404.02202}{{\ttfamily 2404.02202}}].

\bibitem{Brdar:2022kpu}
V.~Brdar, P.~S.~B. Dev, R.~Plestid and A.~Soni, \emph{{A new probe of relic neutrino clustering using cosmogenic neutrinos}}, \href{https://doi.org/10.1016/j.physletb.2022.137358}{\emph{Phys. Lett. B} {\bfseries 833} (2022) 137358} [\href{https://arxiv.org/abs/2207.02860}{{\ttfamily 2207.02860}}].

\bibitem{Weiler:1982qy}
T.~J. Weiler, \emph{{Resonant Absorption of Cosmic Ray Neutrinos by the Relic Neutrino Background}}, \href{https://doi.org/10.1103/PhysRevLett.49.234}{\emph{Phys. Rev. Lett.} {\bfseries 49} (1982) 234}.

\bibitem{Budker:2013hfa}
D.~Budker, P.~W. Graham, M.~Ledbetter, S.~Rajendran and A.~Sushkov, \emph{{Proposal for a Cosmic Axion Spin Precession Experiment (CASPEr)}}, \href{https://doi.org/10.1103/PhysRevX.4.021030}{\emph{Phys. Rev. X} {\bfseries 4} (2014) 021030} [\href{https://arxiv.org/abs/1306.6089}{{\ttfamily 1306.6089}}].

\bibitem{JacksonKimball:2017elr}
D.~F. Jackson~Kimball et~al., \emph{{Overview of the Cosmic Axion Spin Precession Experiment (CASPEr)}}, \href{https://doi.org/10.1007/978-3-030-43761-9_13}{\emph{Springer Proc. Phys.} {\bfseries 245} (2020) 105} [\href{https://arxiv.org/abs/1711.08999}{{\ttfamily 1711.08999}}].

\bibitem{Walter:2025ldb}
J.~Walter et~al., \emph{{Search for axionlike dark matter using liquid-state nuclear magnetic resonance}}, \href{https://doi.org/10.1103/39nc-vr9m}{\emph{Phys. Rev. D} {\bfseries 112} (2025) 052008} [\href{https://arxiv.org/abs/2504.16044}{{\ttfamily 2504.16044}}].

\bibitem{bohr1998nuclear}
A.~Bohr and B.~R. Mottelson, \emph{Nuclear structure}, vol.~1. World scientific, 1998.

\bibitem{Esteban:2020cvm}
I.~Esteban, M.~C. Gonzalez-Garcia, M.~Maltoni, T.~Schwetz and A.~Zhou, \emph{{The fate of hints: updated global analysis of three-flavor neutrino oscillations}}, \href{https://doi.org/10.1007/JHEP09(2020)178}{\emph{JHEP} {\bfseries 09} (2020) 178} [\href{https://arxiv.org/abs/2007.14792}{{\ttfamily 2007.14792}}].

\bibitem{deSalas:2020pgw}
P.~F. de~Salas, D.~V. Forero, S.~Gariazzo, P.~Mart{\'\i}nez-Mirav{\'e}, O.~Mena, C.~A. Ternes, M.~T{\'o}rtola and J.~W.~F. Valle, \emph{{2020 global reassessment of the neutrino oscillation picture}}, \href{https://doi.org/10.1007/JHEP02(2021)071}{\emph{JHEP} {\bfseries 02} (2021) 071} [\href{https://arxiv.org/abs/2006.11237}{{\ttfamily 2006.11237}}].

\bibitem{Capozzi:2017ipn}
F.~Capozzi, E.~Di~Valentino, E.~Lisi, A.~Marrone, A.~Melchiorri and A.~Palazzo, \emph{{Global constraints on absolute neutrino masses and their ordering}}, \href{https://doi.org/10.1103/PhysRevD.95.096014}{\emph{Phys. Rev. D} {\bfseries 95} (2017) 096014} [\href{https://arxiv.org/abs/2003.08511}{{\ttfamily 2003.08511}}]. [Addendum: Phys.Rev.D 101, 116013 (2020)].

\bibitem{1925AnP...381...29G}
R.~{Gans}, \emph{{Strahlungsdiagramme ultramikroskopischer Teilchen}}, \href{https://doi.org/10.1002/andp.19253810103}{\emph{Annalen der Physik} {\bfseries 381} (1925) 29}.

\bibitem{Lindblad:1975ef}
G.~Lindblad, \emph{{On the Generators of Quantum Dynamical Semigroups}}, \href{https://doi.org/10.1007/BF01608499}{\emph{Commun. Math. Phys.} {\bfseries 48} (1976) 119}.

\bibitem{Gorini:1975nb}
V.~Gorini, A.~Kossakowski and E.~C.~G. Sudarshan, \emph{{Completely Positive Dynamical Semigroups of N Level Systems}}, \href{https://doi.org/10.1063/1.522979}{\emph{J. Math. Phys.} {\bfseries 17} (1976) 821}.

\bibitem{Manzano:2020yyw}
D.~Manzano, \emph{{A short introduction to the Lindblad master equation}}, \href{https://doi.org/10.1063/1.5115323}{\emph{AIP Adv.} {\bfseries 10} (2020) 025106}.

\bibitem{PhysRevA.98.063815}
N.~Shammah, S.~Ahmed, N.~Lambert, S.~De~Liberato and F.~Nori, \emph{Open quantum systems with local and collective incoherent processes: Efficient numerical simulations using permutational invariance}, \href{https://doi.org/10.1103/PhysRevA.98.063815}{\emph{Phys. Rev. A} {\bfseries 98} (2018) 063815}.

\bibitem{PhysRev.93.99}
R.~H. Dicke, \emph{Coherence in spontaneous radiation processes}, \href{https://doi.org/10.1103/PhysRev.93.99}{\emph{Phys. Rev.} {\bfseries 93} (1954) 99}.

\bibitem{Gross:1982dkt}
M.~Gross and S.~Haroche, \emph{{Superradiance: An essay on the theory of collective spontaneous emission}}, \href{https://doi.org/10.1016/0370-1573(82)90102-8}{\emph{Phys. Rept.} {\bfseries 93} (1982) 301}.

\bibitem{Masson:2021eyk}
S.~J. Masson and A.~Asenjo-Garcia, \emph{{Universality of Dicke superradiance in arrays of quantum emitters}}, \href{https://doi.org/10.1038/s41467-022-29805-4}{\emph{Nature Commun.} {\bfseries 13} (2022) 2285} [\href{https://arxiv.org/abs/2106.02042}{{\ttfamily 2106.02042}}].

\bibitem{Shammah_2017}
N.~Shammah, N.~Lambert, F.~Nori and S.~De~Liberato, \emph{Superradiance with local phase-breaking effects}, \href{https://doi.org/10.1103/physreva.96.023863}{\emph{Physical Review A} {\bfseries 96} (2017) }.

\bibitem{Boyers:2025qgc}
E.~Boyers, G.~Goldstein and A.~O. Sushkov, \emph{{Spin squeezing of macroscopic nuclear spin ensembles}}, \href{https://doi.org/10.1103/PhysRevD.111.052004}{\emph{Phys. Rev. D} {\bfseries 111} (2025) 052004} [\href{https://arxiv.org/abs/2502.14103}{{\ttfamily 2502.14103}}].

\bibitem{Bondarenko:2023ukx}
K.~Bondarenko, A.~Boyarsky, J.~Pradler and A.~Sokolenko, \emph{{Best-case scenarios for neutrino capture experiments}}, \href{https://doi.org/10.1088/1475-7516/2023/10/026}{\emph{JCAP} {\bfseries 10} (2023) 026} [\href{https://arxiv.org/abs/2306.12366}{{\ttfamily 2306.12366}}].

\bibitem{carmichael1}
H.~J. Carmichael, \emph{Statistical Methods in Quantum Optics}, vol.~1. Springer-Verlag, Berlin, 1999.

\bibitem{QRT}
M.~Lax, \emph{{Formal Theory of Quantum Fluctuations from a Driven State}}, \href{https://doi.org/10.1103/PhysRev.129.2342}{\emph{Physical Review} {\bfseries 129} (1963) 2342}.

\bibitem{Prohaska}
T.~Prohaska et~al., \emph{Standard atomic weights of the elements 2021 (iupac technical report)}, \href{https://doi.org/doi:10.1515/pac-2019-0603}{\emph{Pure and Applied Chemistry} {\bfseries 94} (2022) 573}.

\bibitem{PhysRevLett.87.067601}
M.~V. Romalis and M.~P. Ledbetter, \emph{Transverse spin relaxation in liquid $^{129}xe$ in the presence of large dipolar fields}, \href{https://doi.org/10.1103/PhysRevLett.87.067601}{\emph{Phys. Rev. Lett.} {\bfseries 87} (2001) 067601}.

\bibitem{RevModPhys.69.629}
T.~G. Walker and W.~Happer, \emph{Spin-exchange optical pumping of noble-gas nuclei}, \href{https://doi.org/10.1103/RevModPhys.69.629}{\emph{Rev. Mod. Phys.} {\bfseries 69} (1997) 629}.

\bibitem{doi:10.1021/jp501493k}
P.~Nikolaou et~al., \emph{Multidimensional mapping of spin-exchange optical pumping in clinical-scale batch-mode 129xe hyperpolarizers}, \href{https://doi.org/10.1021/jp501493k}{\emph{The Journal of Physical Chemistry B} {\bfseries 118} (2014) 4809}.

\bibitem{PhysRevB.94.094309}
M.~E. Limes, Z.~L. Ma, E.~G. Sorte and B.~Saam, \emph{Robust solid $^{129}\mathrm{Xe}$ longitudinal relaxation times}, \href{https://doi.org/10.1103/PhysRevB.94.094309}{\emph{Phys. Rev. B} {\bfseries 94} (2016) 094309}.

\bibitem{Galanis:2025amc}
M.~Galanis, O.~Hosten, A.~Arvanitaki and S.~Dimopoulos, \emph{{Toward 48 dB Spin Squeezing and 96 dB Signal Magnification for Cosmic Relic Searches with Nuclear Spins}},  \href{https://arxiv.org/abs/2508.20520}{{\ttfamily 2508.20520}}.

\bibitem{Eills2023SpinHyperpolarization}
J.~Eills, D.~Budker, S.~Cavagnero, E.~Y. Chekmenev, S.~J. Elliott, S.~Jannin, A.~Lesage, J.~Matysik, T.~Meersmann, T.~Prisner, J.~A. Reimer, H.~Yang and I.~V. Koptyug, \emph{Spin hyperpolarization in modern magnetic resonance}, \href{https://doi.org/10.1021/acs.chemrev.2c00534}{\emph{Chemical Reviews} {\bfseries 123} (2023) 1417}.

\bibitem{Walker_2011}
T.~G. Walker, \emph{Fundamentals of spin-exchange optical pumping}, \href{https://doi.org/10.1088/1742-6596/294/1/012001}{\emph{Journal of Physics: Conference Series} {\bfseries 294} (2011) 012001}.

\end{thebibliography}\endgroup
\bibliographystyle{bibi}


\newpage

\onecolumngrid
\appendix

\section{Low energy non-relativistic spin Hamiltonian}
\label{app:ham}

We outline the derivation of the non-relativistic Hamiltonian for the neutrino-fermion spin interaction. We begin with the Lagrangian for a fermion~$\Psi$ in the presence of an external electromagnetic 4-potential  $A_{\mu}=(A_{0},\vec{A})$ and a 4-fermi interaction with a neutrino $\nu_i$ in the mass basis, 
\begin{align}
  \mathcal{L} = \bar{\Psi}({\rm i}\slashed{\partial} - m)\Psi - q\bar{\Psi}\gamma^\mu\Psi A_\mu 
    - \sum_{ij} \frac{G_F}{\sqrt{2}} \bar{\Psi}\gamma^\mu(g^{i,j}_V - g^{i,j}_A \gamma^5)\Psi \, \bar{\nu}_i \gamma_\mu (1 - \gamma^5) \nu_j \, ,
  \label{eq:lagmodapp}
\end{align}
where $i,j$ label the neutrino mass eigenstate, and $q$ is the electric charge of the fermion. 
The corresponding equation of motion is
\begin{equation}
\left({\rm i} \gamma^\mu \partial_\mu-m-q\gamma^\mu A_\mu - \sum_{ij}\frac{G_F}{\sqrt{2}}\gamma^\mu(g^{i,j}_V-g^{i,j}_A\gamma^5) \bar{\nu}_i\gamma_\mu(1-\gamma^5)\nu_j  \right)\Psi=0\, ,
\end{equation}
or, equivalently,
\begin{equation}
    \Bigg((p^{0}\gamma_{0} -q A^{0}\gamma_{0}) - \vec{\gamma}.(\vec{p} -q \vec{A}) - \sum_{ij} \frac{G_{F}}{\sqrt{2}} \left(g_{V}^{i,j} -g_{A}^{i,j}  \right)\gamma^{0}\Sigma_{0}^{i,j}+ \sum_{ij} \frac{G_{F}}{\sqrt{2}}\left(g_{V}^{i,j} -g_{A}^{i,j}  \right)\vec{\gamma}\cdot \vec{\Sigma}^{i,j}   \Bigg)\Psi =0\, ,
    \label{eq:dirac}
    \end{equation}
where we have used $\vec{p}= -{\rm i} \vec{\nabla}$, and $p^0={\rm i}\partial_{t}$, and 
defined $\Sigma^{i,j}_\mu \equiv \bar{\nu}_i\gamma_\mu(1-\gamma^5)\nu_j$. 

We write the fermion spinor in 2-component notation, 
\begin{equation}
\Psi= \begin{bmatrix}
           \phi_{0} \\
           \chi_{0} \\
         \end{bmatrix},
\end{equation}
such that in the non-relativistic limit, we can make the {\it ansatz} 
\begin{equation}
\Psi(\vec{x},t)= \begin{bmatrix}
           \phi_{0} \\
           \chi_{0} \\
         \end{bmatrix}e^{-{\rm i} m t} ,
\end{equation}
which yields the solution $\chi_0\approx [\vec{\sigma} \cdot (\vec{p} - q\vec{A})/(2m)]\phi_0$, where $m$ is the mass of the fermion.  
Substituting this solution back into the equation of motion~\eqref{eq:dirac} in 2-component form, we find a Schr\"{o}dinger-Pauli-like equation with additional neutrino-fermion interaction terms,
\begin{equation}
    \left(\frac{(\vec{p}-q\vec{A})^2}{2m} -\frac{q}{2m}\vec{B}\cdot \vec{\sigma}+qA_0 + \sum_{ij} \frac{G_F}{\sqrt{2}}g^{i,j}_V \Sigma^{i,j}_0+\sum_{ij}\frac{G_F}{\sqrt{2}}g^{i,j}_A\vec{\Sigma}^{i,j}\cdot\vec{\sigma}  \right) \phi_0 = E \phi_{0}\, ,
    \label{eq:eom2comp}
\end{equation}    
where $\vec{B}=\vec{\nabla}\times \vec{A}$ is the external magnetic field, and $p^{0}=E$ is the energy.

In the form~\eqref{eq:eom2comp}, the total Hamiltonian can be immediately identified with the terms in parentheses on the left-hand side of the equation.  We concentrate exclusively on the spin interaction of the fermion with the C$\nu$B. Assuming the fermion (spin) is at rest, this interaction is described by the spin Hamiltonian, 
\begin{equation}
   H = -\vec{\mu}\cdot \vec{B}+\sqrt{2}G_F  \sum_{ij} g^{i,j}_A\vec{J}\cdot\vec{\Sigma}^{i,j}\, ,
   \label{eq:spinint1}
\end{equation}
where we have used $\vec{J}=\vec{\sigma}/{2}$, and the magnetic moment is defined as $\vec{\mu}=(q/m)\vec{J}$. Although this result has been obtained assuming a charged fermion, an analogous spin-magnetic field interaction is also present for neutral particles like the neutron or nuclei.  Here, the interaction of a magnetic field with the magnetic moment due to the particle's internal structure can be described by an effective Lagrangian of the form $\bar{\psi} \sigma^{\mu\nu} \psi F_{\mu\nu}$. In the non-relativistic limit, this interaction term also reduces to a Hamiltonian of the form $H = -\vec{\mu}\cdot \vec{B}$. For nuclei, the convention is to define $\vec{\mu}=g\mu_N\vec{J}$, where $g$ is the $g$-factor of the nucleus under consideration, and $\mu_N=e/2m_p$ is the nuclear magneton.

 Supposing without any loss of generality that the magnetic field points along the negative $z$-axis, i.e., $\vec{B}=-B \hat{z}$,
 the spin Hamiltonian can be recast into a more suggestive form,
\begin{equation}
     H =\omega_0J_z+  \Sigma_zJ_z+\Sigma_-J_-+\Sigma_+J_+\, ,
\end{equation}
where the Larmor frequency $\omega_0=\gamma B$ is also the energy splitting between two adjacent spin states, with $\gamma$ the gyromagnetic ratio, and we have defined $J_\pm \equiv J_1\pm {\rm i}J_2$, and 
\begin{equation}
\begin{aligned}
  \Sigma_z =&\; \sqrt{2}G_F  \sum_{ij} g^{i,j}_A\bar{\nu}_i\gamma^3(1-\gamma^5)\nu_j \, , \\ \Sigma_\pm  =&\;  \frac{G_F}{\sqrt{2}} \sum_{ij} g^{i,j}_A\bar{\nu}_i(\gamma^1\mp {\rm i}\gamma^2)(1-\gamma^5)\nu_j \, .
  \label{eq:Sigma}
  \end{aligned}
\end{equation}
Then, splitting the  Hamiltonian into a spin ensemble and an interaction term as 
\begin{equation}
    H=H_S+H_I\,,
    \label{eq:ham}
\end{equation}
we identify $H_S \equiv \omega_0J_z$, $H_I \equiv \Sigma_zJ_z+\Sigma_-J_-+\Sigma_+J_+$, and 
\begin{equation}
\begin{aligned}
   H^I_I(t)= &\; e^{{\rm i}H_st} H_I e^{-{\rm i}H_st} \\
   =&\; \Sigma_zJ_z+
   e^{-{\rm i}\omega_0 t}\Sigma_-J_-+e^{{\rm i}\omega_0t}\Sigma_+J_+  
   \label{hint}
   \end{aligned}
\end{equation}
gives the interaction Hamiltonian in the interaction picture.

In the case of a system with $N$ spins, the Hamiltonians of the spin ensemble and the interaction can be straightforwardly generalised from Eqs.~\eqref{eq:ham} and~\eqref{hint} to 
\begin{equation}
\begin{aligned}
    H_S= &\; \omega_0\sum_\alpha J^\alpha_z\, , \\
   H^I_I(t)= &\; \sum_\alpha\Sigma^\alpha_zJ^\alpha_z+e^{-{\rm i}\omega_0t}\sum_\alpha\Sigma^\alpha_-J^\alpha_-+e^{{\rm i}\omega_0t}\sum_\alpha\Sigma^\alpha_+J^\alpha_+\, ,  
   \label{eq:nham}
   \end{aligned}
\end{equation}
where the operators $J^\alpha_{\pm,z}$ act only on the spin $\alpha$ at the spatial position $\vec{x}_\alpha$, and $\Sigma^\alpha_{\pm,z}$ are defined as per Eq.~\eqref{eq:Sigma} but with the understanding that the neutrino wavefunctions must now be specified as $\nu_i(t,\vec{x}_\alpha)$.


\section{Excitation and de-excitation rates}\label{app:rates}

\subsection{Single spin}

Consider first a single spin immersed in the cosmic neutrino bath.  Following Fermi's golden rule the excitation and de-excitation rates are given by 
\begin{equation}
\gamma_{\pm}=   |T^{\pm}_{\rm fi}|^2 
  2\pi\delta^{(1)}(E_{\rm f}-E_{\rm i}) \, \rho_\pm(E_{\rm f})\, ,
\label{eq:golden}
\end{equation}
where the superscripts ``$+$'' and ``$-$'' refer respectively to excitation and de-excitation, $|T^+_{\rm fi}|^2$ is the transition probability from the initial state $|{\rm i}\rangle = |\downarrow \rangle \otimes |{\rm i}_{\nu}\rangle$ to the final state $|{\rm f}\rangle= |\uparrow \rangle \otimes |{\rm f}_{\nu}\rangle$, while $|T^-_{\rm fi}|^2$ is its counterpart for the transition $|{\rm i}\rangle = |\uparrow \rangle \otimes |{\rm i}_{\nu}\rangle$ to $|{\rm f}\rangle= |\downarrow \rangle \otimes |{\rm f}_{\nu}\rangle$.  The quantity $\rho(E_{\rm f})$ is the density of states at the final energy $E_{\rm f}$, which in our case can be written as, 
\begin{equation}
\rho_\pm(E_{\rm f}) =  \sum_{jk} \sum_{a,c} \int \frac{{\rm d}^3\vec{p}}{(2 \pi)^3 }  \frac{{\rm d}^3 \vec{p}\,'}{(2 \pi)^3 } \frac{{\rm d}^3\vec{n}}{(2 \pi)^3 }   \, (2 \pi)^3 \delta^{(3)} ( \vec{p}-\vec{p}\,'\mp \vec{n})\, P_{{\rm i}_\nu} \, P_{{\rm f}_\nu}\,,
\label{eq:densityofstates}
\end{equation}
where $\vec{p}$ and $\vec{p}\,'$ are the 3-momenta of the incoming and outgoing neutrinos respectively, $\vec{n}$ is the momentum of the final-state fermion (assuming it is initially stationary),  $P_{{\rm i}_\nu}$ and $P_{{\rm f}_\nu}$ represent respectively the thermal weight of the initial-state neutrino and any Pauli blocking factor associated with the final state, $j$ and $k$ are mass indices of the initial- and final-state neutrinos, while $a$ and $c$ index their helicities.  
Since the recoil of the fermion is expected to be tiny, the angle of the outgoing neutrino is essentially unrestricted.  We therefore proceed to use the Dirac delta in Eq.~\eqref{eq:densityofstates} to eliminate the integral over the nuclear momentum, to arrive at
\begin{equation}
\rho_\pm(E_{\rm f}) =  \sum_{jk} \sum_{a,c} \int \frac{{\rm d}^3\vec{p}}{(2 \pi)^3 }  \frac{{\rm d}^3\vec{p}\,'}{(2 \pi)^3 }   P_{{\rm i}_\nu} \, P_{{\rm f}_\nu}\, ,
\label{eq:densityofstates2}
\end{equation}
where it is now understood that the relation between $|\vec{p}|$ and $|\vec{p}\,'|$ is determined by energy conservation alone.

\begin{figure}
    \centering
    \includegraphics[width=0.9\linewidth]{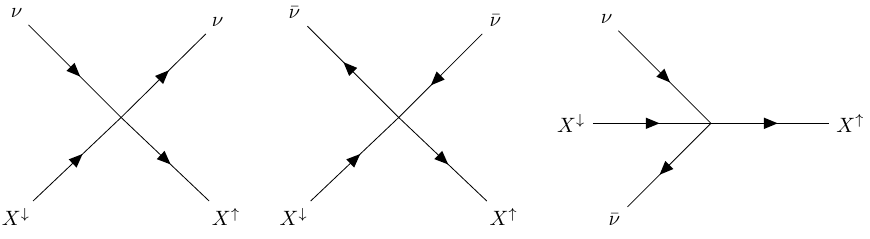}
    \caption{Feynman diagrams of the processes that contribute to the excitation of a spin in a 4-fermi effective interaction.}
    \label{fig:feynman}
\end{figure}

At first order in perturbation theory (i.e., the Born approximation), the transition amplitude is given by 
\begin{equation}
    T^{\pm}_{\rm fi}=\langle {\rm f}|\Sigma_{\pm}J_{\pm}|{\rm i}\rangle \, .
\end{equation}
Focussing on excitation, consider an initial-state neutrino specified by a wavefunction 
\begin{equation}
\nu_j(t,\vec{x}) = u_{j,a}(\vec{p})  e^{-{\rm i} (p^0t-\vec{p}\cdot\vec{x})}\, ,
\end{equation}
where $a$ is a helicity index.  Suppose it scatters off a spin inelastically, thereby transferring energy $\omega_0$ to it, and then exits the interaction in a final-state plane wave given by $\nu_k(t,\vec{x}) = u_{k,c}(\vec{p}\,')  e^{-{\rm i}(p'^0t- \vec{p}' \cdot\vec{x})}$.  This process is represented by the leftmost diagram in Fig.~\ref{fig:feynman}, and its transition probability can be written as 
\begin{equation}
\begin{aligned}
    |T^{+}_{\rm fi}|^2= &\; |\langle \uparrow |J_{+}|\downarrow \rangle|^2\, |\langle {\rm f}_\nu | \Sigma_+ | {\rm i}_\nu \rangle|^2 \\
    =& \;   \, \frac{G_F^2}{2} g_A^{j,k} g_A^{k,j} \frac{1}{2p^{0}2p'^{0}}   \left| \bar{u}_{k,c}(\vec{p}\,') (\gamma^1 - {\rm i} \gamma^2)(1- \gamma^5) u_{j,a}(\vec{p})\right|^2\, ,
\end{aligned}
\end{equation}
where we have used $|\langle\uparrow|J_+|\downarrow \rangle|^2=1$ at the second equality, and the factors of $1/2 p^{0}$ originates from the normalisation of the $u$-spinors to $2p^0$.
The thermal weight factors in Eq.~\eqref{eq:densityofstates} are then $P_{{\rm i}_\nu} = N_{j,a}(p^0)$ and  $P_{{\rm f}_\nu} =1 -N_{k,a}(p'^0)$, where $N_{i,a}(p^0)$ is the neutrino occupation number, while the Dirac delta $\delta^{(1)}(E_{\rm f}-E_{\rm i})$ becomes $\delta^{(1)}(\omega_0+p'^0-p^0)$, 
enforcing the change in energy between the initial- and final-state neutrinos to be equal to $\omega_{0}$, the energy splitting due to the external magnetic field.   A similar calculation can also be performed in the case of an antineutrino exciting a spin, as represented by the middle diagram in Fig.~\ref{fig:feynman}.

Putting it altogether, we then find the excitation rate of one spin to be
\begin{equation}
\begin{aligned}
\gamma_{+}= & \; G_F^2\pi \sum_{jk} g_A^{j,k} g_A^{k,j} \sum_{ac}  
    \int {\rm d} \tilde{p} \int {\rm d} \tilde{p}' \; \delta^{(1)}(\omega_0+p'^0-p^0) \\
  &\; \times \Bigg[(1-N_{k,c}(p'^0)) \, N_{j,a}(p^0)\, \bar{u}_{k,c}(\vec{p}\,') (\gamma^1- {\rm i} \gamma^2)(1-\gamma^5) u_{j,a}(\vec{p}) 
    \bar{u}_{j,a}(\vec{p}) (\gamma^1+ {\rm i} \gamma^2)(1-\gamma^5) u_{k,c}(\vec{p}\,') \\
&\; \qquad +(1-\bar{N}_{k,c}(p'^0))\, \bar{N}_{j,a}(p^0)\bar{v}_{j,a}(\vec{p}) (\gamma^1- {\rm i} \gamma^2)(1-\gamma^5) v_{k,c}(\vec{p}\,') \bar{v}_{k,c}(\vec{p}\,') (\gamma^1+ {\rm i} \gamma^2)(1-\gamma^5) v_{j,a}(\vec{p})  \Bigg]\,,
    \label{eq:rate1heuristic}
    \end{aligned}
\end{equation}
where we have defined the phase space measure ${\rm d} \tilde{p} \equiv {\rm d}^3\vec{p}/((2 \pi)^3 2 p^0)$.  Ignoring the relative motion between the spin and the C$\nu$B, and assuming the neutrino occupation numbers to be independent of helicity, the expression simplifies to 
\begin{equation}
\begin{aligned}
\gamma_{+}= & \;8  \, G_F^2 \pi \sum_{jk} g_A^{j,k} g_A^{k,j} 
    \int {\rm d} \tilde{p} \int {\rm d} \tilde{p}' \; \delta^{(1)}(\omega_0+p'^0-p^0) \, p^0 p'^0\, \left[(1-N_{k}(p'^0)) \, N_{j}(p^0)  +(1-\bar{N}_{k}(p'^0))\, \bar{N}_{j}(p^0)\right]\,,
    \end{aligned}
\end{equation}
which, using the Dirac delta, can be further reduced to 
\begin{equation}
    \gamma_{+}= \frac{4G_F^2}{(2 \pi)^3} \sum_{jk} g_A^{j,k} g_A^{k,j}  
    \int_{\sqrt{(\omega_0+m_k)^2-m_j^2}}^\infty  {\rm d} |\vec{p}|\;|\vec{p}|^2\, p_{k+}\, \sqrt{m_k^2+p^2_{k+}} \,  
    \left[(1-N_{k}(p^0_{k+})) \, N_{j}(p^0)  +(1-\bar{N}_{k}(p^0_{k+}))\, \bar{N}_{j}(p^0)\right]\,,
    \label{eq:gamma+heuristic}
\end{equation}
with $p_{k+}=[(\sqrt{|\vec{p}|^2+m_j^2}-\omega_0)^2-m_k^2]^{1/2}$, and  $p^0_{k+}=\sqrt{m_k^2+p^2_{k+}}$.

Computation of the de-excitation rate from inelastic neutrino-spin scattering proceeds in a similar fashion, and yields
\begin{equation}
\begin{aligned}
\gamma_{-}= & \; G_F^2\pi \sum_{jk} g_A^{j,k} g_A^{k,j} \sum_{ac}  
    \int {\rm d} \tilde{p} \int {\rm d} \tilde{p}' \; \delta^{(1)}(\omega_0+p^0-p'^0) \\
  &\; \times \Bigg[(1-N_{k,c}(p'^0)) \, N_{j,a}(p^0)\, \bar{u}_{k,c}(\vec{p}\,') (\gamma^1+ {\rm i} \gamma^2)(1-\gamma^5) u_{j,a}(\vec{p}) 
    \bar{u}_{j,a}(\vec{p}) (\gamma^1- {\rm i} \gamma^2)(1-\gamma^5) u_{k,c}(\vec{p}\,') \\
&\; \qquad +(1-\bar{N}_{k,c}(p'^0))\, \bar{N}_{j,a}(p^0)\bar{v}_{j,a}(\vec{p}) (\gamma^1+ {\rm i} \gamma^2)(1-\gamma^5) v_{k,c}(\vec{p}\,') \bar{v}_{k,c}(\vec{p}\,') (\gamma^1- {\rm i} \gamma^2)(1-\gamma^5) v_{j,a}(\vec{p})  \Bigg]\,.
    \label{eq:rate2heuristic}
    \end{aligned}
\end{equation}
Again, ignoring any neutrino-spin relative motion and assuming  helicity-independent neutrino occupation numbers, we find
\begin{equation}
\begin{aligned}
\gamma_{-}= & \;8  \, G_F^2 \pi \sum_{jk} g_A^{j,k} g_A^{k,j} \sum_{ac}  
    \int {\rm d} \tilde{p} \int {\rm d} \tilde{p}' \; \delta^{(1)}(\omega_0+p^0-p'^0) \, p^0 p'^0\, \left[(1-N_{k}(p'^0)) \, N_{j}(p^0)  +(1-\bar{N}_{k}(p'^0))\, \bar{N}_{j}(p^0)\right] \\
    =&\;  \frac{4G_F^2}{(2 \pi)^3} \sum_{jk} g_A^{j,k} g_A^{k,j}  
    \int_{\sqrt{(\omega_0-m_k)^2-m_j^2}}^\infty  {\rm d} |\vec{p}| \;|\vec{p}|^2\, p_{k-}\, \sqrt{m_k^2+p^2_{k-}} \,  
    \left[(1-N_{k}(p^0_{k-})) \, N_{j}(p^0)  +(1-\bar{N}_{k}(p^0_{k-}))\, \bar{N}_{j}(p^0)\right]\,,
    \label{eq:gamma-heuristic}
    \end{aligned}
\end{equation}
with  $p_{k-}=[(\sqrt{|\vec{p}|^2+m_j^2}+\omega_0)^2-m_k^2]^{1/2}$, and $p^0_{k-}=\sqrt{m_k^2+p^2_{k-}}$.    The lower integration holds if it evaluates to a real number.  Otherwise, it needs to be set to zero.

Suppose for simplicity $m_j = m_k=m$ and $\omega_0 \ll m$.  The lower integration limits of the $\gamma_+$ and $\gamma_-$ integrals, Eqs.~\eqref{eq:gamma+heuristic} and~\eqref{eq:gamma-heuristic}, are $\sqrt{(\omega_0+m)^2-m^2} \approx \sqrt{2 m\omega_0}$ and zero, respectively.  We can take the former to zero too if $\omega_0 \ll |\vec{p}|^2/(2 m) \ll m$ is satisfied by the bulk of the neutrino distribution.  The same condition also allows us to approximate $p_{k\pm} \approx \sqrt{ |\vec{p}|^2\mp 2m \omega_0}\approx|\vec{p}| \mp m\omega_0/|\vec{p}|$. Then, to leading order, we find 
\begin{equation}
\begin{aligned}
\gamma_+ \approx \gamma_- & \propto \int_0^\infty {\rm d}|\vec{p}|\; |\vec{p}|^2 \; p_{k+}\sqrt{m^2+p^2_{k+}}  \left[N(p^0) + \bar{N}(p^0)\right] \\
& \approx \int_0^\infty {\rm d}|\vec{p}|\; |\vec{p}|^2\;  m \, |\vec{p}|  \left[N(p^0) + \bar{N}(p^0)\right] \, ,
\end{aligned}
\end{equation}
and
\begin{equation}
\begin{aligned}
\gamma_{\rm net} \equiv \gamma_--\gamma_+&  \propto \int^\infty_0 {\rm d}|\vec{p}| \; |\vec{p}|^2\; \left[p_{k-}\, \sqrt{m^2+p^2_{k-}}- p_{k+}\, \sqrt{m^2+p^2_{k+}}
\right] \left[N(p^0) + \bar{N}(p^0)\right] \\
& \approx   \int^\infty_0 {\rm d}|\vec{p}| \; |\vec{p}|^2\; \frac{2 m^2 \omega_0}{|\vec{p}|} \; \left[N(p^0) + \bar{N}(p^0)\right] \, ,
\end{aligned}
\end{equation}
where we have omitted for brevity the Pauli-blocking terms in both expressions.  Thus, we see immediately that $\gamma_{+}< \gamma_-$, and $\gamma_{\rm net}$ is suppressed by a factor $\sim 2 m \omega_0/|\vec{p}|^2$ relative to $\gamma_\pm$.


\subsection{\texorpdfstring{$N$}{N} spins}

Consider now a system of $N$ spins described by the interaction Hamiltonian~\eqref{eq:nham}.  We are interested in the excitation and de-excitation rates of the whole spin ensemble, which can be defined via
\begin{equation}
\Gamma_\pm = \sum_{{\rm f}_{\rm spin}}|T_{\rm fi}^{\pm}|^2
2 \pi \delta^{(1)}(E_{\rm f}-E_{\rm i})\, \rho_\pm (E_{\rm f})\, ,
\label{eq:Gammapm}
\end{equation}
where 
\begin{equation}
\begin{aligned}
T^{\pm}_{\rm fi} = &\; \sum_\alpha \langle {\rm f}_{\rm spin} |  J_\pm^\alpha |{\rm i}_{\rm spin}\rangle \langle {\rm f}_\nu | \Sigma_\pm^\alpha | {\rm i}_\nu \rangle
\end{aligned}
\end{equation}
are the excitation and de-excitation amplitudes of the $N$-spin ensemble from an initial state $|{\rm i}_{\rm spin} \rangle$ to an unspecified final state $|{\rm f}_{\rm spin} \rangle$. The summation over ${\rm f}_{\rm spin}$ encompasses a complete basis of  final states; the ladder operators will project out the final states that have one unit of energy $\omega_0$ more ($+$) or less ($-$) than the initial state.

Again, focussing first on spin excitation due to inelastic scattering of a plane-wave neutrino from $\nu_j(t,\vec{x}_\alpha) = u_{j,a}(\vec{p})  e^{-{\rm i} (p^0t-\vec{p}\cdot\vec{x}_\alpha)}$ to $\nu_k(t,\vec{x}_\beta) = u_{k,c}(\vec{p}\,')  e^{-{\rm i} (p'^0t-\vec{p}\,'\cdot\vec{x}_\beta)}$, we find
\begin{equation}
\begin{aligned}
 |T^{+}_{\rm fi} |^2 = &\;  \sum_{\alpha \beta} \langle {\rm f}_{\rm spin}|  J_+^\alpha |{\rm i}_{\rm spin}\rangle
 \langle {\rm i}_{\rm spin}|  J_-^\beta |{\rm f}_{\rm spin} \rangle \\
 &\qquad \times \frac{G_F^2}{2} g_A^{j,k} g_A^{k,j}  \frac{1}{2p^{0}2p'^{0}} \left| \bar{u}_{k,c}(\vec{p}\,') (\gamma^1 - {\rm i} \gamma^2)(1- \gamma^5) u_{j,a}(\vec{p})\right|^2\, e^{-{\rm i}(\vec{p}\,' -\vec{p})\cdot(\vec{x}_\alpha-\vec{x}_\beta)} \, .
    \end{aligned}
    \end{equation}
A similar expression can be found for $|T^{-}_{\rm fi}|^2$.
Then, following the same procedure as for a single spin, it is straightforward to arrive at the collective excitation and  de-excitation rates
\begin{equation}
\begin{aligned}
\Gamma_{+}
    =&\;\frac{2G_F^2}{(2 \pi)^3} \sum_{jk} g_A^{j,k} g_A^{k,j}  
    \int_{\sqrt{(\omega_0+m_k)^2-m_j^2}}^\infty  {\rm d} |\vec{p}| \; |\vec{p}|^2\, p_{k+}\, \sqrt{m_k^2+p^2_{k+}} \,  
    \left[(1-N_{k}(p^0_{k+})) \, N_{j}(p^0)  +(1-\bar{N}_{k}(p^0_{k+}))\, \bar{N}_{j}(p^0)\right]\\
    &\; \hspace{30mm} \times\int_{-1}^1 {\rm d} \mu \; F_+(\vec{p}_{k+}-\vec{p})\, , \\
\Gamma_{-}= &\;\frac{2G_F^2}{(2 \pi)^3} \sum_{jk} g_A^{j,k} g_A^{k,j}  
    \int_{\sqrt{(\omega_0-m_k)^2-m_j^2}}^\infty  {\rm d} |\vec{p}|\; |\vec{p}|^2\, p_{k-}\, \sqrt{m_k^2+p^2_{k-}} \,  
    \left[(1-N_{k}(p^0_{k-})) \, N_{j}(p^0)  +(1-\bar{N}_{k}(p^0_{k-}))\, \bar{N}_{j}(p^0)\right]\\
    &\; \hspace{30mm} \times\int_{-1}^1 {\rm d} \mu \; F_-(\vec{p}_{k-}-\vec{p})\, ,
    \label{eq:BigGamma}
    \end{aligned}
\end{equation}
where $\mu \equiv \hat{p}' \cdot \hat{p}=\cos\theta_{\vec{p}\vec{p}\,'}$, with $\theta_{\vec{p}\vec{p}\,'}$ the angle between the incoming and outgoing momenta, and
\begin{equation}
F_\pm(\vec{p}\,'-\vec{p}) \equiv \sum_{\rm f_{spin}}\Big|\langle {\rm f}_{\rm spin}| \sum_\alpha  J_\pm^\alpha e^{-{\rm i} (\vec{p}\,' - \vec{p}) \cdot\vec{x}_\alpha}|{\rm i}_{\rm spin}\rangle \Big|^2
\label{eq:formfactors}
\end{equation}
are form factors that depend on the initial state of the whole $N$ spin ensemble.
 Since the operator $ \sum_\alpha  J_\pm^\alpha$ already projects the initial state into the subspace of final states that has the correct energy requirements,  we can use the completeness relation of all possible final states $\sum_{\rm f'_{spin}}|{\rm f'}_{\rm spin}\rangle\langle {\rm f'}_{\rm spin}|=1$ to simplify the form factors to
\begin{equation}
 F_\pm(\vec{p}\,'-\vec{p}) = \langle {\rm i}_{\rm spin}| \sum_\beta  J_\mp^\beta e^{+{\rm i} (\vec{p}\,' - \vec{p})\cdot \vec{x}_\beta} \sum_\alpha  J_\pm^\alpha e^{-{\rm i} (\vec{p}\,' - \vec{p}) \cdot \vec{x}_\alpha}|{\rm i}_{\rm spin}\rangle \,,
\label{eq:formfactorsAA}
\end{equation}
where it is understood that $F_\pm(\vec{p}\,'-\vec{p})$ are to be evaluated for $\vec{p}\,'=\vec{p}_{k+}$ or $\vec{p}\,'=\vec{p}_{k-}$.

Suppose the spin ensemble is initially in the ground state, i.e., $|{\rm i}_{\rm spin} \rangle = |\downarrow_1,\downarrow_2,\cdots,\downarrow_\gamma, \cdots, \downarrow_N \rangle$.   We see immediately that the de-excitation form factor must always yield $F^G (\vec{p}\,'-\vec{p})=0$, while the excitation form factor evaluates to
\begin{equation}
 F^G_+(\vec{p}\,'-\vec{p}) = \langle\downarrow_1,\downarrow_2, \cdots, \downarrow_N | \sum_\beta  J_-^\beta e^{+{\rm i} (\vec{p}\,' - \vec{p})\cdot \vec{x}_\beta} \sum_\alpha  J_+^\alpha e^{-{\rm i} (\vec{p}\,' - \vec{p})\cdot \vec{x}_\alpha}|\downarrow_1,\downarrow_2, \cdots, \downarrow_N \rangle=N\, ,
\end{equation}
irrespective of the momentum transfer. Thus, in this case there is no coherent effect from the neutrino-spin interaction.  If, on the other hand, the spin ensemble is initially in the product state $|{\rm i}_{\rm spin} \rangle = |\rightarrow_1, \rightarrow_2, \cdots, \rightarrow_\gamma, \cdots, \rightarrow_N \rangle$, where $|\rightarrow \rangle \equiv (|\uparrow \rangle + |\downarrow \rangle)/\sqrt{2}$, then the excitation and de-excitation form factors evaluate to
\begin{equation}
\begin{aligned}
 F^P_\pm(\vec{p}\,'-\vec{p}) & = \langle \rightarrow_1,\rightarrow_2, \cdots, \rightarrow_N | \sum_\beta  J_\mp^\beta e^{+{\rm i} (\vec{p}\,' - \vec{p})\cdot \vec{x}_\beta} \sum_\alpha  J_\pm^\alpha e^{-{\rm i} (\vec{p}\,' - \vec{p})\cdot \vec{x}_\alpha} |\rightarrow_1,\rightarrow_2, \cdots, \rightarrow_N \rangle\\
 & =\frac{N}{2}+\frac{1}{4}\sum_{\alpha, \beta \neq \alpha} e^{-{\rm i} (\vec{p}\,' - \vec{p})\cdot (\vec{x}_{\alpha}-\vec{x}_\beta)}\, ,
\end{aligned}
\end{equation}
which features possible coherent effects in the second term.  When the spatial separation between spins $|\vec{x}_{\alpha}-\vec{x}_\beta|$ is much smaller than $V^{1/3}$, where $V$ is the volume of the sample, we can take the continuum limit and, for a constant spin density $n_s=N/V$, approximate the last expression as
\begin{equation}
\begin{aligned}
F^P_\pm(\vec{p}\,'-\vec{p}) &= \frac{N}{2} +\frac{N^2}{4 V^2} \left| \int {\rm d}^3 x \; e^{{\rm i}(\vec{p}\,'-\vec{p})\cdot\vec{x}} \right|^2 \\
&= \frac{N}{2} + \frac{N^2}{4} \frac{9}{|\vec{p}\,'-\vec{p}|^2 R^2} \left|j_1\left(|\vec{p}\,'-\vec{p}| R\right) \right|^2\, ,
\label{eq:formP}
\end{aligned}
\end{equation}
where $j_1(x)$ is the spherical Bessel function of order one, and we have assumed at the second equality the spin sample to be a sphere of radius $R$.  Clearly, the second, $N^2$ term dominates if $|\vec{p}-\vec{p}\,'|R\ll1$,  with $|j_1(x)|^2/x^2 \to 1/9$ as $x \to 0$.
In this limit, the system exhibits coherent behaviours. 

Given the typical momentum of the neutrino bath $\vec{p}_\nu$, in order for the coherence condition to hold over the whole length $R$ of the spin sample, we must demand $|\vec{p}_\nu-\vec{p}\,'(\vec{p}_\nu)|R\ll1$. Assuming for simplicity $m_i=m_j=m$, the typical momentum transfers during excitation and de-excitation are 
\begin{equation}
    |\vec{p}_\nu-\vec{p}\,'(\vec{p}_\nu)|^2_\pm=p_\nu^2+(\sqrt{p_\nu^2+m^2}\mp\omega_0)^2-m^2-2p_\nu\left[(\sqrt{p_\nu^2+m^2}\mp\omega_0)^2-m^2\right]^{1/2}\cos{\theta_{\vec{p}\vec{p}\,'}}\,,
    \label{eq:co}
\end{equation}
with $p_{\nu} \equiv|\vec{p}_{\nu}|$.
 If we want coherence over the whole object, we must demand that $|\vec{p}_\nu-\vec{p}\,'(\vec{p}_\nu)|^2 \ll 1/R^2$ even for $\cos \theta_{\vec{p}\vec{p}\,'} = -1$.  
For non-relativistic neutrinos, where $\sqrt{p_\nu^2+m^2} \approx m + p_\nu^2/(2 m)$, 
this yields the condition
\begin{equation}
    p_\nu^2+(m+\epsilon_\pm)^2-m^2+2p_\nu\sqrt{(m+\epsilon_\pm)^2-m^2}\ll1/R^2\,,
\end{equation}
or, equivalently,
\begin{equation}
    p_\nu+\sqrt{(m+\epsilon_\pm)^2-m^2}\ll1/R\,,
    \label{eq:co2}
\end{equation}
with $\epsilon_\pm \equiv p_\nu^2/(2m) \mp \omega_0$.  This in turn leads to
\begin{equation}
    p_\nu \left[1+ \sqrt{1\mp \omega_0 \frac{2 m}{p_\nu^2}} \right]\ll1/R\,,
    \label{eq:co3}
\end{equation}
when expanded to linear order in $\epsilon_\pm$.

From Eq.~\eqref{eq:co3}, we see that excitation can only happen if $\omega_0< p_\nu^2/(2m)$.  But, if excitation happens at all, the coherence condition is also automatically fulfilled for any incoming momentum satisfying $p_\nu R\ll 1$. 
On the other hand, while de-excitation has no kinematic threshold, coherence also does not come for free. Rather, in order to have coherent de-excitation, the condition  $\omega_0\ll p_\nu^2/(2m)$ must be satisfied if $p_\nu R \ll 1$.  Physically, this means that the energy splitting must be much smaller than the incoming neutrino kinetic energy. If this is not the case, the momentum transfer of the neutrino is large and coherence is lost.    For a typical C$\nu$B momentum of $p_\nu \sim 5.3 \times 10^{-4}~{\rm eV} \sim 27~{\rm cm}^{-1}$ and a neutrino mass of $m=0.05$~eV, this condition amounts to demanding that $\omega_0 \ll 3 \times 10^{-6}$~eV. 

For typical spin samples of size $R \sim 1$~cm, the condition $p_\nu R \ll 1$ cannot be realised by the C$\nu$B across the whole sample volume. However, even in the case of $p_\nu R \gg1$, coherence is still possible under limited conditions.  Specifically, assuming $\omega_0\ll p_\nu^2/(2m)$, Eq.~\eqref{eq:co} evaluates to 
\begin{equation}
    |\vec{p}_\nu-\vec{p}\,'(\vec{p}_\nu)|^2\approx4p_\nu^2\sin^2{\frac{\theta_{\vec{p}\vec{p}\,'}}{2}}\,.
\end{equation}
Demanding $|\vec{p}_\nu-\vec{p}\,'(\vec{p}_\nu)|^2R^2 \ll 1$, we see that 
 coherence can be achieved even in the case $p_\nu R \gg 1$ for small angles $\theta_{\vec{p}\vec{p}\,'}\ll1/(p_\nu R)$ between the incoming and outgoing neutrino momenta. Under this condition, we can approximate the coherent part of the form factors as
\begin{equation}
\int_{-1}^1 {\rm d} \mu\; F_\pm^P(\vec{p}\,'-\vec{p}) \approx \frac{N^2}{4} \int_0^{(p_\nu R)^{-1}} {\rm d} \theta_{\vec{p}\vec{p}\,'}\, \sin\theta_{\vec{p}\vec{p}\,'} \approx \frac{N^2}{4} \frac{1}{2(p_\nu R)^{2}} \, .
\end{equation}
In other words, the collective excitation and de-excitation rates $\Gamma_\pm$ in the $p_\nu R \gg 1$ regime are suppressed by a factor $(2 p_\nu R)^{-2}$ relative to the fully coherent rates.  Evaluating Eq.~\eqref{eq:co} for the momentum transfer squared $q^2 \equiv|\vec{p}_\nu-\vec{p}\,'(\vec{p}_\nu)|^2$ in the forward direction (i.e., $\cos{\theta_{\vec{p}\vec{p}\,'}} = 1$) and expanding in $\epsilon_\pm$ as above, it is straightforward to show that the momentum transfer in this regime is $q \sim \omega_0 m/p_\nu$.


\section{Derivation of the Lindblad equation}\label{app:derlindblad}

The Lindblad formalism assumes unitary dynamics of the total neutrino-spin system described by a Hamiltonian that can be decomposed into three parts,
\begin{equation}
    H=H_S+H_B+H_I\, .
\end{equation}
Here, $H_S$ is the free Hamiltonian of the spin ensemble, $H_B$ the free Hamiltonian of the neutrino bath, and $H_I$ specifies the interaction between the spin ensemble and the neutrino bath. The density matrix $\rho$ acts on the Hilbert space of the whole neutrino-spin system, and we assume it can be split into a direct product of the spin-system density matrix $\rho_S$ and the bath density matrix $\rho_B$, i.e.,
\begin{equation}
    \rho=\rho_S\otimes\rho_B\, .
\end{equation}
Such a separation can be made in the limit of weak system-bath interactions and large baths (relative to the spin ensemble), where no significant correlations develop between the spin ensemble and the bath. This is the so-called Born approximation.

The dynamics of the total neutrino-spin system is governed by the Liouville-von-Neumann equation,
\begin{equation}
    \frac{\mathrm{d} \rho}{\mathrm{d}t}=-{\rm i}\, [H,\rho]\, ,
\end{equation}
where the density matrix $\rho$ is understood to be in the Schr\"{o}dinger picture.
In the interaction picture this equation can be equivalently written as
\begin{equation}
    \frac{\mathrm{d} \rho^I}{\mathrm{d}t}=-{\rm i}[H^I_I,\rho^I]\, ,
    \label{vonneuman}
\end{equation}
where $H^I_I$ is given by Eq.~\eqref{hint}, and has the formal solution 
\begin{equation}
    \rho^I(t)=\rho^I(0)-{\rm i}\int_0^t[H^I_I(s),\rho^I(s)]\, {\rm d}s\, .
    \label{sol}
\end{equation}
Plugging \eqref{sol} into \eqref{vonneuman} and doing a partial trace over the bath, we arrive at an equation of motion for the spin ensemble alone that reads
\begin{equation}
     \frac{\mathrm{d} \rho_S}{\mathrm{d}t}=-{\rm i} \, \text{tr}_B\{[H_I(t),\rho(0)]\}-\int_0^t\text{tr}_B\{[H_I(t),[H_I(s),\rho(s)]]\}\, {\rm d}s\, ,
     \label{sol2}
\end{equation}
where we have omitted the interaction label (superscript ``$I$'') for notational simplicity. 

Observe that,  implicit in the interaction Hamiltonian $H_{I}$, the first term is formally $\mathcal{O} (G_{F})$, while the second term is $\mathcal{O} (G_{F}^{2})$, and  repeatedly plugging Eq.~\eqref{sol} into \eqref{sol2} would yield in principle a perturbative series in $G_F$. Considering however that the interactions of interest are very weak, it suffices to iterate only once to capture the leading-order effects. Physically, the first, $\mathcal{O}(G_F)$ term in Eq.~\eqref{sol2} corresponds to the so-called  Stodolsky effect~\cite{Stodolsky:1974aq}, which can be most easily seen in the Schr\"{o}dinger picture where it manifests as a shift in the energy levels of the spin ensemble, {\it provided} that a non-zero neutrino-antineutrino asymmetry is present.  We shall therefore not be concerned with this shift and will henceforth drop the $\mathcal{O}(G_F)$ term.

To discern the role of the second, $\mathcal{O}(G_F^2)$ term in Eq.~\eqref{sol2}, we apply the Markovian approximation, which is valid under two assumptions: (i) The bath's correlation time is much shorter than the spin ensemble's relaxation time. This ensures the bath quickly loses memory of the interaction, thereby preventing information from flowing back to the spin ensemble; and (ii) the bath does not change much over time. This is the case for the relatively weak interactions under consideration.   Implementing the Markovian approximation consists formally in changing $\rho(s)\rightarrow\rho(t)$ on the right-hand side of Eq.~\eqref{sol2}, i.e., 
\begin{equation}
     \frac{\mathrm{d} \rho_S}{\mathrm{d}t}=  -\int_0^t\text{tr}_B\{[H_I(t),[H_I(s),\rho(t)]]\}\, {\rm d}s \,,
     \label{eq:redfield}
\end{equation}
such that any step taken by the spin ensemble at any time $t$ depends only on the state of the system at that time.  Equation~\eqref{eq:redfield} is the so-called Redfield equation.
Next, we define a new integration variable $s'= t-s$ and take the integration limit to $\infty$ (or, equivalently in terms of the original integration variable, we take the lower limit to $s \to - \infty$).  This approximation leads us to
\begin{equation}
     \frac{\mathrm{d} \rho_S}{\mathrm{d}t}  =-\int_0^\infty\text{tr}_B\{[H_I(t),[H_I(t-s),\rho(t)]]\} \, {\rm d}s\, ,
     \label{eq:changevariable}
\end{equation}
and corresponds to assuming that the correlations decay fast enough that the dynamics of the system remains the same irrespective of what happened at earlier times.

At this point we can simply evaluate Eq.~\eqref{eq:changevariable} for the Hamiltonian~\eqref{hint}.  A brute-force evaluation will yield an intractably long expression.  However, because the neutrino-spin interaction rate is expected to be much smaller than the Larmor frequency $\omega_0=g\mu_{N}B$ (i.e., many oscillations can occur over 
a time scale dictated by the interaction rate), we may adopt the so-called rotating-wave/secular approximation, which amounts to neglecting fast-oscillation terms of the form $e^{\pm {\rm i}\omega_0t}$ or $e^{\pm 2{\rm i}\omega_0t}$. Doing so yields 
\begin{equation}
\begin{aligned}
\frac{\mathrm{d} \rho_S}{\mathrm{d}t}  = &\,  (J_- \rho_S J_+ -J_+ J_- \rho_S ) \Gamma_{+-}^{(2)}(-\omega_0) 
 + (J_+ \rho_S J_- - \rho_S J_- J_+) \Gamma_{-+}^{(1)}(-\omega_0) 
 + (J_+ \rho_S J_- - J_- J_+ \rho_S) \Gamma_{-+}^{(2)}(\omega_0)  \\
& + (J_- \rho_S J_+ - \rho_S J_+ J_-) \Gamma_{+-}^{(1)}(\omega_0) 
 + (J_z \rho_S J_z - J_z J_z \rho_S) \Gamma_{zz}^{(2)}(0) 
 + (J_z \rho_S J_z - \rho_S J_z J_z) \Gamma_{zz}^{(1)}(0)\, ,
\label{lind}
\end{aligned}
\end{equation}
where 
\begin{equation}
\begin{aligned}
\Gamma_{\alpha \beta}^{(1)}(\omega_0)\equiv &\; \int_0^\infty e^{-{\rm i}\omega_0 s} \,\text{tr}_B\{\Sigma_\alpha(t-s)\Sigma_\beta(t)\rho_B\}\, {\rm d}s\, , \\
\Gamma_{\alpha \beta}^{(2)}(\omega_0)\equiv &\; \int_0^\infty e^{-{\rm i}\omega_0 s} \,\text{tr}_B\{\Sigma_\alpha(t)\Sigma_\beta(t-s)\rho_B\}\, {\rm d} s\, .
\end{aligned}
\end{equation}
Note that $\Sigma_\alpha$ is an operator in the interaction picture and hence depends on time.

Rearranging Eq.~\eqref{lind} to separate the hermitian and non-hermitian terms yields the Lindblad equation for our system:
\begin{equation}
\begin{aligned}
\frac{\mathrm{d} \rho_S}{\mathrm{d}t} = & \, 
-{\rm i}\, [G_{-}J_+J_-+G_{+}J_-J_++G_{z}J_zJ_z,\rho_S] \\
& + \gamma_{-}(J_- \rho_S J_+ - \frac{1}{2} \{J_+ J_-,\rho_S\} )  + \gamma_{+}(J_+ \rho_S J_- - \frac{1}{2} \{J_- J_+,\rho_S\} )
 + \gamma_{z}(J_z \rho_S J_z - \frac{1}{2} \{J_z J_z,\rho_S\} ) \, ,
\label{linblad}
\end{aligned}
\end{equation}
with
\begin{equation}
\begin{aligned}
\gamma_{+} \equiv & \; \Gamma_{+-}^{(1)}(\omega_0)+\Gamma_{+-}^2(-\omega_0)=\int_{-\infty}^{\infty}e^{{\rm i}\omega_0 s}\, \text{tr}_B\{\Sigma_+(0)\Sigma_-(s)\rho_B\}\, {\rm d}s\, , \\
\gamma_{-} \equiv  &\; \Gamma_{-+}^{(2)}(\omega_0)+\Gamma_{-+}^{(1)}(-\omega_0)=\int_{-\infty}^{\infty}e^{{\rm i}\omega_0 s}\text{tr}_B\{\Sigma_-(s)\Sigma_+(0)\rho_B\}\, {\rm d}s\, , \\
\gamma_{z}\equiv &\; \Gamma_{zz}^{(1)}(0)+\Gamma_{zz}^{(2)}(0)=\int_{-\infty}^{\infty}\text{tr}_B\{\Sigma_z(0)\Sigma_z(s)\rho_B\}\, {\rm d}s,\,
\label{eq:gammas}
\end{aligned}
\end{equation}
and
\begin{equation}
\begin{aligned}
G_{-}\equiv &\; \frac{\Gamma_{+-}^{(2)}(-\omega_0)-\Gamma_{+-}^{(1)}(\omega_0)}{2{\rm i}}\, , \\
G_{+} \equiv &\; \frac{\Gamma_{-+}^{(2)}(\omega_0)-\Gamma_{-+}^{(1)}(-\omega_0)}{2{\rm i}}\, ,\\
G_{z} \equiv &\; \frac{\Gamma_{zz}^{(2)}(0)-\Gamma_{zz}^{(1)}(0)}{2{\rm i}}\,,
\end{aligned}
\end{equation}
and we have assumed that the bath is constant in time.  Observe that the first, Hermitian term in Eq.~\eqref{linblad} commutes with $H_S$. Its form therefore remains unchanged in Schr\"{o}dinger picture and represents an energy shift in the free Hamiltonian that is suppressed by $G_F^2$ through the bath-dependent coefficients $G_{\pm,z}$. The remaining terms are non-hermitian and characterise dissipative effects in the spin ensemble due to the bath via the $\gamma_{\pm,z}$ coefficients.
Note that if the $g$-factor was negative, e.g., in the case of electrons, the Larmor frequency $\omega_0$ would be positive, which would effectively exchange $\gamma_{+}$ and $\gamma_{-}$ in the Lindblad equation.


\section{Computation of \texorpdfstring{$\gamma_{\pm},\gamma_z$}{gamma+-,gammaz}}\label{app:gammas}

Our objective now is to compute $\gamma_{+}$, $\gamma_{-}$ and $\gamma_{z}$, which requires that we calculate correlators of $\Sigma_\alpha$ that depend on the neutrino fields. To this end, we assume that neutrinos are Dirac fermions, whose expansion in modes is given by
\begin{equation}
    \nu_i(t,\vec{x})= \sum_s \int {\rm d}\tilde{p} \; \left(a^s_{{\rm i},\vec{p}}u_{i,s}(\vec{p})e^{-{\rm i}(p^0 t-\vec{p}\cdot\vec{x})}+b^{s \dag}_{i,\vec{p}}v_{i,s}(\vec{p})e^{{\rm i} (p^0t-\vec{p}\cdot\vec{x})}\right)\, ,
    \label{modes}
\end{equation}
where ${\rm d}\tilde{p}\equiv {\rm d}^3\vec{p}/((2\pi)^32p^0)$, $u_s(\vec{p})$ and $v_s(\vec{p})$ are the spinor solutions of the Dirac equation, and the creation/annihilation operators obey the anticommutation relations
\begin{equation}
   \{a^s_{i,\vec{p}},a^{r \dag}_{j,\vec{q}}\}=\{b^s_{i,\vec{p}},b^{r \dag}_{j,\vec{q}}\}=(2\pi)^32p^0\delta^{(3)}(\vec{p}-\vec{q})\delta_{rs}\delta_{ij}\,.
   \label{anticom}
\end{equation}
All other anticommutators of $a^s_{i,\vec{p}}, a^{r \dag}_{j,\vec{q}},b^s_{i,\vec{p}},b^{r \dag}_{j,\vec{q}}$ are zero.

We start with $\gamma_+$ given in Eq.~\eqref{eq:gammas}, which evaluates using~\eqref{eq:Sigma} to
\begin{equation}
\gamma_+=2 G_F\sum_{ijkl}\int_{-\infty}^{\infty} {\rm d} s\; e^{{\rm i}\omega_0 s}\, \text{tr}_B\{g_A^{i,j}\bar{\nu}_i(0,\vec{x})\; \Upsilon_-\;  P_L\nu_j(0,\vec{x})g_A^{k,l}\bar{\nu}_k(s,\vec{x})\; \Upsilon_+\; P_L\nu_l(s,\vec{x})\rho_B\}\,,
\end{equation}
where have introduced the notations $P_L  \equiv (1-\gamma^5)/2$ and $\Upsilon_\pm  \equiv \gamma^1 \pm {\rm i} \gamma^2$.
Substituting Eq.~\eqref{modes} into this expression, the non-vanishing terms read
\begin{equation}
\begin{aligned}
 \gamma_+    = &\; 2G_F \sum_{ijkl} g_A^{i,j} g_A^{k,l} 
    \sum_{abcd} \int_{-\infty}^{\infty} {\rm d} s\; e^{{\rm i}\omega_0 s} 
    \int{\rm  d} \tilde{p} \int {\rm d}\tilde{q} \int {\rm d}\tilde{p}' \int {\rm d} \tilde{q}'  \\
    &\times \Bigg[ e^{{\rm i}[(q^0-q'^0)s-(\vec{p} - \vec{p}\,' + \vec{q} - \vec{q}\,')\cdot\vec{x}]}\,
    \bar{u}_{i,a}(\vec{p}) \; \Upsilon_-\; P_L u_{j,b}(\vec{p}\,') 
    \bar{u}_{k,c}(\vec{q}) \;\Upsilon_+ \; P_L u_{l,d}(\vec{q}\,') \,
    \text{tr}_B\left\{ 
        a^{a \dag}_{i,\vec{p}} a^b_{j,\vec{p}'} 
        a^{c \dag}_{k,\vec{q}} a^{d}_{l,\vec{q}'} \rho_B 
    \right\} \\
    &\;\quad  +e^{{\rm i}[(-q^0+q'^0)s-(\vec{p} - \vec{p}\,' - \vec{q}+ \vec{q}\,')\cdot\vec{x}]}\,
    \bar{u}_{i,a}(\vec{p}) \;\Upsilon_-\; P_L u_{j,b}(\vec{p}\,') 
    \bar{v}_{k,c}(\vec{q})\; \Upsilon_+\; P_L v_{l,d}(\vec{q}\,') \,  \text{tr}_B\left\{ 
        a^{a \dag}_{i,\vec{p}} a^b_{j,\vec{p}'} 
    b^{c}_{k,\vec{q}} b^{d  \dag}_{l,\vec{q}'} \rho_B 
    \right\} \\
    &\; \quad  +e^{{\rm i}[(-q^0-q'^0)s-(\vec{p} + \vec{p}\,' - \vec{q} - \vec{q}\,)\cdot\vec{x}]}\,
    \bar{u}_{i,a}(\vec{p}) \; \Upsilon_- \;P_L v_{j,b}(\vec{p}\,') 
    \bar{v}_{k,c}(\vec{q}) \; \Upsilon_+\;P_L u_{l,d}(\vec{q}\,') \,  \text{tr}_B\left\{ 
        a^{a \dag}_{i,\vec{p}} b^{b \dag}_{j,\vec{p}'} 
    b^{c}_{k,\vec{q}} a^{d }_{l,\vec{q}'} \rho_B 
    \right\} \\
    &\;\quad  +e^{{\rm i} [(q^0+q'^0)s-(-\vec{p} -\vec{p}\,' + \vec{q} + \vec{q}\,')\cdot\vec{x}]}\,
    \bar{v}_{i,a}(\vec{p}) \; \Upsilon_- \;P_L u_{j,b}(\vec{p}\,') 
    \bar{u}_{k,c}(\vec{q}) \; \Upsilon_+\;P_L v_{l,d}(\vec{q}\,')\, \text{tr}_B\left\{ 
        b^{a}_{i,\vec{p}} a^{b }_{j,\vec{p}'} 
    a^{c \dag}_{k,\vec{q}} b^{d \dag}_{l,\vec{q}'} \rho_B 
    \right\} \\
    &\; \quad +e^{{\rm i}[(q^0-q'^0)s-(-\vec{p} + \vec{p}\,' + \vec{q} - \vec{q}\,)\cdot\vec{x}]}\,
    \bar{v}_{i,a}(\vec{p}) \; \Upsilon_- \;P_Lv_{j,b}(\vec{p}\,') 
    \bar{v}_{k,c}(\vec{q}) \; \Upsilon_+\;P_Lv_{l,d}(\vec{q}\,') \,  \text{tr}_B\left\{  b^{a}_{i,\vec{p}} b^{b \dag }_{j,\vec{p}'} 
    a^{c \dag}_{k,\vec{q}} a^{d }_{l,\vec{q}'} \rho_B 
    \right\} \\
    &\;\quad  +e^{{\rm i}[(-q^0+q'^0)s-(-\vec{p} + \vec{p}\,' - \vec{q} + \vec{q})\cdot\vec{x})]}\,
    \bar{v}_{i,a}(\vec{p}) \; \Upsilon_- \;P_L v_{j,b}(\vec{p}\,') 
    \bar{v}_{k,c}(\vec{q}) \; \Upsilon_+\;P_L v_{l,d}(\vec{q}\,') \, \text{tr}_B\left\{ 
        b^{a}_{i,\vec{p}} b^{b \dag }_{j,\vec{p}'} 
    b^{c}_{k,\vec{q}} b^{d \dag }_{l,\vec{q}'} \rho_B 
    \right\} \Bigg].
    \label{exp1}
\end{aligned}
\end{equation}
To compute the thermal traces of the bath, we assume that each neutrino species assumes a thermal distribution, such that the background density matrix is given by
\begin{equation}
    \rho_B=\frac{e^{-\beta H_B}}{\text{tr}_B\left\{ e^{-\beta H_B}\right\}}\, ,
\end{equation}
where $T=1/\beta$ is the neutrino temperature,  and the bath Hamilton can be expanded as
\begin{equation}
    H_B= \sum_i \sum_s \int {\rm d} \tilde{p}\; p^0\, (a^{s \dag}_{i,\vec{p}} a^s_{i,\vec{p}} + b^{s \dag}_{i,\vec{p}} b^s_{i,\vec{p}})\, .
\end{equation}
Using this information and the anticommutation relations~\eqref{anticom}, the traces evaluate to 
\begin{equation}
\begin{aligned}
 & \text{tr}_B\left\{ 
        a^{a \dag}_{i,\vec{p}} a^b_{j,\vec{p}'} 
        a^{c \dag}_{k,\vec{q}} a^{d}_{l,\vec{q}'} \rho_B 
    \right\}=\delta_{pp'}\delta_{qq'}N_{i,a}(p^0)N_{k,c}(q^0)+\delta_{q'p}\delta_{p'q}(1-N_{k,c}(q^0))N_{i,a}(p^0)\,, \\
    & \text{tr}_B\left\{ 
        a^{a \dag}_{i,\vec{p}} a^b_{j,\vec{p}'} 
    b^{c}_{k,\vec{q}} b^{d  \dag}_{l,\vec{q}'} \rho_B 
    \right\}=\delta_{pp'}\delta_{qq'}N_{i,a}(p^0)(1-\bar{N}_{k,c}(q^0))\, , \\
    &\text{tr}_B\left\{ 
        a^{a \dag}_{i,\vec{p}} b^{b \dag}_{j,\vec{p}'}
    b^{c}_{k,\vec{q}} a^{d }_{l,\vec{q}'} \rho_B 
    \right\}=\delta_{pq'}\delta_{p'q}N_{i,a}(p^0)\bar{N}_{k,c}(q^0)\,, \\
    &  \text{tr}_B\left\{ 
        b^{a}_{i,\vec{p}} a^{b }_{j,\vec{p}'} 
    a^{c \dag}_{k,\vec{q}} b^{d \dag}_{l,\vec{q}'} \rho_B 
    \right\}=\delta_{pq'}\delta_{p'q}(1-N_{k,c}(q^0))(1-\bar{N}_{i,a}(p^0))\,, \\
    &\text{tr}_B\left\{ 
        b^{a}_{i,\vec{p}} b^{b \dag }_{j,\vec{p}'} 
    a^{c \dag}_{k,\vec{q}} a^{d }_{l,\vec{q}'} \rho_B 
    \right\}=\delta_{pp'}\delta_{qq'}N_{k,c}(q^0)(1-\bar{N}_{i,a}(p^0))\, , \\
    & \text{tr}_B\left\{ 
        b^{a}_{i,\vec{p}} b^{b \dag }_{j,\vec{p}'} 
    b^{c}_{k,\vec{q}} b^{d \dag }_{l,\vec{q}'} \rho_B 
    \right\}=\delta_{pp'}\delta_{qq'}(1-\bar{N}_{i,a}(p^0))(1-\bar{N}_{k,c}(q^0))+\delta_{qp'}\delta_{pq'}(1-\bar{N}_{i,a}(p^0))\bar{N}_{k,c}(q^0)\,,
\end{aligned}
\end{equation}
where we have introduced the notation $\delta_{pq} \equiv (2\pi)^3 2p^0\delta^{(3)}(\vec{p}-\vec{q})\delta_{rs}\delta_{ij}$, and $N_{i,a}(p^0)$ ($\bar{N}_{i,a}(p^0)$) represents the occupation number of the $i$th neutrino (antineutrino) mass eigenstate with helicity~$a$.

Substituting the thermal traces into Eq.~\eqref{exp1} and integrating over $s$, we find 
\begin{equation}
 \begin{aligned}
\gamma_{+}& =  G_F^2\pi \sum_{ik} g_A^{i,k} g_A^{k,i} 
    \sum_{ac}  
    \int {\rm d}\tilde{p} \int {\rm d} \tilde{q}  \\
    & \;\times \Bigg[\delta^{(1)}(\omega_0+q^0-p^0)
    \bar{u}_{i,a}(\vec{p})  (\gamma^1- {\rm i} \gamma^2) (1-\gamma^5) u_{k,c}(\vec{q}) 
    \bar{u}_{k,c}(\vec{q}) (\gamma^1+ {\rm i} \gamma^2)(1-\gamma^5) u_{i,a}(\vec{p}) (1- N_k(q^0))N_i(p^0) \\
    &\quad \;+\delta^{(1)}(\omega_0-q^0-p^0)
    \bar{u}_{i,a}(\vec{p})  (\gamma^1- {\rm i} \gamma^2)(1-\gamma^5) v_{k,c}(\vec{q}) 
    \bar{v}_{k,c}(\vec{q}) (\gamma^1+ {\rm i} \gamma^2)(1-\gamma^5) u_{i,a}(\vec{p})N_i(p^0)\bar{N}_k(q^0)\\
    & \quad\; +\delta^{(1)}(\omega_0-q^0+p^0)
    \bar{v}_{i,a}(\vec{p})  (\gamma^1- {\rm i} \gamma^2)
    (1-\gamma^5) v_{k,c}(\vec{q}) 
    \bar{v}_{k,c}(\vec{q}) (\gamma^1+ {\rm i} \gamma^2)(1-\gamma^5) v_{i,a}(\vec{p}) (1- \bar{N}_i(p^0))\bar{N}_k(q^0) \Bigg]\,,
    \label{exp2}
\end{aligned}
\end{equation}
where we have assumed the neutrino occupation numbers to be independent of helicity. Compared with the single-spin excitation rate~\eqref{eq:rate1heuristic} computed in Appendix~\ref{app:rates}, we recognise immediately that the first and third terms correspond to inelastic scattering of a neutrino and an antineutrino, respectively, off a spin, thereby transferring energy to it.
The second term describes the absorption of a neutrino-antineutrino pair, which we did not include in the heuristic calculation of Appendix~\ref{app:rates}, but can also excite a spin if kinematically allowed; the rightmost diagram in Fig~\ref{fig:feynman} represents this process.

Neglecting the relative movement between the Earth and the C$\nu$B and using the properties of the spinors, Eq.~\eqref{exp2} can be further simplified to 
\begin{equation}
\begin{aligned}
\gamma_+ =\; & 8\pi G_F^2 \sum_{ik} g_A^{i,k} g_A^{k,i}  
    \int \mathrm{d} \tilde{p} \int \mathrm{d} \tilde{q} \; 
    p^0 q^0 \; \Big[ 
 \delta^{(1)}(\omega_0 + q^0 - p^0)\,\big(1 - N_k(q^0)\big) N_i(p^0) \\
&\qquad \qquad + \delta^{(1)}(\omega_0 - q^0 - p^0)\, N_i(p^0)\, \bar{N}_k(q^0) + \delta^{(1)}(\omega_0 - q^0 + p^0)\, \big(1 - \bar{N}_i(p^0)\big)\bar{N}_k(q^0) 
\Big]\,.
\end{aligned}
\label{eq:gamma_plus}
\end{equation}
Using the Dirac deltas to eliminate one integral, the final expression reads
\begin{equation}
\begin{aligned}
    \gamma_{+}=&\; \frac{4G_F^2}{(2 \pi)^3} \sum_{ik} g_A^{i,k} g_A^{k,i}  \Bigg[
    \int_{\sqrt{(\omega_0+m_k)^2-m_i^2}}^\infty  {\rm d} |\vec{p}|\; |\vec{p}|^2q_{k+}\sqrt{m_k^2+q_{k+}^2} \,  \left((1- N_{k}(q_{k+}^0))N_i(p^0)+(1- \bar{N}_k(q_{k+}^0))\bar{N}_i(p^0) \right) \\
   &\hspace{35mm} + \int_0^{\sqrt{(\omega_0-m_k)^2-m_i^2}} {\rm d} |\vec{p}|\;   |\vec{p}|^2q_{k+}\sqrt{m_k^2+q_{k+}^2} \; N_i(p^0)\bar{N}_k(q_{k+}^0) \Bigg],
    \label{exp3}
\end{aligned}
\end{equation}
with $q_{k+}=[(\sqrt{|\vec{p}|^2+m_i^2}-\omega_0)^2-m_k^2]^{1/2}$, and $q_{k+}^0=\sqrt{m_k^2+q_{k+}^2}$. Except for the appearance of the second integral, this expression is the same as the heuristic rate~\eqref{eq:gamma+heuristic} estimated earlier.
Observe also that the second integral contributes only if $m_k+m_i<\omega_0$. This is the kinematic requirement for neutrino-antineutrino pair absorption alluded to above. For typical values of the magnetic field considered in this work, $\omega_0\sim10^{-9} \rm eV$. Thus this process plays a role only for extremely small neutrino masses $m_i\lesssim10^{-9} \rm eV$.

The computations of $\gamma_{-}$ and $\gamma_z$ proceed in an analogous fashion. Contrary to $\gamma_+$, however, $\gamma_-$ describes de-excitation of a spin via 
inelastic scattering with a neutrino or an antineutrino, i.e., the spin loses energy.  In addition, a spin (up) can de-excite (to a spin-down) by emitting a neutrino-antineutrino pair if kinematically permitted. 
The final expression for $\gamma_-$ reads
\begin{equation}
\begin{aligned}
\gamma_{-}= &\; \frac{4G_F^2}{(2 \pi)^3} \sum_{ik} g_A^{i,k} g_A^{k,i}  \Bigg[
    \int_{\sqrt{(\omega_0-m_k)^2-m_i^2}}^\infty {\rm d}|\vec{p}| \; |\vec{p}|^2q_{k-}\sqrt{m_k^2+q_{k-}^2}\; \left((1- N_k(q_{k-}^0))N_i(p^0)+(1- \bar{N}_k(q_{k-}^0))\bar{N}_i(p^0)\right) \\
    &\hspace{30mm}+ \int_0^{\sqrt{(\omega_0-m_k)^2-m_i^2}} {\rm d} |\vec{p}|\; |\vec{p}|^2q_{k'}\sqrt{m_k^2+q_{k'}^2} \,  \left(1-\bar{N}_i(p^0)\right) \left(1-\bar{N}_k(q_{k'}^0)\right) \Bigg]\,,
    \label{exp4}
\end{aligned}
\end{equation}
with $q_{k-}=[(\sqrt{|\vec{p}|^2+m_i^2}+\omega_0)^2-m_k^2]^{1/2}$, $q_{k-}^0=\sqrt{m_k^2+q_{-}^2}$, $q_{k'}=[(\sqrt{|\vec{p}|^2+m_i^2}-\omega_0)^2-m_k^2]^{1/2}$, and $q_k'^0=\sqrt{m_k^2+q_k'^2}$. Again, this expression can be contrasted with the heuristic rate~\eqref{eq:gamma-heuristic}, which does not contain the second integral that contributes only if $m_k+m_i<\omega_0$.

For $\gamma_{z}$ we find
\begin{equation}
   \gamma_{z}=\frac{16G_F^2}{(2 \pi)^3} \sum_{ik} g_A^{i,k} g_A^{k,i}  \int_{\sqrt{m_k^2-m_i^2}}^\infty {\rm d} |\vec{p}|\; |\vec{p}|^2\sqrt{|\vec{p}|^2+m_i^2-m_k^2}\sqrt{m_i^2+|\vec{p}|^2}\, \left(N_i(p^0)(1-N_k(p^0))+\bar{N}_i(p^0)(1-\bar{N}_k(p^0))\right)\,.
   \label{eq:gammaz}
\end{equation}
Contrary to $\gamma_\pm$, this term does not describe energy exchange processes.  Rather, it describes the collective dephasing of the spin ensemble due to elastic scattering.

In the case of Majorana neutrinos, the main difference in the computation of $\gamma_{\pm,z}$ is that the expansion in modes of the neutrino fields is given by
\begin{equation}
    \nu_i(t,\vec{x})= \sum_s \int {\rm d}\tilde{p} \; \left(a^s_{{\rm i},\vec{p}}u_{i,s}(\vec{p})e^{-{\rm i}(p^0t-\vec{p}\cdot\vec{x})}+a^{s \dag}_{i,\vec{p}}v_{i,s}(\vec{p})e^{{\rm i}(p^0t-\vec{p}\cdot\vec{x})}\right)\, ,
\end{equation}
and the corresponding bath Hamiltonian is
\begin{equation}
    H_B= \sum_i \sum_s \int {\rm d} \tilde{p}\; p^0\, a^{s \dag}_{i,\vec{p}} a^s_{i,\vec{p}}\, .
\end{equation}
If we neglect pair absorption and emission (which, as discussed above, are normally not kinematically allowed in NMR experiments except for close-to-zero neutrino masses), the expressions for $\gamma_{\pm,z}$---Eqs.~\eqref{exp3}, \eqref{exp4}, and \eqref{eq:gammaz}---are exactly identical for  Dirac and Majorana neutrinos. The only difference between the two cases occurs in the value of the occupation number $N_{i,a}(p^0)$.  In the Dirac case, when ultra-relativistic neutrinos decouple from the early universe plasma, they are always in a left-helicity state. However, as the universe cools down and the neutrinos become non-relativistic, we would generically expect them to relax via gravitational interactions into an approximately equal number of left- and right-helicity states with an average occupation number (assuming no neutrino-antineutrino asymmetry)  given by
$N_{i,a} (p^0) = \bar{N}_{i,a}(p^0)\approx n_F(|\vec{p}|)/2$, where 
\begin{equation}
    n_F(|\vec{p}|)= \frac{1}{e^{|\vec{p}|/T+1}}
\end{equation}
is the relativistic Fermi-Dirac distribution.
On the other hand, Majorana neutrinos are their own antiparticles. Thus, relaxation into left- and right-helicity states merely turn an neutrino into an antineutrino and vice versa, 
leading to an occupation number of $N_{i,a} (p^0) = \bar{N}_{i,a}(p^0)\approx n_F(|\vec{p}|)$.


\section{Lindblad equation for \texorpdfstring{$N$}{N} spins}
\label{app:lindn}

Consider now a system comprising $N$ $1/2$-spins, where the Hilbert space is spanned by $\rho_S=\otimes^N\rho_{1/2}$, and the Hamiltonian of neutrino-spin system is given in Eq.~\eqref{eq:nham}.
The corresponding Lindblad equation can be generalised from Eq.~\eqref{linblad}. In the interaction picture (and omitting the hermitian terms) it reads 
\begin{equation}
\begin{aligned}
\frac{\mathrm{d} \rho_S}{\mathrm{d}t} =\sum_{\alpha \beta}\gamma^{\alpha \beta}_{-}(J^\alpha_- \rho_S J^\beta_+ - \frac{1}{2} \{J^\alpha_+ J^\beta_-,\rho_S\} )
 + \gamma^{\alpha \beta}_{+}(J^\alpha_+ \rho_S J^\beta_- - \frac{1}{2} \{J^\alpha_- J^\beta_+,\rho_S\} ) + \gamma^{\alpha \beta}_{z}(J^\alpha_z \rho_S J^\beta_z - \frac{1}{2} \{J^\alpha_z J^\beta_z,\rho_S\} ) \, ,
\end{aligned}
\end{equation}
with
\begin{equation}
\begin{aligned}
\gamma^{\alpha \beta}_{+}= &\; \int_{-\infty}^{\infty}e^{{\rm i}\omega_0 s}\text{tr}_B\{\Sigma^\alpha_+(0)\Sigma^\beta_-(s)\rho_B\}\, {\rm d} s \, ,\\
\gamma^{\alpha \beta}_{-}=&\; \int_{-\infty}^{\infty}e^{{\rm i}\omega_0 s}\text{tr}_B\{\Sigma^\alpha_-(s)\Sigma^\beta_+(0)\rho_B\}\, {\rm d}s\, , \\
\gamma^{\alpha \beta}_{z}=&\; \int_{-\infty}^{\infty}\text{tr}_B\{\Sigma^\alpha_3(0)\Sigma^\beta_3(s)\rho_B\}\, {\rm d}s\,,
\end{aligned}
\end{equation}
which are generalisations of the $\gamma_{\pm,z}$ coeffcients from Eq.~\eqref{eq:gammas}.

Consider first the coefficient $\gamma^{\alpha \beta}_{+}$.  Following the procedure outlined in Appendix~\ref{app:gammas}, we find
\begin{equation}
\begin{aligned}
 \gamma^{\alpha \beta}_{+} &= 2 G_F \sum_{ijkl} g_A^{i,j} g_A^{k,l} 
    \sum_{abcd} \int_{-\infty}^{\infty} {\rm d} s\, e^{{\rm i}\omega_0 s} 
    \int {\rm d}\tilde{p} \int {\rm d}\tilde{q} \int {\rm d}\tilde{p}' \int {\rm d}\tilde{q}' \\
    &\times \Bigg[ e^{{\rm i}[(q^0-q'^0)s-(\vec{p}\cdot\vec{x}_\alpha - \vec{p}\,'\cdot\vec{x}_\alpha + \vec{q}\cdot\vec{x}_\beta - \vec{q}\,'\cdot\vec{x}_\beta)]}
    \bar{u}_{i,a}(\vec{p}) \; \Upsilon_- \;P_L u_{j,b}(\vec{p}\,') 
    \bar{u}_{k,c}(\vec{q})\; \Upsilon_+\;P_L u_{l,d}(\vec{q}\,') \, \text{tr}_B\left\{ 
        a^{a \dag}_{i,\vec{p}} a^b_{j,\vec{p}'} 
        a^{c \dag}_{k,\vec{q}} a^{d}_{l,\vec{q}'} \rho_B 
    \right\} \\
    &\;  +e^{{\rm i}[(-q^0+q'^0)s-(\vec{p}\cdot\vec{x}_\alpha - \vec{p}\,'\cdot\vec{x}_\alpha - \vec{q}\cdot\vec{x}_\beta + \vec{q}\,'\cdot\vec{x}_\beta)]}
    \bar{u}_{i,a}(\vec{p}) \; \Upsilon_- \; P_L u_{j,b}(\vec{p}\,') 
    \bar{v}_{k,c}(\vec{q}) \; \Upsilon_+\;P_L v_{l,d}(\vec{q}\,') \,  \text{tr}_B\left\{ 
        a^{a \dag}_{i,\vec{p}} a^b_{j,\vec{p}'} 
    b^{c}_{k,\vec{q}} b^{d  \dag}_{l,\vec{q}'} \rho_B 
    \right\} \\
    &\;  +e^{{\rm i}[(-q^0-q'^0)s-(\vec{p}\cdot\vec{x}_\alpha + \vec{p}\,'\cdot\vec{x}_\alpha - \vec{q}\cdot\vec{x}_\beta - \vec{q}\,'\cdot\vec{x}_\beta)]}
    \bar{u}_{i,a}(\vec{p}) \; \Upsilon_- \;P_L v_{j,b}(\vec{p}\,') 
    \bar{v}_{k,c}(\vec{q}) \; \Upsilon_+\;P_L u_{l,d}(\vec{q}\,') \,  \text{tr}_B\left\{ 
        a^{a \dag}_{i,\vec{p}} b^{b \dag}_{j,\vec{p}'} 
    b^{c}_{k,\vec{q}} a^{d }_{l,\vec{q}'} \rho_B 
    \right\} \\
    &\;  +e^{{\rm i}[(q^0+q'^0)s-(-\vec{p}\cdot\vec{x}_\alpha - \vec{p}\,'\cdot\vec{x}_\alpha + \vec{q}\cdot\vec{x}_\beta + \vec{q}\,'\cdot\vec{x}_\beta)]}
    \bar{v}_{i,a}(\vec{p}) \; \Upsilon_- \;P_L u_{j,b}(\vec{p}\,') 
    \bar{u}_{k,c}(\vec{q}) \; \Upsilon_+\;P_L v_{l,d}(\vec{q}\,') \,
 \text{tr}_B\left\{ 
        b^{a}_{i,\vec{p}} a^{b }_{j,\vec{p}'} 
    a^{c \dag}_{k,\vec{q}} b^{d \dag}_{l,\vec{q}'} \rho_B 
    \right\} \\
    &\;  +e^{{\rm i}[(q^0-q'^0)s-(-\vec{p}\cdot\vec{x}_\alpha + \vec{p}\,'\cdot\vec{x}_\alpha + \vec{q}\cdot\vec{x}_\beta - \vec{q}\,'\cdot\vec{x}_\beta)]}
    \bar{v}_{i,a}(\vec{p}) \; \Upsilon_- \;P_L v_{j,b}(\vec{p}\,') 
    \bar{v}_{k,c}(\vec{q}) \; \Upsilon_+\;P_L v_{l,d}(\vec{q}\,') \, 
   \text{tr}_B\left\{ 
        b^{a}_{i,\vec{p}} b^{b \dag }_{j,\vec{p}'} 
    a^{c \dag}_{k,\vec{q}} a^{d }_{l,\vec{q}'} \rho_B 
    \right\} \\
    &\;  +e^{{\rm i}[(-q^0+q'^0)s-(-\vec{p}\cdot\vec{x}_\alpha + \vec{p}\,'\cdot\vec{x}_\alpha - \vec{q}\cdot\vec{x}_\beta + \vec{q}\,'\cdot\vec{x}_\beta)]}
    \bar{v}_{i,a}(\vec{p})\; \Upsilon_- \; P_L v_{j,b}(\vec{p}\,') 
    \bar{v}_{k,c}(\vec{q}) \; \Upsilon_+ \; P_L v_{l,d}(\vec{q}\,') \, \text{tr}_B\left\{ 
        b^{a}_{i,\vec{p}} b^{b \dag }_{j,\vec{p}'} 
    b^{c}_{k,\vec{q}} b^{d \dag }_{l,\vec{q}'} \rho_B 
    \right\} \Bigg].
\end{aligned}
\end{equation}
After evaluating the thermal traces, the expression simplifies to 
\begin{equation}
\begin{aligned}
\gamma^{\alpha \beta}_{+} =& \; 4 G_F^2\pi \sum_{ik} g_A^{i,k} g_A^{k,i} 
    \sum_{ac}  
    \int {\rm d}\tilde{p} \int {\rm d}\tilde{q}  \\
    &\times \Bigg[e^{-{\rm i}\Delta\vec{x}_{\alpha \beta}\cdot(\vec{p}-\vec{q})}\delta^{(1)}(\omega_0+q^0-p^0)
    \bar{u}_{i,a}(\vec{p}) \; \Upsilon_- \;P_L u_{k,c}(\vec{q}) 
    \bar{u}_{k,c}(\vec{q}) \; \Upsilon_+\;P_L u_{i,a}(\vec{p})(1\! -\!N_{k,c}(q^0)) N_{i,a}(p^0) \\
    & \quad\;  + e^{-{\rm i}\Delta\vec{x}_{\alpha \beta}\cdot(\vec{p}+\vec{q})} \delta^{(1)}(\omega_0-q^0-p^0)
    \bar{u}_{i,a}(\vec{p}) \; \Upsilon_- \;P_L v_{k,c}(\vec{q}) 
    \bar{v}_{k,c}(\vec{q}) \; \Upsilon_+\;P_L u_{i,a}(p)N_{i,a}(p^0)\bar{N}_{k,c}(q^0) \\
    &\quad \; +e^{{\rm i}\Delta\vec{x}_{\alpha \beta}\cdot(\vec{p}-\vec{q})} \delta^{(1)}(\omega_0-q^0+p^0)
    \bar{v}_{i,a}(\vec{p}) \; \Upsilon_- \;P_L v_{k,c}(\vec{q}) 
    \bar{v}_{k,c}(\vec{q}) \; \Upsilon_+\;P_L v_{i,a}(\vec{p}) (1- \bar{N}_{i,a}(p^0))\bar{N}_{k,c}(q^0)\Bigg],
    \label{coh}
\end{aligned}
\end{equation}
where $\Delta\vec{x}_{\alpha \beta} \equiv \vec{x}_{\alpha}-\vec{x}_{\beta}$.

From Eq.~\eqref{coh} we see immediately that in the limit  $|\vec{p}-\vec{q}|\ll|1/\Delta\vec{x}_{\alpha \beta}|$ and $|\vec{p}+\vec{q}|\ll|1/\Delta\vec{x}_{\alpha \beta}|$,
the exponentials evaluate to unity such that $\gamma^{\alpha \beta}_{+}$ loses its dependence on $\alpha$ and $\beta$, and  approximates $\gamma_{+}$ of Eq.~\eqref{exp2} computed previously for the single-spin system.  The same limiting behaviour is also seen in $\gamma_{-}^{\alpha \beta}$ and $\gamma_{z}^{\alpha \beta}$ (not shown here).  Then, in this coherent regime, the Lindblad equation reads
\begin{equation}
\frac{\mathrm{d} \rho_S}{\mathrm{d}t} = \gamma_{-}({\cal J}_- \rho_S {\cal J}_+ - \frac{1}{2} \{{\cal J}_+ {\cal J}_-,\rho_S\} ) 
+ \gamma_{+}({\cal J}_+ \rho_S {\cal J}_- - \frac{1}{2} \{{\cal J}_- {\cal J}_+,\rho_S\} )
+ \gamma_{z}({\cal J}_z \rho_S {\cal J}_z - \frac{1}{2} \{{\cal J}_z {\cal J}_z,\rho_S\} )\, ,
\end{equation}
where we have defined ${\cal J}_+ \equiv \sum_\alpha J^{\alpha}_+$, ${\cal J}_-\equiv \sum_\alpha J^{\alpha}_-$, and ${\cal J}_z \equiv \sum_\alpha J^{\alpha}_z$.
If on the other hand $|\vec{p}-\vec{q}|\gg|1/\Delta\vec{x}_{\alpha \beta}|$,
the cross (i.e., $\alpha \neq \beta$) terms in Eq.~\eqref{coh} average to zero and only the  $\alpha=\beta$ terms survive.  (The condition on $|\vec{p}+\vec{q}|$ versus $|1/\Delta\vec{x}_{\alpha \beta}|$ is not important, as neutrino-antineutrino pair absorption and emission are not kinematically allowed for neutrino masses $m_i\gtrsim10^{-9} \rm eV$ given typical values of the magnetic field $B\sim 0.1$ T.)
In this limit the Lindblad equation reads
\begin{equation}
\frac{\mathrm{d} \rho_S}{\mathrm{d}t} = 
\sum_{\alpha }\gamma_{-}(J^\alpha_- \rho_S J^\alpha_+ - \frac{1}{2} \{J^\alpha_+ J^\alpha_-,\rho_S\} ) 
 + \gamma_{+}(J^\alpha_+ \rho_S J^\alpha_- - \frac{1}{2} \{J^\alpha_- J^\alpha_+,\rho_S\} )
 + \gamma_{z}(J^\alpha_z \rho_S J^\alpha_z - \frac{1}{2} \{J^\alpha_z J^\alpha_z,\rho_S\} )\,,
\end{equation}
a form that is generally applicable for any ``local'' interaction that is not coherent over the length of the spin sample.


\section{Full numerical solutions}\label{sec:fullsol}

We demonstrate here how we find a full numerical solution of the Lindblad master equation \eqref{eq:lindnum} for the spin ensemble's density matrix $\rho_S$.   We show first the case of collective neutrino interactions only.  This effectively restricts the dynamics to the fully symmetric subspace of total angular momentum $j = N/2$, where numerical solutions can scale with $N^3$ (see Ref.~\cite{PhysRevA.98.063815} for a detailed discussion) and are hence tractable for relatively large values of $N$. We then extend the numerical treatment to include local interactions, where the dynamics can no longer be restricted to the $j = N/2$ manifold.

In the presence of coherent interactions alone, we set $\gamma_{\phi}^{\rm loc}=0$, and solve 
\begin{equation}
\begin{aligned}
    \frac{\mathrm{d} \rho_S}{\mathrm{d}t} &= \gamma_-{\cal D}_{{\cal J}_-}[\rho_S]+\gamma_+{\cal D}_{{\cal J}_+}[\rho_S]\label{eq:lind_coh}\\
    & =L_{D}[\rho_S]\, ,  
\end{aligned} 
\end{equation}
where we have introduced the Liouvillian super-operator at the second equality, expressed in the Dicke basis. We work in Liouville space by vectorising (i.e., column-stacking) the density matrix $\rho_S\to\vert\rho_S\rangle$ and using the Liouvillian super-operator
\begin{equation}
L_D = \sum_{\pm}\gamma_\pm \left( {\cal J}_\mp^\top \otimes {\cal J}_\pm - \frac{1}{2} I \otimes {\cal J}_\mp {\cal J}_\pm - \frac{1}{2} ({\cal J}_\mp {\cal J}_\pm)^\top \otimes I \right)\,,
\end{equation} 
where the action of the operators on $\rho_S$ from the left and the right becomes a Kronecker product, namely, $X \rho_S Y\to(Y^{\top}\otimes X)\vert\rho_S\rangle$.

To numerically integrate the master equation~\eqref{eq:lind_coh}, we implement standard time-integration schemes such as the Euler method, as well as  higher-order solvers such as the fourth-order Runge–Kutta (RK4) algorithm. To ensure numerical stability, especially for such explicit integration schemes,
we compute the spectral radius of the Liouvillian, i.e., the largest eigenvalue of $L_D$,  $\lambda_{\rm max}$, and set the time step $\Delta t$ such that $\Delta t \, \lambda_{\rm max} \ll 1$. This criterion ensures convergence and bounds the error at each step, particularly for stiff systems or rapidly decaying modes. In practice, we choose $\Delta t = h / \lambda_{\rm max}$ with $h\sim 0.1$ depending on the desired accuracy. We have additionally tested implicit integration methods and found no substantial improvement in convergence compared to explicit methods, provided the timestep factor $h$ is sufficiently small. Given their significantly lower computational cost, explicit schemes remain the optimal choice for efficiency in this context.

\begin{figure}
    \centering
    \includegraphics[width=0.48\linewidth]{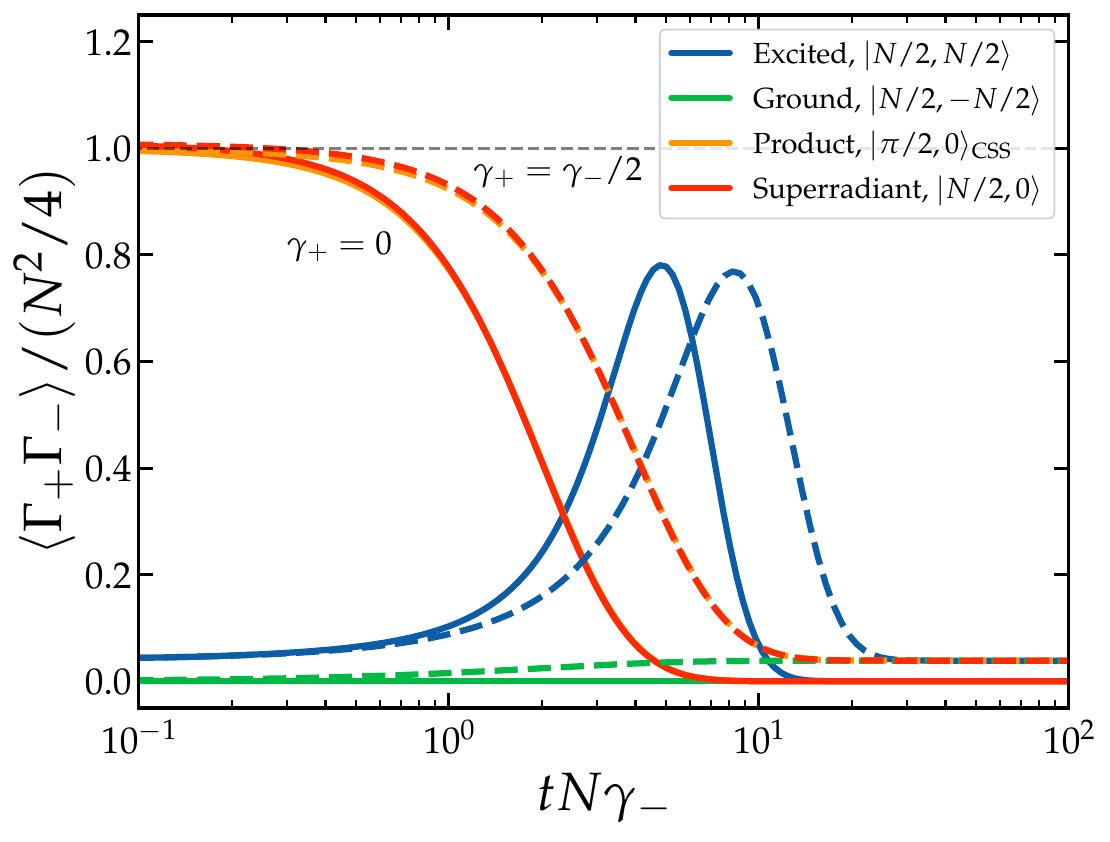}
    \includegraphics[width=0.48\linewidth]{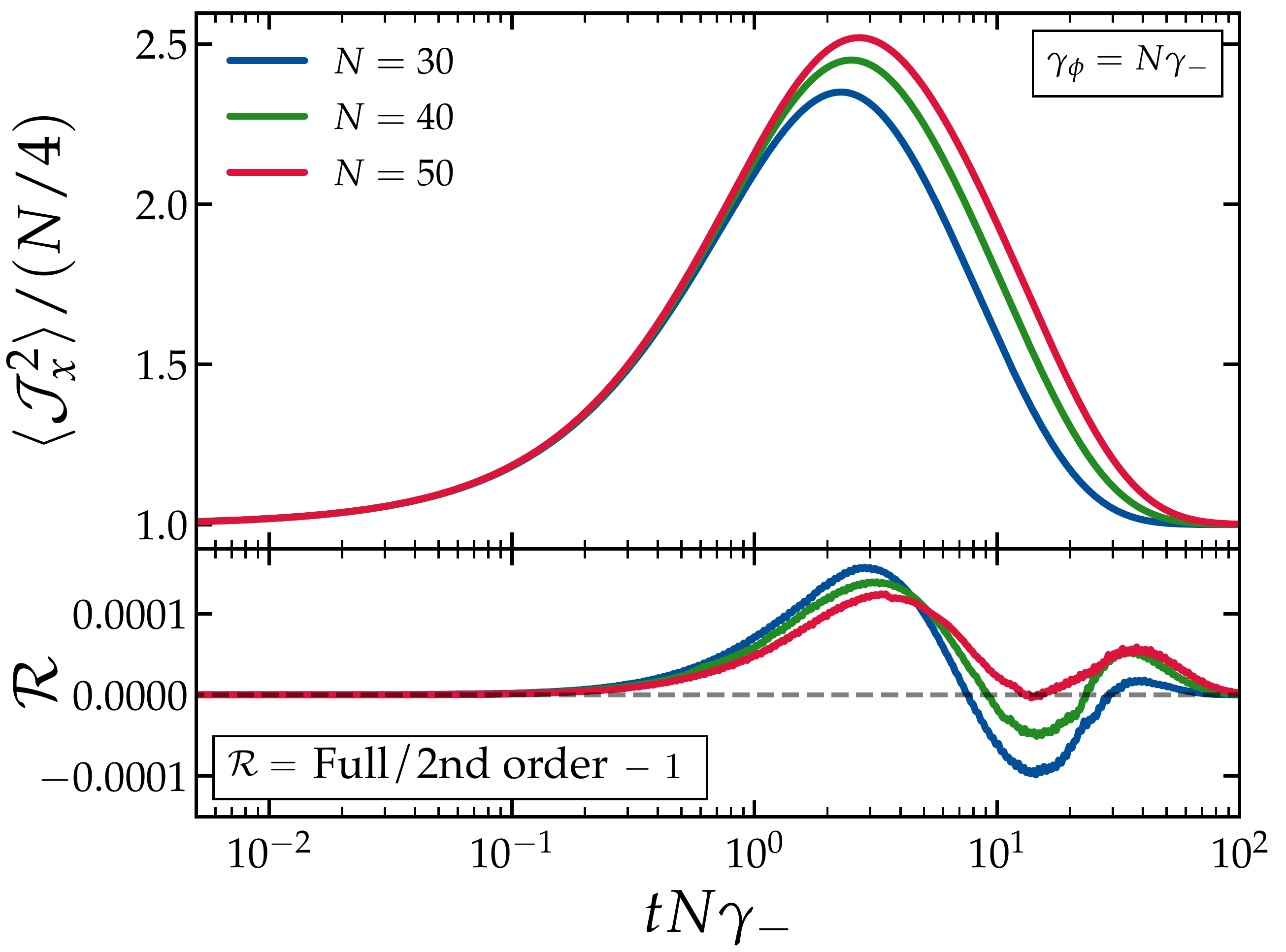}
    \caption{{\it Left}: Time evolution of the emission rate observable $\langle \Gamma_+\Gamma_-\rangle$, showing different initial conditions under the influence of coherent neutrino effects only. Solid lines represent collective de-excitation ($\gamma_+=0$), highlighting the delayed burst associated with the excited state, i.e., Dicke superradiance. These solutions can be reproduced using the public code {\tt OpenNu}~(see text). {\it Right}: Time evolution of the $\langle {\cal J}^2_x\rangle_{\vert G\rangle}$ observable for $N=30, 40, 50$, including both neutrino collective effects ($\gamma_+/\gamma_-=0.997$) and local depahsing ($\gamma_\phi=N\gamma_-$), computed using the ${\tt qutip}$ code (see text). The bottom panel shows the residual between the full solution from ${\tt qutip}$ and the second-order approximate methods, demonstrating that the two methods match at the ${\cal O}(0.01)\%$ level.}
    \label{fig:full_sols}
\end{figure}

Our simulations are implemented in the publicly available code {\tt OpenNu},\footnote{The source code is available at \url{https://github.com/gpierobon/OpenNu}.} written in {\tt C++} and based on the \texttt{Eigen}\footnote{Eigen is a high-performance {\tt C++} library for linear algebra: \url{https://eigen.tuxfamily.org}} library for efficient sparse matrix operations. This implementation achieves high computational performance and a low memory footprint, which is essential for solving large-scale Lindblad dynamics. In particular, it allows us to simulate on a single modern CPU collective spin ensembles with Hilbert space dimensions up to the equivalent of $N \sim 10^4$.  To avoid numerical instabilities due to very large or small values, we normalise time by the characteristic dissipation timescale $t_{\rm diss}$, achieved by rescaling the Liouvillian with the product of the number of spins and the de-excitation rate $\gamma_{-}$. As a result, the ratio $\gamma_+/\gamma_-$ remains as the sole control parameter governing the dynamics in this regime.  The left panel of Fig.~\ref{fig:full_sols} displays our numerical solutions obtained using ${\tt OpenNu}$ for $N=100$ and different initial conditions.

To include local interactions and fully solve Eq.~\eqref{eq:lindnum}, we use the Permutational-Invariant Quantum Solver (PIQS) introduced in Ref.~\cite{Shammah_2017} and implemented in the {\tt qutip} library.%
\footnote{The  {\tt qutip} source code is available at \href{https://github.com/qutip/qutip}{github.com/qutip/qutip}} 
PIQS also reduces the computational complexity of simulating open quantum dynamics from exponential to polynomial in $N$ by exploiting the permutational symmetry of two-level systems, under the assumption that they are identical and identically prepared. However, when the dynamics involve population outside the fully symmetric subspace, i.e., when $j<N/2$ states are included, the symmetric Dicke basis must be extended to account for the action of local operators such as $J^{\alpha}_z$ in Eq.~\eqref{eq:lindnum}. As a result,  with respect to {\tt OpenNu} the size of the Liouvillian super-operator increases by a factor of $N^2$, significantly raising the computational cost and limiting tractable simulations to system sizes of $N\sim {\cal O}(100)$. In the right panel of Fig.~\ref{fig:full_sols}, we present numerical solutions computed with ${\tt qutip}$ for values up to $N=50$, demonstrating excellent agreement between the full density matrix solution and the fast second-order approximation.


\section{Hyperpolarisation}
\label{app:hyperpolarised}

Hyperpolarised spin systems, which are out of thermal equilibrium, enhance signal intensity in NMR experiments~\cite{Eills2023SpinHyperpolarization}. A technique widely used in NMR applications is spin-exchange optical pumping (SEOP)~\cite{RevModPhys.69.629}, particularly for noble gases such as $^{129}$Xe or $^3  $He. SEOP can be broken down into two steps:

\begin{itemize}
    \item Optical pumping of alkali metal atoms. A circularly polarised laser excites alkali atoms (typically Rb or Cs vapour) in a low magnetic field. A large electron spin polarisation $p_{\rm A}\approx 1$ in the alkali ground state develops rapidly via repeated excitation-relaxation cycles and from angular momentum conservation in the atomic transition. The dynamics of this process is governed by a rate equation~\cite{Walker_2011}, 
    \begin{equation}
        \frac{{\rm d}p_{\rm A}}{{\rm d}t} = \Gamma_{\rm pump} (1 - p_{\rm A}) - \Gamma_{{\rm A, rel}} p_{\rm A}\, ,
    \end{equation} 
    where $\Gamma_{\rm pump}$ is the pumping rate, and $\Gamma_{{\rm A, rel}}$ the total alkali spin relaxation rate. The steady-state  $p_A$ is given by
    \begin{equation}
        p_{\rm A}^{\rm ss}=\frac{\Gamma_{\rm pump}}{\Gamma_{\rm pump}+\Gamma_{\rm A, rel}}\, ,
        \label{eq:PA}
    \end{equation} 
    which shows that to obtain a large $p_{\rm A}$ requires $\Gamma_{\rm pump}\gg \Gamma_{\rm A, rel}$.
    
    \item Spin-exchange collisions. The polarised alkali atoms collide with the target noble gas, e.g., Xe, atoms in a gas cell, effectively transferring angular momentum to the noble gas nuclear spins (via a Fermi-contact interaction). The polarisation of the noble gas spins $p_{\rm Xe}$ can be computed from
    \begin{equation}
        \frac{{\rm d}p_{\rm Xe}}{{\rm d}t}=\Gamma_{\rm SE}(p_{\rm A}-p_{\rm Xe})-\Gamma_{\rm rel}~ p_{\rm Xe}\, ,
    \end{equation} 
    where  $\gamma_{\rm SE}$ is the spin-exchange rate, and $\Gamma_{\rm rel}$ is the nuclear spin relaxation rate. Then,
    \begin{equation}
        p^{\rm ss}_{\rm Xe}=\frac{\Gamma_{\rm SE}p_{\rm A}^{\rm ss}}{\Gamma_{\rm SE}+\Gamma_{\rm rel}}
        \label{eq:polXe}
    \end{equation}
    gives the steady-state value of $p_{\rm Xe}$.
\end{itemize}

SEOP typically operates in dilute gas phases, where nuclear spin-spin interactions are negligible and polarisation accumulates through random spin-transfer collisions that do not generate coherences between spins. The overall mechanism is analogous to the local interaction channels described in Sec.~\ref{sec:local}. Matching  $p^{\rm ss}_{\rm Xe}$ of Eq.~\eqref{eq:polXe} to the polarisation parameter in Eq.~\eqref{eq:polhyper} then yields a mapping 
\begin{equation}
    \frac{\gamma^{\rm hyper}_-}{\gamma^{\rm hyper}_+}=\frac{\Gamma_{\rm SE}(1+p_{\rm A}^{\rm ss})+\Gamma_{\rm rel}}{\Gamma_{\rm SE}(1-p_{\rm A}^{\rm ss})+\Gamma_{\rm rel}}
\end{equation}
between the various rates to the hyperpolarisation excitation and de-excitation rates used in Sec.~\ref{sec:2ndorder}. 


\section{Two-time correlation functions} \label{sec:twotime}

We outline here the computation of the two-time correlation functions $\langle C(t) A(t+\tau) \rangle$ via the quantum regression theorem~\cite{QRT}, where $\tau \geq0$, and the operators $A$ and $C$ act on the spin ensemble alone (i.e., not the bath).  See also Ref.~\cite{carmichael1} for a more comprehensive introduction.
This correlation function is defined in the Heisenberg picture via 
\begin{equation}
\langle C(t) A(t+\tau) \rangle \equiv {\rm tr}_T \left[\rho(0) \, C(t) A(t +\tau)\right]\,,
\end{equation}
where $\rho(0)$ is the density matrix of total system of spin and bath at time $t=0$, and the trace ${\rm tr}_T$ pertains to the total system. Applying time evolution to the operators, i.e., $A(t) = e^{{\rm i}Ht} A(0) e^{-{\rm i} Ht}$,  $C(t) = e^{{\rm i}Ht} C(0) e^{-{\rm i} Ht}$, and $\rho(t) = e^{-{\rm i}Ht} \rho(0) e^{{\rm i} Ht}$ assuming a constant $H$, allows us to rewrite the expression in the Schr\"{o}dinger picture, 
\begin{equation}
\langle C(t) A(t+\tau) \rangle = {\rm tr}_T\left[ e^{-{\rm i} H \tau} \rho(t) C(0) e^{{\rm i} H\tau} A(0)\right]\, ,
\end{equation}
where the trace is now carried out over the state at time $t+\tau$, with $A(0)$ and $C(0)$ as time-independent Schr\"{o}dinger-picture operators.

Assuming the total density matrix can be split into a spin ensemble and a bath piece, i.e., $\rho(t) = \rho_S(t) \otimes \rho_B(t)$, at all times allows us to ``swap the order'' of $\rho_B(t)$ and $C(0)$ to get
\begin{equation}
\begin{aligned}
\langle C(t) A(t+\tau) \rangle 
& = {\rm tr}_T\left[ e^{-{\rm i} H \tau} [\rho_S(t) C(0) \otimes \rho_B(t) ] e^{{\rm i} H\tau} A(0)\right] \\
& = {\rm tr}_T\left[ e^{-{\rm i} H \tau} [\rho_{S;C}(t) \otimes \rho_B(t) ] e^{{\rm i} H\tau} A(0)\right]\, ,
\end{aligned}
\end{equation}
where we have defined in the last line a modified spin density matrix $\rho_{S;C}(t) \equiv \rho_S(t) C(0)$.  Writing out the total trace ${\rm tr}_T$ in terms of individual traces over the bath (subscript $B$) and spin ensemble (no subscript, for consistency with the notation of the main text), i.e., ${\rm tr}_T =  {\rm tr}_B \otimes {\rm tr}$, we arrive at
\begin{equation}
\langle C(t) A(t+\tau) \rangle = {\rm tr} \left[ \rho_{S;C}(t+\tau)\, A(0)
\right]\, ,
\end{equation}
where 
\begin{equation}
\rho_{S;C}(t+\tau) = {\rm tr}_B \left[ e^{-{\rm i} H \tau} [\rho_{S;C}(t) \otimes \rho_B(t) ] e^{{\rm i} H\tau}\right]
\label{eq:modified}
\end{equation}
is simply the modified spin matrix $\rho_{S;C}$ forwarded in time in time from $t$ to $t + \tau$ under the total Hamiltonian $H$ with the bath degrees of freedom traced out. 

Using $\tau$ as the time variable, the formal equation of motion for $\rho_{S;C}(\tau)$ is
\begin{equation}
 \frac{{\rm d}}{{\rm d} \tau}  \rho_{S;C}(t+\tau) = {\cal L}\{\rho_{S;C} (t+\tau)  \} \, ,
 \label{eq:eomcorr}
\end{equation}
where ${\cal L}$ is some super-operator,  which in our case would be the Lindbladian. The formal solution reads
$\rho_{S;C}(t+\tau) = e^{{\cal L} \tau} \{ \rho_{S;C}(t)\}$,
such that the two-time correlation  function can also be written as
\begin{equation}
\langle C(t) A(t+\tau) \rangle = {\rm tr}\left[e^{{\cal L} \tau} \{\rho_{S}(t) C(0) \} \; A(0)
\right]\, .
\end{equation}
Note that while we have assumed a time-independent $H$ in the above derivation, the equation of motion~\eqref{eq:eomcorr} applies also to a time-dependent $H$. 
The ${\tt qutip}$ library contains functions that compute numerically two-time correlation functions for a given ${\cal L}$.

We are interested to derive equations of motion directly for the two-time correlation functions $\langle C(t) A_\mu(t+\tau) \rangle$, $A_\mu$ is a set of operators indexed by $\nu= 1,2,3, \ldots$.  Suppose $\langle A_\mu \rangle$ evolves under the Lindblad master equation as 
\begin{equation}
\frac{{\rm d}}{{\rm d} t} \langle A_\mu \rangle = {\rm tr} \left[\frac{{\rm d} \rho_S}{{\rm d} t} A_\mu \right] = \sum_\alpha M_{\mu \alpha} {\rm tr} [\rho_S A_\alpha]   = \sum_\alpha M_{\mu \alpha} \langle A_\alpha \rangle\, ,
\label{eq:corr1}
\end{equation}
where $M_{\mu \alpha}$ is square matrix of constant coefficients.  It then following straightforwardly from Eqs.~\eqref{eq:eomcorr} that
\begin{equation}
\frac{{\rm d}}{{\rm d} \tau} \langle C(t) A_\mu (t+\tau) \rangle = \sum_\alpha M_{\mu \alpha} \langle C(t) A_\alpha(t+\tau) \rangle\, .
\label{eq:corr2}
\end{equation}
In other words, the ``same'' set of equations of motion describe both the single-time expectation values and the two-time correlations.

\end{document}